\documentclass[amsmath,aps,prd,amssymb, tightenlines, nofootinbib ]{revtex4-2}

\usepackage[utf8]{inputenc}
\usepackage{float}
\usepackage{graphics,graphicx}
\usepackage{dcolumn}
\usepackage{bm}
\usepackage{amsmath}
\usepackage{tensor}
\usepackage{slashed}
\usepackage{amsmath,amssymb,amsfonts,mathrsfs,bbold}
\usepackage{yfonts}
\usepackage{xcolor}
\usepackage{mathtools}
\usepackage{hyperref} 
\usepackage[framemethod=TikZ]{mdframed}
\usepackage{color}

\usepackage{multirow}
\usepackage{mathbbol}
\usepackage{mathtools}

\usepackage{slashed}
\usepackage{simpler-wick}

\usepackage[normalem]{ulem}

\usepackage{stackrel}
\usepackage[most]{tcolorbox}

\usepackage{caption}
\usepackage{subcaption}

\def\eq#1{{Eq.~(\ref{#1})}}

\def\fig#1{{Fig.~\ref{#1}}}
\newcommand{\thalf}{\tfrac{1}{2}}

\newcommand{\as}{\alpha_s}
\newcommand{\soz}{s_{10}}
\newcommand{\sto}{s_{21}}
\newcommand{\stt}{s_{32}}
\newcommand{\xoz}{x_{10}}
\newcommand{\xto}{x_{21}}
\newcommand{\xtt}{x_{32}}
\newcommand{\wint}{\int \frac{\mathrm{d}\omega}{2\pi i}}
\newcommand{\gint}{\int \frac{\mathrm{d}\gamma}{2\pi i}}

\newcommand{\dw}{\delta_{\omega}}
\newcommand{\gw}{\gamma_{\omega}}

\newcommand{\bas}{\overline{\alpha}_s}

\newcommand{\gtwg}{G_{2\omega\gamma}}

\usepackage{ragged2e} 
\DeclareCaptionJustification{justified}{\justifying}

\begin{document}

\title{Analytic Solution for the Helicity Evolution Equations at Small \texorpdfstring{$x$}{x} and Large \texorpdfstring{$N_c\&N_f$}{Nc\&Nf}}

\author{Jeremy Borden} 
         \email[Email: ]{borden.75@buckeyemail.osu.edu}
         \affiliation{Department of Physics, The Ohio State
           University, Columbus, OH 43210, USA}
\author{Yuri V. Kovchegov} 
         \email[Email: ]{kovchegov.1@osu.edu}
         \affiliation{Department of Physics, The Ohio State
           University, Columbus, OH 43210, USA}

\begin{abstract}
    We construct an exact analytic solution of the revised small-$x$ helicity evolution equations from~\cite{Borden:2024bxa}, where the contributions of the quark-to-gluon and gluon-to-quark transition operators were newly included. These evolution equations are written in the large-$N_c\&N_f$ limit and are double-logarithmic, resumming powers of $\as\ln^2(1/x)$. Here $N_c$ and $N_f$ are the numbers of quark colors and flavors, respectively, while $\as$ is the strong coupling constant and $x$ is the Bjorken-$x$ variable. Using our solution, we obtain analytic, small-$x$, large-$N_c\&N_f$ expressions for the flavor singlet quark and gluon helicity parton distribution functions (PDFs) and for the $g_1$ structure function as double-inverse Laplace transforms. We also extract analytic expressions for the eigenvalues of the matrix of polarized DGLAP anomalous dimensions and, subsequently, analytic expressions for each of the four individual polarized anomalous dimensions themselves ($\Delta \gamma_{qq}, \Delta \gamma_{qG}, \Delta \gamma_{Gq}$, and $\Delta \gamma_{GG}$): these expressions resum powers of $\as/\omega^2$ to all orders at large-$N_c\&N_f$ (with $\omega$ the Mellin moment variable). We extract the leading small-$x$ asymptotic power-law growth of the helicity distributions, given by
    \begin{align}
        \Delta\Sigma(x,Q^2) \sim \Delta G(x,Q^2)\sim g_1(x,Q^2) \sim \left(\frac{1}{x}\right)^{\alpha_h} \notag,
    \end{align}
    where the intercept $\alpha_h$ satisfies an algebraic equation. Although the algebraic complexity of the equation prevented us from obtaining a general analytic expression for $\alpha_h$, we determine $\alpha_h$ numerically for various values of $N_c$ and $N_f$ (and, for the special case of $N_f = 2 \, N_c$, we determine $\alpha_h$ analytically). We further obtain the explicit asymptotic expressions for the helicity distributions, which yield numerical values for the ratio of the gluon helicity PDF to the flavor singlet quark helicity PDF in the small-$x$ asymptotic limit (for different $N_f/N_c$). Just as for the analytic solution to small-$x$ helicity evolution equations in the large-$N_c$ limit constructed in \cite{Borden:2023ugd} and for the iterative solution to the large-$N_c\&N_f$ helicity evolution equations constructed in \cite{Borden:2024bxa}, we again find that all our predictions for polarized DGLAP anomalous dimensions are fully consistent with the existing finite-order calculations. Similar to the large-$N_c$ case \cite{Borden:2023ugd},  our intercept $\alpha_h$ (evaluated at various $N_c$ and $N_f$ values) exhibits a very slight disagreement with the predictions made within the infrared evolution equations framework in \cite{Bartels:1996wc, Blumlein:1996hb, Boussarie:2019icw}.
\end{abstract}

\maketitle

\tableofcontents


\section{Introduction}

The proton spin puzzle \cite{EuropeanMuon:1987isl, Jaffe:1989jz, Ji:1996ek, Boer:2011fh, Aidala:2012mv, Accardi:2012qut, Leader:2013jra, Aschenauer:2013woa, Aschenauer:2015eha,  Proceedings:2020eah, Ji:2020ena, AbdulKhalek:2021gbh} remains an unsolved problem in our understanding of hadronic structure and represents a fundamental test of our knowledge of Quantum Chromodynamics (QCD). We break the spin puzzle down into spin sum rules, like the Ji sum rule \cite{Ji:1996ek} or the Jaffe-Manohar sum rule \cite{Jaffe:1989jz} below,
\begin{align}\label{JMsumrule}
    S_q+L_q+S_G+L_G = \frac{1}{2}\,,
\end{align}
distinguishing spin $S$ and orbital angular momentum $L$ contributions to the proton spin coming from the quarks $q$ and gluons $G$. Each of these (spin or orbital) angular momentum contributions can be written as an integral over Bjorken $x$ of an appropriate (spin or orbital) angular momentum distribution. For the spin contributions we have
\begin{align}\label{spins}
    S_q(Q^2) = \frac{1}{2}\int\limits_0^1\mathrm{d}x \, \Delta\Sigma(x,Q^2)\,, \quad\quad S_G(Q^2) = \int\limits_0^1 \mathrm{d}x \, \Delta G(x,Q^2) \,,
\end{align}
in terms of the gluon and flavor-singlet quark helicity parton distribution functions (hPDFs), $\Delta G(x,Q^2)$ and $\Delta \Sigma(x,Q^2)$, respectively, with
\begin{align}\label{DeltaSigmageneral}
    \Delta \Sigma(x,Q^2) = \sum_{f} \left[\Delta q_f(x,Q^2) + \Delta\overline{q}(x,Q^2) \right]\,,
\end{align}
and $\Delta q_f(x,Q^2)$ and $\Delta \overline{q}_f(x,Q^2)$ the quark and antiquark helicity distributions of flavor $f$. As usual, $x$ is the parton's longitudinal momentum fraction while $Q$ is the renormalization scale.

To fully constrain $S_q$ and $S_G$, the helicity distributions $\Delta\Sigma(x,Q^2)$ and $\Delta G(x,Q^2)$ need to be known for all values of $x$, all the way down to the lower bound of the integrals in \eq{spins}. {\sl A priori}, this calls for a detailed theoretical understanding, since $x=0$ is experimentally inaccessible and any experiment is and will always be limited to $x>x_{\text{min}}$ with $x_{\text{min}}$ determined by its energy and acceptance. Furthermore, early calculations of the helicity distributions at small-$x$ done by Bartels, Ermolaev, and Ryskin (BER) \cite{Bartels:1995iu,Bartels:1996wc} in the infrared evolution equations (IREE) framework \cite{Gorshkov:1966ht,Kirschner:1983di,Kirschner:1994rq,Kirschner:1994vc,Blumlein:1996hb,Griffiths:1999dj} seemed to suggest that a substantial amount of the proton spin may reside in the small-$x$ region of phase space (see \cite{Blumlein:1995jp,Blumlein:1996hb,Ermolaev:1999jx,Ermolaev:2000sg,Ermolaev:2003zx,Ermolaev:2009cq} for more details on phenomenology developed using the BER IREE framework).

More recently, the $s$-channel/shock wave approach \cite{Mueller:1994rr,Mueller:1994jq,Mueller:1995gb,Balitsky:1995ub,Balitsky:1998ya,Kovchegov:1999yj,Kovchegov:1999ua,Jalilian-Marian:1997dw,Jalilian-Marian:1997gr,Weigert:2000gi,Iancu:2001ad,Iancu:2000hn,Ferreiro:2001qy} has emerged as a powerful tool to study the small-$x$ regime of hadronic structure and to calculate the hPDFs and the $g_1$ structure function \cite{Kovchegov:2015pbl, Hatta:2016aoc, Kovchegov:2016zex, Kovchegov:2016weo, Kovchegov:2017jxc, Kovchegov:2017lsr, Kovchegov:2018znm, Kovchegov:2019rrz, Boussarie:2019icw, Cougoulic:2019aja, Kovchegov:2020hgb, Cougoulic:2020tbc, Chirilli:2021lif, Kovchegov:2021lvz, Cougoulic:2022gbk, Borden:2023ugd, Adamiak:2023okq, Borden:2024bxa}.
To employ the shock wave formalism in the study of small-$x$ helicity, one has to go beyond the eikonal approximation \cite{Altinoluk:2014oxa,Balitsky:2015qba,Balitsky:2016dgz, Kovchegov:2017lsr, Kovchegov:2018znm, Chirilli:2018kkw, Jalilian-Marian:2018iui, Jalilian-Marian:2019kaf, Altinoluk:2020oyd, Boussarie:2020vzf, Boussarie:2020fpb, Kovchegov:2021iyc, Altinoluk:2021lvu, Kovchegov:2022kyy, Altinoluk:2022jkk, Altinoluk:2023qfr,Altinoluk:2023dww, Li:2023tlw, Altinoluk:2024dba, Altinoluk:2024tyx, Altinoluk:2025dwd, Altinoluk:2025ewj} and introduce the relevant helicity-dependent sub-eikonal (energy suppressed) corrections. Following \cite{Altinoluk:2014oxa,Balitsky:2015qba}, in \cite{Kovchegov:2015pbl, Kovchegov:2016zex, Kovchegov:2018znm, Cougoulic:2022gbk} such sub-eikonal corrections were incorporated as insertions of operators coupling to the proton helicity into the usual infinite light-cone Wilson lines of the eikonal approximation (cf. \cite{Balitsky:2016dgz, Kovchegov:2017lsr, Kovchegov:2018znm, Chirilli:2018kkw, Jalilian-Marian:2018iui, Jalilian-Marian:2019kaf, Altinoluk:2020oyd, Boussarie:2020vzf, Boussarie:2020fpb, Kovchegov:2021iyc, Altinoluk:2021lvu, Kovchegov:2022kyy, Altinoluk:2022jkk, Altinoluk:2023qfr,Altinoluk:2023dww, Li:2023tlw, Altinoluk:2024dba, Altinoluk:2024tyx, Altinoluk:2025dwd, Altinoluk:2025ewj}). The resulting objects are called `polarized Wilson lines' and, when combined in a color trace with a regular Wilson line, yield polarized dipole scattering amplitudes. The small-$x$ evolution of these polarized dipole amplitudes has been constructed in the shock wave formalism and studied extensively \cite{Kovchegov:2015pbl,  Kovchegov:2016zex,  Kovchegov:2017lsr, Kovchegov:2018znm, Chirilli:2021lif,  Cougoulic:2022gbk, Borden:2023ugd, Adamiak:2023okq, Borden:2024bxa}. The resulting evolution equations do not close in general, but instead form an infinite hierarchy (cf.~\cite{Balitsky:1995ub}). However, in the large-$N_c$ \cite{tHooft:1977xjm} and large-$N_c\&N_f$ \cite{Veneziano:1976wm} limits (where $N_c$ and $N_f$ are the number of quark colors and flavors, respectively), the infinite hierarchy is replaced with a closed set of integral equations: see \cite{Cougoulic:2022gbk} and \cite{Borden:2024bxa} for the final results for the evolution equations in each of these limits. These equations are double-logarithmic, resumming powers of $\as\ln^2(1/x)$ and $\as\ln(1/x)\ln(Q^2/\Lambda^2)$, with $\as$ the strong coupling and $\Lambda$ the infrared (IR) cutoff.

Early versions of the small-$x$ helicity evolution (KPS) \cite{Kovchegov:2015pbl, Kovchegov:2016zex, Kovchegov:2018znm} yielded full agreement in the flavor-nonsinglet sector when compared to the earlier work of BER \cite{Bartels:1995iu}, but exhibited discrepancies on the order of $30\%$ in the intercept for the flavor-singlet hPDFs \cite{Bartels:1996wc}. In \cite{Cougoulic:2022gbk} an important correction to the KPS evolution was found in the gluon sector which modified the flavor-singlet evolution (the non-singlet evolution was unaffected). The resulting evolution (KPS-CTT) is in full agreement with finite-order calculations if one compares its prediction for the small-$x$, large-$N_c$ $GG$ polarized splitting function of the Dokshitzer-Gribov-Lipatov-Altarelli-Parisi (DGLAP) evolution equations \cite{Gribov:1972ri,Altarelli:1977zs,Dokshitzer:1977sg} to that calculated in \cite{Altarelli:1977zs,Dokshitzer:1977sg,Zijlstra:1993sh,Mertig:1995ny,Moch:1999eb,vanNeerven:2000uj,Vermaseren:2005qc,Moch:2014sna,Blumlein:2021ryt,Blumlein:2021lmf,Davies:2022ofz,Blumlein:2022gpp}. Similarly, the KPS-CTT evolution equations of \cite{Cougoulic:2022gbk} were solved numerically in that same reference, and the resulting small-$x$ intercept appeared to agree numerically with the predictions of BER \cite{Bartels:1996wc}.

However, in \cite{Borden:2023ugd} an analytic solution was constructed for the large-$N_c$ helicity evolution equations of \cite{Cougoulic:2022gbk}. The (arbitrary) precision afforded by this analytic solution, along with the simple difference between the analytic expressions, revealed some numerically small discrepancies with the predictions of BER. The numerical evaluation of the analytic prediction in \cite{Borden:2023ugd} for the small-$x$ intercept $\alpha_h$ based on the large-$N_c$ KPS-CTT evolution is shown below in \eq{intro_largeNcintercepts}, along with the prediction of BER:
\begin{align}\label{intro_largeNcintercepts}
    \alpha_h^{(\text{us})} = 3.66074\sqrt{\bas}\,, \quad\quad \alpha_h^{(\text{BER})} = 3.66394\sqrt{\bas}\,,
\end{align}
with 
\begin{align}\label{alphabar}
    \bas \equiv \frac{\as N_c}{2\pi}.
\end{align}
A similarly small disagreement between KPS-CTT and BER results was found in the prediction for the $GG$ polarized DGLAP anomalous dimension, beginning at the four loop level (here $\omega$ is the Mellin moment space variable, and the expansion is in powers of $\bas/\omega^2$, capturing the leading power of $1/\omega$ as $\omega\to 0$):
\begin{align}\label{intro_largeNcanomdims}
    \Delta\gamma_{GG}^{(\text{us})\,(3)}(\omega) = \frac{496\, \bas^4}{\omega^7}\,, \quad\quad  \Delta\gamma_{GG}^{(\text{BER})\,(3)}(\omega) = \frac{504 \, \bas^4}{\omega^7}\,.
\end{align}
A possible explanation for these discrepancies was explored in Appendix A of \cite{Borden:2023ugd}.

After the solution to the large-$N_c$ evolution was constructed, a similar procedure was employed by the authors of the present paper to construct an analytic solution for the more general but more complicated large-$N_c\&N_f$ limit of the KPS-CTT evolution. However, this solution revealed irreconcilable discrepancies with the predictions of finite-order calculations \cite{Altarelli:1977zs,Dokshitzer:1977sg,Zijlstra:1993sh,Mertig:1995ny,Moch:1999eb,vanNeerven:2000uj,Vermaseren:2005qc,Moch:2014sna,Blumlein:2021ryt,Blumlein:2021lmf,Davies:2022ofz,Blumlein:2022gpp} for the polarized DGLAP anomalous dimensions starting at two loops. Similar discrepancies were observed in \cite{Adamiak:2023okq}, where it was explicitly shown that scheme dependence could not resolve the disagreements with finite-order predictions.

In \cite{Borden:2024bxa}, it was shown that a class of quark-to-gluon and gluon-to-quark transition operators (where a quark transitions to a gluon after traversing the shock wave, and vice versa) generated double-logarithmic contributions that needed to be included in the small-$x$ helicity evolution (only the large-$N_c\&N_f$ evolution was affected since the large-$N_c$ evolution contains only gluons). These operators had been extensively studied in \cite{Chirilli:2021lif}. In \cite{Borden:2024bxa}, these contributions were derived using both the `light-cone operator treatment' (LCOT) method \cite{Kovchegov:2017lsr,Kovchegov:2018znm,Cougoulic:2022gbk,Kovchegov:2021iyc} and light-cone perturbation theory (LCPT) \cite{Lepage:1980fj, Brodsky:1997de}. The result was the inclusion of a new object $\widetilde{Q}$ into helicity evolution equations; this object was shown to be closely related to the quark helicity transverse momentum-dependent (TMD) PDF. The inclusion of this new object amended several existing evolution equations and required a new equation to describe its own evolution. The resulting new set of evolution equations (KPS-CTT-BCL) derived in \cite{Borden:2024bxa} were solved iteratively in that same reference. The iterative solution yielded extractions of the polarized DGLAP splitting functions (at small-$x$ and large-$N_c\&N_f$) order-by-order in $\as$. All four splitting functions ($\Delta P_{qq}, \Delta P_{qG}, \Delta P_{Gq}$, and $\Delta P_{GG}$) agreed with the corresponding limit of the three known loops calculated in the finite-order framework \cite{Altarelli:1977zs,Dokshitzer:1977sg,Mertig:1995ny,Moch:2014sna}, some after a minor scheme transformation \cite{Borden:2024bxa}. The splitting functions extracted in \cite{Borden:2024bxa} also completely agreed with the first three loops predicted by BER \cite{Bartels:1996wc}, but all showed minor disagreements beginning at the four-loop level \cite{Blumlein:1996hb} --- the same situation as encountered in the solution of the large-$N_c$ helicity evolution equations \cite{Borden:2023ugd}. 

In this work our goal is to go beyond the iterative solution for the newly-revised set of large-$N_c\&N_f$ evolution equations of \cite{Borden:2024bxa} and to instead construct a fully analytic solution to all orders in $\as$, following the formalism of \cite{Borden:2023ugd}. Our paper is structured as follows. In Sec.~\ref{sec:equations} we state the full set of small-$x$, large-$N_c\&N_f$ helicity evolution equations that we will solve. In Sec.~\ref{sec:solution} we construct a fully analytic solution to these equations, based on the same double-inverse-Laplace transform methodology as used in \cite{Borden:2023ugd}. The results are analytic expressions, written as double-inverse-Laplace transforms, for all of the polarized dipole amplitudes that evolve under the large-$N_c\&N_f$ evolution. We summarize the full solution in Sec.~\ref{sec:solnsummary}, where we also use our solution to explicitly construct analytic expressions for $\Delta \Sigma(x,Q^2)$, $\Delta G(x,Q^2)$, and the $g_1$ structure function (at small-$x$, in the large-$N_c\&N_f$ limit). In Sec.~\ref{sec:DGLAP} we cross check our solution by comparing to DGLAP. There we obtain analytic resummed (to all orders in the double logarithmic parameter $\as/\omega^2$, with $\omega$ the Mellin/Laplace conjugate of $\ln (1/x)$) predictions for the eigenvalues of the matrix of polarized DGLAP anomalous dimensions, along with analytic resummed predictions for each of the four polarized anomalous dimensions themselves: $\Delta \gamma_{qq}, \Delta \gamma_{qG}, \Delta \gamma_{Gq}$, and $\Delta \gamma_{GG}$. We observe the same agreement with the finite-order calculations, the same agreement with BER to three loops, and the same disagreement with BER at four loops as was seen iteratively in \cite{Borden:2024bxa}.
In Sec.~\eqref{sec:asymptotics} we analyze the behavior of the helicity distributions in the small-$x$ asymptotic limit, first extracting the intercept $\alpha_h$ that controls the power law growth of the distributions at asymptotically small $x$,
\begin{align}\label{intro_asymptotics}
    \Delta \Sigma(x,Q^2) \sim \Delta G(x,Q^2) \sim g_1(x,Q^2) \sim \left(\frac{1}{x}\right)^{\alpha_h} \,,
\end{align}
and subsequently performing a more detailed analysis to obtain explicit expressions for $\Delta\Sigma$ and $\Delta G$ in the vicinity of this asymptotic limit, with the result given in Eqs.~\eqref{preasymptoticexpansion}, \eqref{expansioncoefsds}, and \eqref{hpdfssaddlepoint3}. These results enable us to numerically compute the asymptotic ratio of $\Delta G$ to $\Delta \Sigma$. In both the intercept and the asymptotic ratio $\Delta G/\Delta\Sigma$, we find small disagreements with the predictions made in the BER formalism \cite{Bartels:1996wc, Boussarie:2019icw}. We conclude in Sec.~\ref{sec:conclusion}.


\section{Large-\texorpdfstring{$N_c\&N_f$}{Nc\&Nf} Equations}\label{sec:equations}

The newly revised small-$x$ helicity evolution equations in the large-$N_c\&N_f$ limit are given in Eqs.~(76) of \cite{Borden:2024bxa}. They describe the small-$x$ evolution of the impact-parameter-integrated polarized dipole amplitudes $Q(\xoz^2,zs)$, $\widetilde{G}(\xoz^2,zs)$, and $G_2(\xoz^2,zs)$ along with a slightly different object, $\widetilde{Q}(\xoz^2,zs)$, which is related to the flavor-singlet quark helicity TMD. Operator definitions of these objects are given in \cite{Cougoulic:2022gbk, Borden:2024bxa}. Here, $x_{ij}^2 = |\underline{x}_{ij}|^2$ is the squared transverse size of the dipole consisting of partons labeled $i,j = 0,1,2,\ldots$, with transverse vectors $\underline{x}_{ij} = \underline{x}_i - \underline{x}_j$, where two-dimensional transverse coordinate space vectors are denoted by $\underline{x} = (x^1,x^2)$. The variable $z$ is a (small) longitudinal momentum fraction (which could be smaller than the individual momentum fractions of either line making up the dipole \cite{Kovchegov:2015pbl, Kovchegov:2016zex, Cougoulic:2019aja, Kovchegov:2018znm}), and $zs$ is the center of mass energy squared controlling the next step of evolution. In addition, the three proper dipole amplitudes $Q(\xoz^2,zs)$, $\widetilde{G}(\xoz^2,zs)$, and $G_2(\xoz^2,zs)$ are accompanied by the `neighbor' dipole amplitudes, denoted $\overline{\Gamma}(\xoz^2,\xto^2,zs)$, $\widetilde{\Gamma}(\xoz^2,\xto^2,zs)$, and $\Gamma_2(\xoz^2,\xto^2,zs)$, respectively. These auxiliary functions are necessary to enforce the light-cone lifetime ordering in the double logarithmic approximation (DLA) and have the same operator definitions as their non-neighbor counterparts but with different light-cone lifetime ordering constraints \cite{Kovchegov:2015pbl, Kovchegov:2016zex, Kovchegov:2017lsr, Kovchegov:2018znm, Cougoulic:2019aja, Cougoulic:2022gbk}. Note that the neighbor dipole amplitudes depend on an additional (`neighbor') transverse dipole size squared $x_{21}^2$ and are only defined for $x_{21}\leq x_{10}$. The object $\widetilde{Q}(\xoz^2,zs)$ has no such neighbor dipole amplitude (for details see the discussion at the end of Sec.~IIIB in \cite{Borden:2024bxa}). We re-state here the full set of large-$N_c\&N_f$ equations, as given in \cite{Borden:2024bxa}. Note that $\Lambda$ is explicitly taken to be an IR cutoff such that no dipole size may exceed $1/\Lambda$.
\begin{subequations}\label{evoleqs_IR}\allowdisplaybreaks
    \begin{align}
    &Q(\xoz^2,zs) = Q^{(0)}(\xoz^2,zs) + \frac{\as N_c}{2\pi}\int\limits_{1/s\xoz^2}^{z}\frac{\mathrm{d}z'}{z'} \int\limits_{1/z's}^{\xoz^2} \frac{\mathrm{d}\xto^2}{\xto^2} \bigg[2\widetilde{G}(\xto^2,z's) + 2\widetilde{\Gamma}(\xoz^2,\xto^2,z's) + Q(\xto^2,z's) \\
    &\hspace{8cm} - \overline{\Gamma}(\xoz^2,\xto^2,z's) + 2\Gamma_2(\xoz^2,\xto^2,z's) + 2G_2(\xto^2,z's) \bigg] \notag \\
    &\hspace{4cm}+ \frac{\as N_c}{4\pi} \int\limits_{\Lambda^2/s}^{z}\frac{\mathrm{d}z'}{z'}\int\limits_{1/z's}^{\text{min}\{\xoz^2\tfrac{z}{z'},\tfrac{1}{\Lambda^2} \}} \frac{\mathrm{d}\xto^2}{\xto^2}\bigg[Q(\xto^2,z's) + 2G_2(\xto^2,z's) \bigg] \notag \,,\\
    \label{Gammabar_IR} 
    &\overline{\Gamma}(\xoz^2,\xto^2,z's) = Q^{(0)}(\xoz^2,z's) + 
    \frac{\as N_c}{2\pi}\int\limits_{1/s\xoz^2}^{z'}\frac{\mathrm{d}z''}{z''} \int\limits_{1/z''s}^{\text{min}\{\xoz^2,\xto^2\tfrac{z'}{z''} \}} \frac{\mathrm{d}\xtt^2}{\xtt^2} \bigg[2\widetilde{G}(\xtt^2,z''s) + 2\widetilde{\Gamma}(\xoz^2,\xtt^2,z''s) \\
    &\hspace{6cm}+ Q(\xtt^2,z''s) - \overline{\Gamma}(\xoz^2,\xtt^2,z''s) + 2\Gamma_2(\xoz^2,\xtt^2,z''s) + 2G_2(\xtt^2,z''s) \bigg] \notag \\
    &\hspace{4.5cm}+ \frac{\as N_c}{4\pi} \int\limits_{\Lambda^2/s}^{z'}\frac{\mathrm{d}z''}{z''}  \int\limits_{1/z''s}^{\text{min}\{\xto^2\tfrac{z'}{z''},\tfrac{1}{\Lambda^2} \} }\frac{\mathrm{d}\xtt^2}{\xtt^2} \bigg[Q(\xtt^2,z''s) + 2G_2(\xtt^2,z''s) \bigg] \notag \,,\\
    \label{Gtilde_IR}
    &\widetilde{G}(\xoz^2,zs) = \widetilde{G}^{(0)}(\xoz^2,zs) + \frac{\as N_c}{2\pi}\int\limits_{1/s\xoz^2}^{z}\frac{\mathrm{d}z'}{z'} \int\limits_{1/z's}^{\xoz^2} \frac{\mathrm{d}\xto^2}{\xto^2} \bigg[3\widetilde{G}(\xto^2,z's) + \widetilde{\Gamma}(\xoz^2,\xto^2,z's) + 2G_2(\xto^2,z's) \\
    &\hspace{6cm} + \left(2-\frac{N_f}{2N_c} \right)\Gamma_2(\xoz^2,\xto^2,z's) - \frac{N_f}{4N_c}\overline{\Gamma}(\xoz^2,\xto^2,z's) - \frac{N_f}{2N_c}\widetilde{Q}(\xto^2,z's) \bigg] \notag \\
    &\hspace{4cm} - \frac{\as N_f}{8\pi} \int\limits_{\Lambda^2/s}^{z}\frac{\mathrm{d}z'}{z'} \int\limits_{\text{max}\{\xoz^2,\tfrac{1}{z's}\} }^{\text{min}\{\xoz^2\tfrac{z}{z'},\tfrac{1}{\Lambda^2}\} }\frac{\mathrm{d}\xto^2}{\xto^2} \bigg[Q(\xto^2,z's) + 2G_2(\xto^2,z's) \bigg]\notag \,,\\
    \label{Gammatilde_IR}
    &\widetilde{\Gamma}(\xoz^2,\xto^2,z's) = \widetilde{G}^{(0)}(\xoz^2,z's) + \frac{\as N_c}{2\pi} \int\limits_{1/s\xoz^2}^{z'}\frac{\mathrm{d}z''}{z''} \int\limits_{1/z''s}^{\text{min}\{\xoz^2,\xto^2\tfrac{z'}{z''}\} }\frac{\mathrm{d}\xtt^2}{\xtt^2} 
    \bigg[3\widetilde{G}(\xtt^2,z''s) + \widetilde{\Gamma}(\xoz^2,\xtt^2,z''s) \\
    &\hspace{3.5cm}+ 2G_2(\xtt^2,z''s) + \left(2-\frac{N_f}{2N_c} \right)\Gamma_2(\xoz^2,\xtt^2,z''s) - \frac{N_f}{4N_c}\overline{\Gamma}(\xoz^2,\xtt^2,z''s) -\frac{N_f}{2N_c}\widetilde{Q}(\xtt^2,z''s) \bigg] \notag \\
    &\hspace{4.5cm} - \frac{\as N_f}{8\pi} \int\limits_{\Lambda^2/s}^{z'\tfrac{\xto^2}{\xoz^2}}\frac{\mathrm{d}z''}{z''} \int\limits_{\text{max}\{\xoz^2,\tfrac{1}{z''s} \} }^{\text{min}\{\xto^2\tfrac{z'}{z''}, \tfrac{1}{\Lambda^2} \} }\frac{\mathrm{d}\xtt^2}{\xtt^2} \bigg[Q(\xtt^2,z''s) + 2G_2(\xtt^2,z''s) \bigg]\notag \,,\\
    \label{G2_IR}
    &G_2(\xoz^2,zs) = G_2^{(0)}(\xoz^2,zs) + \frac{\as N_c}{\pi} \int\limits_{\Lambda^2/s}^{z}\frac{\mathrm{d}z'}{z'} \int\limits_{\text{max}\{\xoz^2,\tfrac{1}{z's}\} }^{\text{min}\{\xoz^2\tfrac{z}{z'},\tfrac{1}{\Lambda^2}\} }\frac{\mathrm{d}\xto^2}{\xto^2} \bigg[\widetilde{G}(\xto^2,z's) + 2G_2(\xto^2,z's) \bigg]\,,\\
    \label{Gamma2_IR}
    &\Gamma_2(\xoz^2,\xto^2,z's) = G_2^{(0)}(\xoz^2,z's) + \frac{\as N_c}{\pi} \int\limits_{\Lambda^2/s}^{z'\tfrac{\xto^2}{\xoz^2}}\frac{\mathrm{d}z''}{z''}\int\limits_{\text{max}\{\xoz^2,\tfrac{1}{z''s}\} }^{\text{min}\{\xto^2\tfrac{z'}{z''},\tfrac{1}{\Lambda^2}\} }\frac{\mathrm{d}\xtt^2}{\xtt^2} \bigg[\widetilde{G}(\xtt^2,z''s) + 2G_2(\xtt^2,z''s) \bigg] \,,\\
    \label{Qtilde_IR}
    &\widetilde{Q}(\xoz^2,zs) = \widetilde{Q}^{(0)}(\xoz^2,zs) - \frac{\as N_c}{2\pi} \int\limits_{\Lambda^2/s}^z \frac{\mathrm{d}z'}{z'}\int\limits_{\text{max}\{\xoz^2,\tfrac{1}{z's}\}}^{\text{min}\{\xoz^2\tfrac{z}{z'}, \tfrac{1}{\Lambda^2} \}} \frac{\mathrm{d}\xto^2}{\xto^2} \left[Q(\xto^2,z's) + 2G_2(\xto^2,z's)   \right]\,.
\end{align}
\end{subequations}
The objects with $(0)$ in the superscript are the initial conditions for the dipole amplitudes, which are usually taken at the Born level in DLA. They enter as inhomogeneous terms in the integral equations at hand.

Introducing the rescaled variables
\begin{align}\label{rescaledvars}
   & \eta = \sqrt{\frac{\as N_c}{2\pi}}\ln\frac{zs}{\Lambda^2}\, , \qquad \eta' = \sqrt{\frac{\as N_c}{2\pi}}\ln\frac{z's}{\Lambda^2}\, , \qquad \eta'' = \sqrt{\frac{\as N_c}{2\pi}}\ln\frac{z''s}{\Lambda^2}\, , \, \\
   & \soz = \sqrt{\frac{\as N_c}{2\pi}}\ln\frac{1}{\xoz^2\Lambda^2}\, , \qquad \sto = \sqrt{\frac{\as N_c}{2\pi}}\ln\frac{1}{\xto^2\Lambda^2}\, , \qquad \stt = \sqrt{\frac{\as N_c}{2\pi}}\ln\frac{1}{\xtt^2\Lambda^2}\,, \notag
\end{align}
we can write the large-$N_c\&N_f$ equations \eqref{evoleqs_IR} as 
\begin{subequations}\label{evoleqs_IR_1}\allowdisplaybreaks
\begin{align}
    \label{Q_IR2}
    &Q(\soz,\eta) = Q^{(0)}(\soz,\eta) + \int\limits_{\soz}^{\eta}\mathrm{d}\eta' \int\limits_{\soz}^{\eta'} \mathrm{d}\sto \bigg[2\widetilde{G}(\sto,\eta') + 2\widetilde{\Gamma}(\soz,\sto,\eta') + Q(\sto,\eta') \\
    &\hspace{6.5cm} - \overline{\Gamma}(\soz,\sto,\eta') + 2\Gamma_2(\soz,\sto,\eta') + 2G_2(\sto,\eta') \bigg] \notag \\
    &\hspace{2cm}+ \frac{1}{2}\left[ \int\limits_{0}^{\soz}\mathrm{d}\sto\int\limits_{\sto}^{\eta+\sto-\soz} \mathrm{d}\eta' + \int\limits_{\soz}^{\eta}\mathrm{d}\sto\int\limits_{\sto}^{\eta}\mathrm{d}\eta'\right] \bigg[Q(\sto,\eta') + 2G_2(\sto,\eta') \bigg] \notag\,,\\
    &\overline{\Gamma}(\soz,\sto,\eta') = Q^{(0)}(\soz,\eta') + \Bigg[\ \int\limits_{\soz}^{\sto}\mathrm{d}\stt \int\limits_{\stt}^{\eta'-\sto+\stt}\mathrm{d}\eta'' + \int\limits_{\sto}^{\eta'}\mathrm{d}\stt \int\limits_{\stt}^{\eta'}\mathrm{d}\eta'' \Bigg]
    \bigg[2\widetilde{G}(\stt,\eta'') + 2\widetilde{\Gamma}(\soz,\stt,\eta'') + Q(\stt,\eta'') \notag \\
    &\hspace{8.5cm} - \overline{\Gamma}(\soz,\stt,\eta'') + 2\Gamma_2(\soz,\stt,\eta'') + 2G_2(\stt,\eta'') \bigg] \label{Gammabar_IR2}  \\
    &\hspace{2cm}+ \frac{1}{2} \Bigg[\int\limits_{0}^{\sto}\mathrm{d}\stt  \int\limits_{\stt}^{\eta'-\sto+\stt}\mathrm{d}\eta'' + \int\limits_{\sto}^{\eta'}\mathrm{d}\stt   \int\limits_{\stt}^{\eta'}\mathrm{d}\eta''
    \Bigg]\bigg[Q(\stt,\eta'') + 2G_2(\stt,\eta'') \bigg] \notag \,,\\
    \label{Gtilde_IR2}
    &\widetilde{G}(\soz,\eta) = \widetilde{G}^{(0)}(\soz,\eta) + \int\limits_{\soz}^{\eta}\mathrm{d}\eta' \int\limits_{\soz}^{\eta'} \mathrm{d}\sto \bigg[3\widetilde{G}(\sto,\eta') + \widetilde{\Gamma}(\soz,\sto,\eta') + 2G_2(\sto,\eta') \\
    &\hspace{6.5cm} + \left(2-\frac{N_f}{2N_c} \right)\Gamma_2(\soz,\sto,\eta') - \frac{N_f}{4N_c}\overline{\Gamma}(\soz,\sto,\eta') -\frac{N_f}{2N_c}\widetilde{Q}(\sto,\eta')  \bigg] \notag \\
    &\hspace{2cm} - \frac{N_f}{4N_c} \int\limits_{0}^{\soz}\mathrm{d}\sto \int\limits_{\sto}^{\eta+\sto-\soz}\mathrm{d}\eta' \bigg[Q(\sto,\eta') + 2G_2(\sto,\eta') \bigg]\notag \,,\\
    &\widetilde{\Gamma}(\soz,\sto,\eta') = \widetilde{G}^{(0)}(\soz,\eta') + \Bigg[\ \int\limits_{\soz}^{\sto}\mathrm{d}\stt \int\limits_{\stt}^{\eta'-\sto+\stt}\mathrm{d}\eta'' + \int\limits_{\sto}^{\eta'}\mathrm{d}\stt \int\limits_{\stt}^{\eta'}\mathrm{d}\eta''  \Bigg]
    \bigg[3\widetilde{G}(\stt,\eta'') + \widetilde{\Gamma}(\soz,\stt,\eta'') + 2G_2(\stt,\eta'') \notag \\
    &\hspace{5.75cm} + \left(2-\frac{N_f}{2N_c} \right)\Gamma_2(\soz,\stt,\eta'') - \frac{N_f}{4N_c}\overline{\Gamma}(\soz,\stt,\eta'') -\frac{N_f}{2N_c}\widetilde{Q}(\stt,\eta'')  \bigg]  \label{Gammatilde_IR2} \\
    &\hspace{2cm} - \frac{N_f}{4N_c} \int\limits_{0}^{\soz}\mathrm{d}\stt \int\limits_{\stt}^{\eta'-\sto+\stt}\mathrm{d}\eta'' \bigg[Q(\stt,\eta'') + 2G_2(\stt,\eta'') \bigg]\notag \,,\\
    \label{G2_IR2}
    &G_2(\soz,\eta) = G_2^{(0)}(\soz,\eta) + 2 \int\limits_{0}^{\soz}\mathrm{d}\sto \int\limits_{\sto}^{\eta+\sto-\soz}\mathrm{d}\eta' \bigg[\widetilde{G}(\sto,\eta') + 2G_2(\sto,\eta') \bigg]\,,\\
    \label{Gamma2_IR2}
    &\Gamma_2(\soz,\sto,\eta') = G_2^{(0)}(\soz,\eta') + 2 \int\limits_{0}^{\soz}\mathrm{d}\stt \int\limits_{\stt}^{\eta'-\sto+\stt}\mathrm{d}\eta'' \bigg[\widetilde{G}(\stt,\eta'') + 2G_2(\stt,\eta'') \bigg]\,,\\
    \label{Qtilde_IR2}
    &\widetilde{Q}(\soz,\eta) = \widetilde{Q}^{(0)}(\soz,\eta) - \int\limits_0^{\soz}\mathrm{d}\sto \int\limits_{\sto}^{\eta+\sto-\soz}\mathrm{d}\eta' \bigg[Q(\sto,\eta')+2G_2(\sto,\eta')\bigg]\,,
    \end{align}
\end{subequations}
where we assume the ordering $0\leq\soz\leq\sto\leq\eta'$ in Eqs.~\eqref{Gammabar_IR2}, \eqref{Gammatilde_IR2}, and \eqref{Gamma2_IR2}.

Once analytic expressions for the dipole amplitudes\footnote{Since its structure is like that of a sum of TMDs with the forward- and past pointing Wilson-line staples (i.e., with the semi-inclusive deep inelastic scattering (SIDIS) and Drell-Yan (DY) staples), $\widetilde{Q}(\xoz^2,zs)$ cannot be properly called a dipole amplitude. However, for simplicity, we will often refer to the collection of seven objects that evolve under Eqs.~\eqref{evoleqs_IR_1} as `dipole amplitudes'.} are known, one can obtain the flavor-singlet quark and gluon hPDFs using Eqs.~(77) from \cite{Borden:2024bxa}, restated below:
\begin{subequations}\label{pdfsfromdipoles}
\begin{align}
    \label{DeltaGfromdipoles}
    &\Delta G(x,Q^2) = \frac{2N_c}{\as \pi^2} \, G_2\left( \xoz^2 = \frac{1}{Q^2}, s=\frac{Q^2}{x}\right), \\
    \label{DeltaSigmafromdipoles}
    &\Delta \Sigma(x,Q^2) = \frac{N_f}{\as \pi^2} \, \widetilde{Q} \left( \xoz^2 = \frac{1}{Q^2}, s=\frac{Q^2}{x}\right).
\end{align}
\end{subequations}
Note that \eq{DeltaGfromdipoles} is consistent with previous versions of the small-$x$ helicity evolution \cite{Kovchegov:2015pbl, Kovchegov:2017lsr, Kovchegov:2018znm,Cougoulic:2022gbk,Kovchegov:2016zex}, while \eq{DeltaSigmafromdipoles} is a slight modification from the previous results and was derived in \cite{Borden:2024bxa}, ultimately representing a scheme transformation relative to the previous result. In addition, the $g_1$ structure function is given in terms of the dipole amplitudes as \cite{Kovchegov:2015pbl, Cougoulic:2022gbk}
\begin{align}\label{g1_DLA}
g_1 (x, Q^2)  = - \sum_f \frac{N_c \, Z^2_f}{4 \pi^3} \int\limits_{\Lambda^2/s}^1 \frac{dz}{z} \,  \int\limits^{\min \left\{ \frac{1}{z Q^2} , \frac{1}{\Lambda^2} \right\}}_\frac{1}{zs} \frac{d x^2_{10}}{x_{10}^2} \, \left[ Q (x_{10}^2, zs) + 2 \, G_2 (x_{10}^2, zs) \right],
\end{align}
with $Z_f$ the fractional electric charge of the quark. For simplicity, we assume here that the objects $\widetilde Q$ and $Q$, whose operator definitions include quark fields of a fixed flavor \cite{Borden:2024bxa}, are independent of the quark flavors: to bring back the flavor dependence, one needs to replace $\widetilde Q \to \widetilde Q_f$, $Q \to Q_f$, and $N_f \to \sum_f$ in Eqs.~\eqref{evoleqs_IR}, \eqref{evoleqs_IR_1}, \eqref{DeltaSigmafromdipoles}, and \eqref{g1_DLA} (cf. \cite{Adamiak:2021ppq, Adamiak:2023okq, Adamiak:2023yhz}).\footnote{While the definition of the dipole amplitude $\widetilde G$ also includes quark fields, the sum over all flavors is implied in the definition itself, making $\widetilde G$ flavor-independent.}


\section{Solution}\label{sec:solution}

\subsection{Double Inverse Laplace Representations for \texorpdfstring{$G_2\,,\,\Gamma_2, \widetilde{G}, Q,\widetilde{Q}$}{G2,Gamma2,Gtilde,Q,Qtilde}}

The solution we construct here follows very closely that constructed in \cite{Borden:2023ugd} for the large-$N_c$ evolution equations. We begin by introducing the following double-inverse Laplace transforms for $G_2(\soz^2,\eta)$, $\widetilde{G}(\soz^2,\eta)$, $Q(\soz^2,\eta)$, $\widetilde{Q}(\soz^2,\eta)$ and their initial conditions/inhomogeneous terms:
\begin{subequations}\label{doubleLaplacestart}
    \begin{align} \label{doubleLaplaceG2}
    &G_2(\soz,\eta) = \wint \gint e^{\omega(\eta-\soz)}e^{\gamma\soz}G_{2\omega\gamma}\,,
    \\ 
    \label{doubleLaplaceG20}
    &G_2^{(0)}(\soz,\eta) = \wint \gint e^{\omega(\eta-\soz)}e^{\gamma\soz}G^{(0)}_{2\omega\gamma}\,,
    \\ 
    \label{doubleLaplaceGtilde}
    &\widetilde{G}(\soz,\eta) = \wint \gint e^{\omega(\eta-\soz)}e^{\gamma\soz}\widetilde{G}_{\omega\gamma}\,,
    \\ 
    \label{doubleLaplaceGtilde0}
    &\widetilde{G}^{(0)}(\soz,\eta) = \wint \gint e^{\omega(\eta-\soz)}e^{\gamma\soz}\widetilde{G}^{(0)}_{\omega\gamma}\,,
    \\ 
    \label{doubleLaplaceQ}
    &Q(\soz,\eta) = \wint \gint e^{\omega(\eta-\soz)}e^{\gamma\soz}Q_{\omega\gamma}\,,
    \\ 
    \label{doubleLaplaceQ0}
    &Q^{(0)}(\soz,\eta) = \wint \gint e^{\omega(\eta-\soz)}e^{\gamma\soz}Q^{(0)}_{\omega\gamma}\,,
    \\ 
    \label{doubleLaplaceQtilde}
    &\widetilde{Q}(\soz,\eta) = \wint \gint e^{\omega(\eta-\soz)}e^{\gamma\soz}\widetilde{Q}_{\omega\gamma}\,,
    \\ 
    \label{doubleLaplaceQtilde0}
    &\widetilde{Q}^{(0)}(\soz,\eta) = \wint \gint e^{\omega(\eta-\soz)}e^{\gamma\soz}\widetilde{Q}^{(0)}_{\omega\gamma}.
    \end{align}
\end{subequations}
As usual, these integrals are taken along vertical contours parallel to the imaginary axes in the $\omega$ and $\gamma$ planes, with all singularities of the integrands located to the left of the contours. 

As can be seen from Eqs.~\eqref{G2_IR2} and \eqref{Gamma2_IR2}, the dipole amplitudes $G_2$ and $\Gamma_2$ obey the following property:
\begin{align}\label{G2Gamma2scaling}
    \Gamma_2(\soz,\sto,\eta') - G_2^{(0)}(\soz,\eta') = G_2(\soz,\eta=\eta'+\soz-\sto) - G_{2}^{(0)}(\soz,\eta=\eta'+\soz-\sto).
\end{align}
Then using Eqs.~\eqref{doubleLaplaceG2}, \eqref{doubleLaplaceG20}, and \eqref{G2Gamma2scaling} we straightforwardly have
\begin{align}\label{doubleLaplaceGamma2}
    &\Gamma_2(\soz,\sto,\eta') = \wint \gint \left[e^{\omega(\eta'-\sto)}e^{\gamma\soz}\left(G_{2\omega\gamma} - G^{(0)}_{2\omega\gamma}  \right)  + e^{\omega(\eta'-\soz)}e^{\gamma\soz} \, G^{(0)}_{2\omega\gamma}  \right].
\end{align}
Next, we substitute our double Laplace transforms from Eqs.~\eqref{doubleLaplaceG2} and \eqref{doubleLaplaceGtilde} into the evolution equation \eqref{G2_IR2}. Carrying out the integrals over $\sto$ and $\eta'$ and then inverting the Laplace transforms, we find
\begin{align}\label{G2minusG20}
    G_{2\omega\gamma} - G^{(0)}_{2\omega\gamma} = \frac{2}{\omega\gamma}\left(\widetilde{G}_{\omega\gamma} + 2 \, G_{2\omega\gamma}\right),
\end{align}
or equivalently
\begin{align}\label{Gtildeomegagamma}
    \widetilde{G}_{\omega\gamma} = \frac{\omega\gamma}{2}\left(G_{2\omega\gamma} - G^{(0)}_{2\omega\gamma} \right) - 2 \, G_{2\omega\gamma}.
\end{align}
Since all double-Laplace images ($G_{2\omega\gamma}, G^{(0)}_{2\omega\gamma}, \widetilde{G}_{\omega\gamma}, \widetilde{G}^{(0)}_{\omega\gamma}, Q_{\omega\gamma}, Q^{(0)}_{\omega\gamma},\widetilde{Q}_{\omega\gamma},\widetilde{Q}^{(0)}_{\omega\gamma}$) must go to zero as $\omega$ or $\gamma$ go to infinity, Eq.~\eqref{G2minusG20} implies that the difference between $G_{2\omega\gamma}$ and $G^{(0)}_{2\omega\gamma}$ goes to zero faster than $1/\omega$ or $1/\gamma$ as $\omega$ or $\gamma$, respectively, go to infinity. We can write
\begin{subequations}\label{G2minusG20atinfty}
    \begin{align}
    &\wint \left(G_{2\omega\gamma} - G^{(0)}_{2\omega\gamma}\right) = \wint \frac{2}{\omega\gamma}\left(\widetilde{G}_{\omega\gamma} + 2 \, G_{2\omega\gamma}\right) = 0 
    \end{align}
    and
    \begin{align}
    &\gint \left(G_{2\omega\gamma} - G^{(0)}_{2\omega\gamma}\right) = \gint \frac{2}{\omega\gamma}\left(\widetilde{G}_{\omega\gamma} + 2G_{2\omega\gamma}\right) = 0
    \end{align}
\end{subequations}
where the last equality in each line follows from closing the $\omega$- or $\gamma$-contour to the right. This fact can be used along with the double Laplace representations in Eqs.~\eqref{doubleLaplacestart} to straightforwardly show that the boundary conditions implied by Eqs.~\eqref{G2_IR2} and \eqref{Gamma2_IR2},
\begin{align}
    &G_2(\soz=0,\eta) = G^{(0)}_2(\soz=0,\eta)\,,\\
    &G_2(\soz,\eta=\soz) = G^{(0)}_2(\soz,\eta=\soz)\,, \\
    &\Gamma_2(\soz=0,\sto,\eta') = G^{(0)}_2(\soz=0,\eta')\,, \\
    &\Gamma_2(\soz,\sto,\eta'=\sto) = G^{(0)}_2(\soz,\eta'=\sto)\,,
\end{align}
are automatically satisfied. All the above steps in this Subsection closely follow those in \cite{Borden:2023ugd}.

Next, we can substitute the double Laplace expressions Eqs.~\eqref{doubleLaplaceG2}, \eqref{doubleLaplaceQ}, and \eqref{doubleLaplaceQtilde} into the evolution equation \eqref{Qtilde_IR2}. Doing this, carrying out the integrals over $\sto$ and $\eta'$, and then inverting the Laplace transforms, we find
\begin{align}\label{QtildeminusQtilde0}
    \widetilde{Q}_{\omega\gamma}- \widetilde{Q}^{(0)}_{\omega\gamma} = -\frac{1}{\omega\gamma}\left(Q_{\omega\gamma} + 2G_{2\omega\gamma}\right),
\end{align}
or equivalently
\begin{align}\label{Qtildeomegagamma}
    Q_{\omega\gamma} = -\omega\gamma\left(\widetilde{Q}_{\omega\gamma} - \widetilde{Q}^{(0)}_{\omega\gamma}\right) - 2G_{2\omega\gamma}.
\end{align}
Again we see that \eq{QtildeminusQtilde0} implies that the difference between $\widetilde{Q}_{\omega\gamma}$ and $\widetilde{Q}^{(0)}_{\omega\gamma}$ goes to zero faster than $1/\omega$ or $1/\gamma$ as $\omega\rightarrow \infty$ or $\gamma\rightarrow\infty$, which allows us to write
\begin{subequations}\label{QtildeminusQtilde0atinfty}
    \begin{align}
    &\wint \left(\widetilde{Q}_{\omega\gamma} - \widetilde{Q}^{(0)}_{\omega\gamma}\right) = -\wint \frac{1}{\omega\gamma} \left(Q_{\omega\gamma} + 2G_{2\omega\gamma}\right) = 0
    \end{align}
    and
    \begin{align}
    &\gint \left(\widetilde{Q}_{\omega\gamma} - \widetilde{Q}^{(0)}_{\omega\gamma}\right) = -\gint \frac{1}{\omega\gamma}\left(Q_{\omega\gamma} + 2G_{2\omega\gamma}\right) = 0\,,
\end{align}
\end{subequations}
where again the last equality in each line follows from closing the contour to the right. Eqs.~\eqref{QtildeminusQtilde0atinfty} can be used along with the double-Laplace representations in \eq{doubleLaplacestart} to show that the two boundary conditions for $\widetilde{Q}$ implied by \eq{Qtilde_IR2},
\begin{align}\label{QtildeBCs}
    &\widetilde{Q}(\soz=0,\eta) = \widetilde{Q}^{(0)}(\soz=0,\eta)\,,\\
    &\widetilde{Q}(\soz,\eta=\soz) = \widetilde{Q}^{(0)}(\soz,\eta=\soz)\,,
\end{align}
are automatically satisfied.

At this point, the evolution equations \eqref{G2_IR2}, \eqref{Gamma2_IR2}, and \eqref{Qtilde_IR2} are completely satisfied, and we have obtained expressions for the double-inverse Laplace transforms of  $\Gamma_2$, $\widetilde{G}$, and $Q$ in terms of the yet unknown double-Laplace images $G_{2\omega\gamma}$ and $\widetilde{Q}_{\omega\gamma}$. It remains to satisfy Eqs.~\eqref{Q_IR2}, \eqref{Gammabar_IR2}, \eqref{Gtilde_IR2}, and \eqref{Gammatilde_IR2}, obtain double-Laplace expressions for the remaining dipole amplitudes $\overline{\Gamma}$ and $\widetilde{\Gamma}$, and ultimately solve for the double-Laplace images $G_{2\omega\gamma}$ and $\widetilde{Q}_{\omega\gamma}$.


\subsection{Double Inverse Laplace Representations for \texorpdfstring{$\overline{\Gamma}$, $\widetilde{\Gamma}$}{Gammabar, Gammatilde}}

Upon differentiating Eqs.~\eqref{Gammabar_IR2} and \eqref{Gammatilde_IR2}, one can show that $\overline{\Gamma}$ and $\widetilde{\Gamma}$ satisfy the following second-order partial differential equations:
\begin{subequations}\label{secondorderpdes}
\begin{align}
    \label{GammabarPDE}
    &\frac{\partial^2 \overline{\Gamma}(\soz,\sto,\eta')}{\partial\sto\partial\eta'} + \frac{\partial^2 \overline{\Gamma}(\soz,\sto,\eta')}{\partial\sto^2} = -2\widetilde{G}(\sto,\eta') - 2\widetilde{\Gamma}(\soz,\sto,\eta') - \frac{3}{2}Q(\sto,\eta') + \overline{\Gamma}(\soz,\sto,\eta') \\
    &\hspace{5.75cm}- 2\Gamma_2(\soz,\sto,\eta') - 3G_2(\sto,\eta') \,,\notag \\
    \label{GammatildePDE}
    &\frac{\partial^2 \widetilde{\Gamma}(\soz,\sto,\eta')}{\partial\sto\partial\eta'} + \frac{\partial^2 \widetilde{\Gamma}(\soz,\sto,\eta')}{\partial\sto^2} = -3\widetilde{G}(\sto,\eta') - \widetilde{\Gamma}(\soz,\sto,\eta') - 2G_2(\sto,\eta') \\
    &\hspace{5.75cm}- \left(2-\frac{N_f}{2N_c} \right)\Gamma_2(\soz,\sto,\eta') + \frac{N_f}{4N_c}\overline{\Gamma}(\soz,\sto,\eta') +\frac{N_f}{2N_c}\widetilde{Q}(\sto,\eta'). \notag
\end{align}
\end{subequations}
Similar to \cite{Borden:2023ugd}, we proceed to solve these two partial differential equations by constructing their homogeneous and particular solutions. 
We begin with the homogeneous solutions, employing the following ansatz:
\begin{subequations}
\begin{align}\label{Gammatildehansatz}
    \widetilde{\Gamma}^{(h)}(\soz,\sto,\eta') = \wint\gint e^{\omega(\eta'-\sto)}e^{\gamma\sto}\widetilde{\Gamma}_{\omega\gamma}(\soz), \\
    \overline{\Gamma}^{(h)}(\soz,\sto,\eta') = \wint\gint e^{\omega(\eta'-\sto)}e^{\gamma\sto}\overline{\Gamma}_{\omega\gamma}(\soz) .
\end{align}
\end{subequations}
Plugging these into the homogeneous part of Eqs.~\eqref{secondorderpdes} we obtain
\begin{subequations}\label{Gamma_eqs2}
\begin{align}
& [\gamma (\gamma - \omega) - 1] \, \overline{\Gamma}_{\omega\gamma}(\soz) = - 2 \, \widetilde{\Gamma}_{\omega\gamma}(\soz), \\
& [\gamma (\gamma - \omega) + 1] \,  \widetilde{\Gamma}_{\omega\gamma}(\soz) = \frac{N_f}{4 N_c} \, \overline{\Gamma}_{\omega\gamma}(\soz) . 
\end{align}
\end{subequations}
Solving these gives
\begin{align}\label{gammafourthorder}
\gamma^2 \, (\gamma - \omega)^2 = 1 - \frac{N_f}{2 \, N_c} \,,
\end{align}
which can readily be solved for $\gamma$, giving
\begin{align}\label{deltapmpm}
    \gamma = \dw^{\pm\pm} \equiv \tfrac{1}{2}\left(\omega \pm \sqrt{\omega^2 \pm 4\sqrt{1-\tfrac{N_f}{2N_c}}}\right).
\end{align}
The notation here is such that the $\pm$ indices on $\dw^{\pm\pm}$ should be read left to right as they are encountered on the right hand side of \eq{deltapmpm}.
In addition, Eqs.~\eqref{Gamma_eqs2} give
\begin{align}\label{relationbtwgammatildeandbar}
  \overline{\Gamma}_{\omega\gamma}(\soz) = \frac{1 \pm \sqrt{ 1 -  \frac{N_f}{2 \, N_c} }}{N_f / (4 N_c) }  \,  \widetilde{\Gamma}_{\omega\gamma}(\soz) ,  
\end{align}
where the $\pm$ here is the same as the second index in $\dw^{\pm\pm}$ . 
We thus have the homogeneous solutions, written as linear combinations of the solutions corresponding to each of the four solutions $\gamma = \dw^{\pm\pm}$ from \eq{deltapmpm},
\begin{subequations}\label{homogeneous_sols}
\begin{align}\label{Gammatildeh}
    & \widetilde{\Gamma}^{(h)}(\soz,\sto,\eta') = \wint e^{\omega(\eta'-\sto)} \sum_{\alpha,\beta = +,-} e^{\dw^{\alpha\beta}\sto} \, \widetilde{\Gamma}_\omega^{(\alpha\beta)}(\soz) , \\
    \label{Gammabarh}
     & \overline{\Gamma}^{(h)}(\soz,\sto,\eta') = \wint e^{\omega(\eta'-\sto)} \sum_{\alpha,\beta = +,-} e^{\dw^{\alpha\beta}\sto}  \frac{1 + \beta \, \sqrt{ 1 - \frac{N_f}{2 \, N_c} }}{N_f / (4 N_c) }  \,  \widetilde{\Gamma}_\omega^{(\alpha\beta)}(\soz) ,
\end{align}
\end{subequations}
where we have also employed \eq{relationbtwgammatildeandbar} in writing \eq{Gammabarh}.

Moving on to the particular solutions of Eqs.~\eqref{secondorderpdes}, we look for them in the following form: 
\begin{subequations}
\begin{align}\label{Gammatildepansatz}
   &  \widetilde{\Gamma}^{(p)}(\soz,\sto,\eta') = \wint\gint \left[e^{\omega(\eta'-\sto)}e^{\gamma\sto}A_{\omega\gamma} + e^{\omega(\eta'-\sto)}e^{\gamma\soz}B_{\omega\gamma} + e^{\omega(\eta'-\soz)}e^{\gamma\soz}C_{\omega\gamma}  \right] , \\
   &  \overline{\Gamma}^{(p)}(\soz,\sto,\eta') = \wint\gint \left[e^{\omega(\eta'-\sto)}e^{\gamma\sto}  \overline{A}_{\omega\gamma} + e^{\omega(\eta'-\sto)}e^{\gamma\soz}  {\overline B}_{\omega\gamma} + e^{\omega(\eta'-\soz)}e^{\gamma\soz}  {\overline C}_{\omega\gamma} \right] .
\end{align}
\end{subequations}
Plugging those into the full Eqs.~\eqref{secondorderpdes}, and employing Eqs.~\eqref{doubleLaplacestart},  \eqref{doubleLaplaceGamma2}, \eqref{Gtildeomegagamma}, and \eqref{Qtildeomegagamma}, yields
\begin{subequations}\label{ABC3}\allowdisplaybreaks
\begin{align}
& A_{\omega\gamma} =  \frac{1}{\gamma^2 \, (\gamma - \omega)^2 - 1 + \tfrac{N_f}{2 N_c}} \, \Bigg\{  \frac{\omega\gamma}{2} \, \left[ 3 - \tfrac{N_f}{2 N_c} - 3 \,  \gamma \, (\gamma - \omega) \right] \, \left(G_{2\omega\gamma} - G^{(0)}_{2\omega\gamma} \right) + \tfrac{3 N_f}{8 N_c} \, \omega \, \gamma \, \left( \widetilde{Q}_{\omega\gamma} - \widetilde{Q}^{(0)}_{\omega\gamma} \right) \\ 
&\hspace{5cm} + \left[ 4 \, \gamma \, (\gamma - \omega) - 4 + \tfrac{N_f}{N_c} \right] \, G_{2\omega\gamma} + \tfrac{N_f}{2 N_c} \, \left[ \gamma \, (\gamma - \omega) - 1 \right] \, \widetilde{Q}_{\omega\gamma} \Bigg\}, \notag \\
& \overline{A}_{\omega\gamma} =  \frac{1}{\gamma^2 \, (\gamma - \omega)^2 - 1 + \tfrac{N_f}{2 N_c}} \, \Bigg\{ \omega\gamma \, [2 - \gamma \, (\gamma - \omega)] \, \left(G_{2\omega\gamma} - G^{(0)}_{2\omega\gamma} \right) + 4 \, [\gamma \, (\gamma - \omega) -1] \, G_{2\omega\gamma}   \notag \\
&\hspace{5cm} + \tfrac{3}{2} \, \omega\gamma \, [\gamma \, (\gamma - \omega) + 1] \, \left( \widetilde{Q}_{\omega\gamma} - \widetilde{Q}^{(0)}_{\omega\gamma} \right) - \tfrac{N_f}{N_c} \, \widetilde{Q}_{\omega\gamma}  \Bigg\}, \\
& B_{\omega\gamma} = {\overline B}_{\omega\gamma} = - 2 \, \left(G_{2\omega\gamma} - G^{(0)}_{2\omega\gamma} \right) , \\
& C_{\omega\gamma} = {\overline C}_{\omega\gamma} = - 2 \, G^{(0)}_{2\omega\gamma} .
\end{align}
\end{subequations}

Since
\begin{subequations}
\begin{align}
    & \widetilde{\Gamma} (\soz,\sto,\eta') = \widetilde{\Gamma}^{(h)}(\soz,\sto,\eta') +\widetilde{\Gamma}^{(p)}(\soz,\sto,\eta') , \\
    &  \overline{\Gamma} (\soz,\sto,\eta') =   \overline{\Gamma}^{(h)}(\soz,\sto,\eta') +  \overline{\Gamma}^{(p)}(\soz,\sto,\eta'), 
\end{align}
\end{subequations}
at this point we have double-Laplace expressions for all the dipole amplitudes, which we collect together here:
\begin{subequations}\label{all}\allowdisplaybreaks
\begin{align}
    \label{all_G2}
    &G_2(\soz,\eta) = \wint \gint e^{\omega(\eta-\soz)}e^{\gamma \soz}\gtwg , \\
    \label{all_Gamma2}
    &\Gamma_2(\soz,\sto,\eta') = \wint \gint \left[e^{\omega(\eta'-\sto)}e^{\gamma\soz}\left(\gtwg - \gtwg^{(0)}\right) + e^{\omega(\eta'-\soz)}e^{\gamma\soz}\gtwg^{(0)} \right] , \\
    \label{all_Gtilde}
    &\widetilde{G}(\soz,\eta) = \wint\gint e^{\omega(\eta-\soz)}e^{\gamma\soz}\left[ \frac{\omega\gamma}{2}\left(G_{2\omega\gamma} - G_{2\omega\gamma}^{(0)}\right) - 2G_{2\omega\gamma}\right] , \\
    \label{all_Q}
    &Q(\soz,\eta) = \wint \gint e^{\omega(\eta-\soz)}e^{\gamma\soz} \, \left[ - \omega\gamma \, \left( \widetilde{Q}_{\omega\gamma} - \widetilde{Q}^{(0)}_{\omega\gamma} \right) - 2G_{2\omega\gamma} \right] , \\
    \label{all_Qtilde}
    &\widetilde{Q}(\soz,\eta) = \wint \gint e^{\omega(\eta-\soz)}e^{\gamma\soz} \, \widetilde{Q}_{\omega\gamma} , \\
    \label{all_Gammatilde}
    &\widetilde{\Gamma}(\soz,\sto,\eta') =  \wint e^{\omega(\eta'-\sto)}\sum_{\alpha,\beta = +,-} e^{\dw^{\alpha\beta}\sto} \widetilde{\Gamma}_\omega^{(\alpha\beta)}(\soz) \\
    &\hspace{2cm} + \wint\gint \left[e^{\omega(\eta'-\sto)}e^{\gamma\sto}A_{\omega\gamma} -2e^{\omega(\eta'-\sto)}e^{\gamma\soz}\left(G_{2\omega\gamma} - G^{(0)}_{2\omega\gamma}\right) -2e^{\omega(\eta'-\soz)}e^{\gamma\soz}G^{(0)}_{2\omega\gamma} \right] , \notag \\
    \label{all_Gammabar}
    &\overline{\Gamma}(\soz,\sto,\eta') = \wint \, e^{\omega(\eta'-\sto)}\sum_{\alpha,\beta = +,-} e^{\dw^{\alpha\beta}\sto} \frac{1 + \beta \, \sqrt{ 1 - \frac{N_f}{2 \, N_c} }}{N_f / (4 N_c) }  \,  \widetilde{\Gamma}_\omega^{(\alpha\beta)}(\soz) \\
    &\hspace{2cm} + \wint\gint \left[ e^{\omega(\eta'-\sto)}e^{\gamma\sto} \, {\overline A}_{\omega\gamma} - 2e^{\omega(\eta'-\sto)}e^{\gamma\soz}\left(G_{2\omega\gamma} - G^{(0)}_{2\omega\gamma}\right) - 2e^{\omega(\eta'-\soz)}e^{\gamma\soz}G^{(0)}_{2\omega\gamma} \right] , \notag \\
    &\text{with} \notag \\
    \label{all_deltapms}
    & \delta_{\omega}^{\pm\pm} \equiv \frac{1}{2}\left[\omega \pm \sqrt{\omega^2 \pm 4\sqrt{1-\frac{N_f}{2N_c} } } \right] , \\
    \label{Aeq}
    & A_{\omega\gamma} =  \frac{1}{\left(\gamma-\dw^{++} \right)\left(\gamma-\dw^{+-} \right)\left(\gamma-\dw^{-+} \right)\left(\gamma-\dw^{--} \right)} \, \Bigg\{  \frac{\omega\gamma}{2} \, \left[ 3 - \tfrac{N_f}{2 N_c} - 3 \,  \gamma \, (\gamma - \omega) \right] \, \left(G_{2\omega\gamma} - G^{(0)}_{2\omega\gamma} \right) \\
    &\hspace{7.5cm} + \tfrac{3 N_f}{8 N_c} \, \omega \, \gamma \, \left( \widetilde{Q}_{\omega\gamma} - \widetilde{Q}^{(0)}_{\omega\gamma} \right) \notag \\ 
&\hspace{7.5cm} + \left[ 4 \, \gamma \, (\gamma - \omega) - 4 + \tfrac{N_f}{N_c} \right] \, G_{2\omega\gamma} + \tfrac{N_f}{2 N_c} \, \left[ \gamma \, (\gamma - \omega) - 1 \right] \, \widetilde{Q}_{\omega\gamma} \Bigg\}, \notag \\
\label{Abar_eq}
& \overline{A}_{\omega\gamma} =  \frac{1}{\left(\gamma-\dw^{++} \right)\left(\gamma-\dw^{+-} \right)\left(\gamma-\dw^{-+} \right)\left(\gamma-\dw^{--} \right)} \, \Bigg\{ \omega\gamma \, [2 - \gamma \, (\gamma - \omega)] \, \left(G_{2\omega\gamma} - G^{(0)}_{2\omega\gamma} \right) + 4 \, [\gamma \, (\gamma - \omega) -1] \, G_{2\omega\gamma}   \notag \\
& \hspace{7.5cm}+ \tfrac{3}{2} \, \omega\gamma \, [\gamma \, (\gamma - \omega) + 1] \, \left( \widetilde{Q}_{\omega\gamma} - \widetilde{Q}^{(0)}_{\omega\gamma} \right) - \tfrac{N_f}{N_c} \, \widetilde{Q}_{\omega\gamma}  \Bigg\} . 
\end{align}
\end{subequations}

Note that we have used Eqs.~\eqref{gammafourthorder} and \eqref{deltapmpm} to rewrite the denominators of $A_{\omega\gamma}$ and $\overline{A}_{\omega\gamma}$ in Eqs.~\eqref{Aeq} and \eqref{Abar_eq}. This makes it clear that our procedure to solve the partial differential equations in Eqs.~\eqref{secondorderpdes} has introduced additional poles in the integrand. For large $\omega$, we note the following behavior of the functions $\dw^{\alpha\beta}$:
\begin{subequations}\label{polescalingslargew}
    \begin{align}
        &\dw^{++} \sim \omega \, , \\
        &\dw^{+-} \sim \omega \, , \\
        &\dw^{-+} \sim -\sqrt{1-\tfrac{N_f}{2N_c}}\frac{1}{\omega} \, , \\
        &\dw^{--} \sim \sqrt{1-\tfrac{N_f}{2N_c}}\frac{1}{\omega}.
    \end{align}
\end{subequations}
Since $\dw^{++},\dw^{+-} \sim \omega$ for large $\omega$, these poles cannot lie to the left of both the $\omega$ and $\gamma$ contours. This is the same situation encountered in the analytic solution of the large-$N_c$ evolution equations constructed in \cite{Borden:2023ugd}, although there was only one such pole in that solution, while here we have two. Nevertheless we can follow the procedure of \cite{Borden:2023ugd} and declare that the poles at $\gamma = \dw^{++}$ and $\gamma = \dw^{+-}$ lie to the left of the $\omega$-contour but to the right of the $\gamma$-contour. This implies that we are choosing Re~$\omega >$~Re~$\gamma$ along the integration contours.

Although we have solved the partial differential equations \eqref{secondorderpdes}, these solutions are not yet solutions of the full integral evolution equations~\eqref{Gammabar_IR2} and \eqref{Gammatilde_IR2} from which we obtained those PDEs. We have also yet to satisfy the non-neighbor partners of these equations, Eqs.~\eqref{Q_IR2} and \eqref{Gtilde_IR2}. The next step is to substitute the double-Laplace results from Eqs.~\eqref{all} back into the evolution equations for $\widetilde{\Gamma}$ and $\overline{\Gamma}$ (Eqs.~\eqref{Gammatilde_IR2} and \eqref{Gammabar_IR2}, respectively) in order to obtain the remaining constraints necessary to ensure the full evolution equations are satisfied. Then, since the evolution equations for $\widetilde{G}$ and $Q$ (Eqs.~\eqref{Gtilde_IR2} and \eqref{Q_IR2}) are special cases of those for $\widetilde{\Gamma}$ and $\overline{\Gamma}$, respectively, we can ensure the evolution equations for $\widetilde{G}$  and $Q$ are also satisfied by setting
\begin{subequations}
\begin{align}
    &\widetilde{\Gamma}(\soz,\sto=\soz,\eta') = \widetilde{G}(\soz,\eta') , 
    \\
    &\overline{\Gamma}(\soz,\sto=\soz,\eta') = Q(\soz,\eta').
\end{align}
\end{subequations}
This will ultimately allow us to solve for the unknown functions $\widetilde{\Gamma}_\omega^{(++)}(\soz)$, $\widetilde{\Gamma}_\omega^{(+-)}(\soz)$, $\widetilde{\Gamma}_\omega^{(-+)}(\soz)$, $\widetilde{\Gamma}_\omega^{(--)}(\soz)$, $G_{2\omega\gamma}$, and $\widetilde{Q}_{\omega\gamma}$. We will do this in the next Sections.


\subsection{Obtaining the Remaining Constraints}

We begin with the evolution equation \eqref{Gammatilde_IR2} for $\widetilde{\Gamma}(\soz,\sto,\eta')$. Substituting all the relevant double-Laplace expressions from Eqs.~\eqref{all}, carrying out all the integrals over $\stt$ and $\eta''$, and performing the forward Laplace transform over $\eta'$ yields    
\begin{align}\label{simplifying2}
    &0 =  e^{- \omega \sto} \sum_{\alpha,\beta = +,-} \widetilde{\Gamma}_\omega^{(\alpha\beta)}(\soz) \, \frac{\dw^{\alpha\beta}-\omega}{\omega}e^{\dw^{\alpha\beta}\soz} \\
    &\hspace{.3cm} + e^{- \omega \, \sto} \gint \, e^{\gamma\soz} \, \left[ A_{\omega\gamma} \frac{\gamma-\omega}{\omega} + \frac{N_f}{4N_c} \,  \left( \widetilde{Q}_{\omega\gamma} - \widetilde{Q}^{(0)}_{\omega\gamma} \right) + 2 \, \left(G_{2\omega\gamma} - G^{(0)}_{2\omega\gamma}\right) \right] \notag \\
    &\hspace{.3cm} + \int \frac{d \omega'}{2\pi i } \sum_{\alpha,\beta = +,-}\widetilde{\Gamma}_{\omega'}^{(\alpha\beta)}(\soz) \left[ \frac{1}{\delta_{\omega'}^{\alpha\beta} - \omega} \,  +  
   \frac{1}{\omega} \frac{\omega'-\delta_{\omega'}^{\alpha\beta}}{\omega'}\, e^{\delta_{\omega'}^{\alpha\beta}\soz} \right]
   + e^{- \omega \soz} \gint  e^{\gamma\soz} \, \left[ \widetilde{G}^{(0)}_{\omega\gamma} + 2 \, G^{(0)}_{2\omega\gamma} \right]. \notag  
\end{align}
Along the way, we have dropped several terms which are zero, as can be shown by closing either the $\omega$ or the $\gamma$ integration contour to the right. Now we observe that two of the terms in \eq{simplifying2} have the same $\sto$ dependence, $e^{-\omega\sto}$, whereas the other two terms are independent of $\sto$. Since the equation is valid for any $\sto$, we conclude that both sets of terms must separately equal zero and arrive at the following two constraints:
\begin{subequations}\label{conditions1}
\begin{align}\label{condition1a}
& \gint \, e^{\gamma\soz} \, \left[ A_{\omega\gamma} \frac{\gamma-\omega}{\omega} + \frac{N_f}{4N_c} \,  \left( \widetilde{Q}_{\omega\gamma} - \widetilde{Q}^{(0)}_{\omega\gamma} \right) + 2 \, \left(G_{2\omega\gamma} - G^{(0)}_{2\omega\gamma}\right) \right] = \sum_{\alpha,\beta = +,-} \widetilde{\Gamma}_\omega^{(\alpha\beta)}(\soz) \, \frac{\omega-\dw^{\alpha\beta}}{\omega} e^{\dw^{\alpha\beta}\soz} \, ,  \\
\label{condition1b}
& 0 = \int \frac{d \omega'}{2\pi i } \sum_{\alpha,\beta = +,-}\widetilde{\Gamma}_{\omega'}^{(\alpha\beta)}(\soz) \left[ \frac{1}{\delta_{\omega'}^{\alpha\beta} - \omega} \,  +  
   \frac{1}{\omega} \frac{\omega'-\delta_{\omega'}^{\alpha\beta}}{\omega'}\, e^{\delta_{\omega'}^{\alpha\beta}\soz} \right]
   + e^{- \omega \soz} \gint  e^{\gamma\soz} \, \left[ \widetilde{G}^{(0)}_{\omega\gamma} + 2 \, G^{(0)}_{2\omega\gamma} \right].
\end{align}
\end{subequations}

We can obtain a third constraint and satisfy the evolution equation \eqref{Gtilde_IR2} by requiring that 
\begin{align}\label{thirdconstraint}
    \widetilde{\Gamma}(\soz,\sto=\soz,\eta') = \widetilde{G}(\soz,\eta').
\end{align}
Using our double-Laplace expressions \eqref{all_Gammatilde} and \eqref{all_Gtilde} and applying the inverse transform over $\eta'-\soz$ (here treating $\eta'-\soz$ and $\soz$ as independent variables), \eq{thirdconstraint} gives
\begin{align}\label{simplifying3}
    \sum_{\alpha,\beta=+,-} e^{\dw^{\alpha\beta}\soz} \widetilde{\Gamma}_{\omega}^{(\alpha\beta)}(\soz) = \gint e^{\gamma\soz}\left[\frac{\omega\gamma}{2}\left(G_{2\omega\gamma} - G^{(0)}_{2\omega\gamma} \right) - A_{\omega\gamma} \right].
\end{align}
In Eqs.~\eqref{conditions1} and \eqref{simplifying3} we have thus obtained the three constraints necessary to fully satisfy the evolution equations for $\widetilde{\Gamma}$ and $\widetilde{G}$. Next we can apply this same procedure to the evolution equations for $\overline{\Gamma}$ and $Q$. 

Beginning with \eq{Gammabar_IR2}, we substitute all the relevant double-Laplace expressions from Eqs.~\eqref{all}, carry out the integrals over $\stt$ and $\eta'$, perform the forward Laplace transform over $\eta'$, and again drop several terms which can be shown to be zero. The result is
\begin{align}\label{simplifying5}
    &0 = -e^{-\omega\sto}\sum_{\alpha,\beta = +,-}e^{\dw^{\alpha\beta}\soz}\widetilde{\Gamma}^{(\alpha\beta)}_\omega(\soz) \frac{\omega-\dw^{\alpha\beta}}{\omega}\frac{4N_c}{N_f}\left(1+\beta\sqrt{1-\tfrac{N_f}{2N_c}}\right) \\
    &\hspace{0.4cm}  + e^{-\omega\sto}\gint e^{\gamma\soz}\left[ {\overline A}_{\omega\gamma} \frac{\gamma-\omega}{\omega} - \frac{1}{2} \,  \left( \widetilde{Q}_{\omega\gamma} - \widetilde{Q}^{(0)}_{\omega\gamma} \right) + 2 \, \left(G_{2\omega\gamma} - G^{(0)}_{2\omega\gamma}\right) \right] \notag \\
    &\hspace{0.4cm} + e^{- \omega \soz} \gint  e^{\gamma\soz} \, \left[ Q^{(0)}_{\omega\gamma} + 2 \, G^{(0)}_{2\omega\gamma} \right] - \int \frac{d \omega'}{2\pi i } \sum_{\alpha,\beta = +,-} \frac{1}{\omega - \delta_{\omega'}^{\alpha\beta}} \, \widetilde{\Gamma}_{\omega'}^{(\alpha\beta)}(\soz) \, \frac{4N_c}{N_f}\left(1 + \beta \, \sqrt{ 1 - \frac{N_f}{2 \, N_c} }\right)  \notag \\
    &\hspace{0.4cm}  +  
   \frac{1}{\omega} \int \frac{d \omega'}{2\pi i } \sum_{\alpha,\beta = +,-} \widetilde{\Gamma}_{\omega'}^{(\alpha\beta)}(\soz) \frac{\omega'-\delta_{\omega'}^{\alpha\beta}}{\omega'}\, \frac{4N_c}{N_f}\left(1 + \beta \, \sqrt{ 1 - \frac{N_f}{2 \, N_c} }\right) \, e^{\delta_{\omega'}^{\alpha\beta}\soz} \notag
\end{align}
Just as with \eq{simplifying2}, two of the terms here share the same $\sto$ dependence, $e^{-\omega\sto}$, whereas the other terms are independent of $\sto$, giving us two separate constraints,
\begin{subequations}\label{conditions2}
    \begin{align}\label{conditions2a}
    &\sum_{\alpha,\beta = +,-}e^{\dw^{\alpha\beta}\soz}\widetilde{\Gamma}^{(\alpha\beta)}_\omega(\soz) \frac{\omega-\dw^{\alpha\beta}}{\omega}\frac{4N_c}{N_f}\left(1+\beta\sqrt{1-\tfrac{N_f}{2N_c}}\right) \\
    &\hspace{3cm} = \gint e^{\gamma\soz}\left[ {\overline A}_{\omega\gamma} \frac{\gamma-\omega}{\omega} - \frac{1}{2} \,  \left( \widetilde{Q}_{\omega\gamma} - \widetilde{Q}^{(0)}_{\omega\gamma} \right) + 2 \, \left(G_{2\omega\gamma} - G^{(0)}_{2\omega\gamma}\right) \right]\notag \, ,  \\
    \label{conditions2b}
    &0 = e^{- \omega \soz} \gint  e^{\gamma\soz} \, \left[ Q^{(0)}_{\omega\gamma} + 2 \, G^{(0)}_{2\omega\gamma} \right] - \int \frac{d \omega'}{2\pi i } \sum_{\alpha,\beta = +,-} \frac{1}{\omega - \delta_{\omega'}^{\alpha\beta}} \, \widetilde{\Gamma}_{\omega'}^{(\alpha\beta)}(\soz) \, \frac{4N_c}{N_f}\left(1 + \beta \, \sqrt{ 1 - \frac{N_f}{2 \, N_c} }\right)  \\
    &\hspace{0.4cm}  +  
   \frac{1}{\omega} \int \frac{d \omega'}{2\pi i } \sum_{\alpha,\beta = +,-} \widetilde{\Gamma}_{\omega'}^{(\alpha\beta)}(\soz) \frac{\omega'-\delta_{\omega'}^{\alpha\beta}}{\omega'}\, \frac{4N_c}{N_f}\left(1 + \beta \, \sqrt{ 1 - \frac{N_f}{2 \, N_c} }\right) \, e^{\delta_{\omega'}^{\alpha\beta}\soz} .\notag
    \end{align}
\end{subequations}

For one final constraint, and to satisfy the evolution equation \eqref{Q_IR2} for $Q$, we require
\begin{align}\label{GammabartoQconstraint}
    \overline{\Gamma}(\soz,\sto=\soz,\eta') = Q(\soz,\eta').
\end{align}
Using the double-Laplace expressions from Eqs.~\eqref{all} and performing the forward transform over $\eta'-\soz$ (again, while treating $\eta'-\soz$ and $\soz$ as independent variables) gives
\begin{align}\label{condition2c}
\gint\, e^{\gamma \soz} \, \left[ - {\overline A}_{\omega\gamma} - \omega\gamma \, \left(\widetilde{Q}_{\omega\gamma} - \widetilde{Q}^{(0)}_{\omega\gamma}  \right) \right] = \sum_{\alpha,\beta = +,-} \widetilde{\Gamma}_\omega^{(\alpha\beta)}(\soz) \, \frac{1 + \beta \, \sqrt{ 1 - \frac{N_f}{2 \, N_c} }}{N_f / (4 N_c) } \, e^{\dw^{\alpha\beta}\soz}.
\end{align}

To summarize, we note that in Eqs.~\eqref{conditions1}, \eqref{simplifying3}, \eqref{conditions2}, and \eqref{condition2c} we have a total of six constraints that must be satisfied by our double-Laplace constructions in order to ensure that all the evolution equations in \eqref{evoleqs_IR_1} are satisfied. What remains is to solve those constraints for the currently unknown functions $\widetilde{\Gamma}^{(\alpha\beta)}_{\omega}(\soz)$, $G_{2\omega\gamma}$, and $\widetilde{Q}_{\omega\gamma}$.


\subsection{Solving the Constraints}

\subsubsection{Four Straightforward Constraints}
The four equations~\eqref{condition1a}, \eqref{simplifying3}, \eqref{conditions2a}, \eqref{condition2c} constitute a system of equations we can straightforwardly solve for all four of the $\widetilde{\Gamma}^{(\alpha\beta)}_{\omega}(\soz)$ functions as integrals over $\gamma$. First we rewrite the common structure shared by each as
\begin{align}\label{Gammatildestructure}
    \widetilde{\Gamma}_{\omega}^{(\alpha\beta)}(\soz) = e^{-\dw^{\alpha\beta}\soz}\gint \, e^{\gamma\soz} \, \widetilde{\Gamma}_{\omega\gamma}^{(\alpha\beta)}
\end{align}
(note the distinction of $\widetilde{\Gamma}^{(\alpha\beta)}_{\omega\gamma}$ on the right-hand side written now as a function of $\omega$ and $\gamma$ and not of $\soz$).
Then we undo the inverse Laplace transform over $\gamma$ in all terms of Eqs.~\eqref{condition1a}, \eqref{simplifying3}, \eqref{conditions2a}, \eqref{condition2c}, solving the resulting linear system of equations algebraically for the functions $\widetilde{\Gamma}^{(\alpha\beta)}_{\omega\gamma}$. The results can be written compactly as 
\begin{align}\label{Gammatildescompact}
& \widetilde{\Gamma}_{\omega\gamma}^{(\alpha\beta)} = \frac{\beta}{8 (\dw^{\alpha\beta} - \dw^{-\alpha, \beta}) \, \sqrt{4 - 2 n}} \, \Big\{ \left[ \, {\overline A}_{\omega\gamma} 2 n - 4 A_{\omega\gamma} (2- \beta \sqrt{4 - 2 n}) \right] (\dw^{-\alpha, \beta} - \gamma) \\ 
& + n \omega  \left( \widetilde{Q}_{\omega\gamma} - \widetilde{Q}^{(0)}_{\omega\gamma} \right) (3 - \beta \sqrt{4 - 2 n} - 2 \gamma \dw^{\alpha, \beta}  ) + 2 \omega \left(G_{2\omega\gamma} - G^{(0)}_{2\omega\gamma}\right) \left[ (2- \beta \sqrt{4 - 2 n}) (4 - \gamma \dw^{\alpha, \beta}  ) - 2 n  \right] \Big\}. \notag 
\end{align}
In writing \eq{Gammatildescompact} we have defined 
\begin{align}\label{littlen}
n \equiv \frac{N_f}{N_c}
\end{align}
and have also used the fact that 
\begin{align}\label{deltasumrule}
\dw^{\alpha\beta} + \dw^{-\alpha, \beta} = \omega.
\end{align}
Combining Eqs.~\eqref{Gammatildestructure} and \eqref{Gammatildescompact} we have
\begin{align}\label{Gamma_final}
  &  \widetilde{\Gamma}_{\omega}^{(\alpha\beta)}(\soz) = e^{-\dw^{\alpha\beta}\soz}\gint e^{\gamma\soz} \frac{\beta}{8 (\dw^{\alpha\beta} - \dw^{-\alpha, \beta}) \, \sqrt{4 - 2 n}} \, \Big\{ \left[ \, {\overline A}_{\omega\gamma} 2 n - 4 A_{\omega\gamma} (2- \beta \sqrt{4 - 2 n}) \right] (\dw^{-\alpha, \beta} - \gamma) \\ 
& + n \omega  \left( \widetilde{Q}_{\omega\gamma} - \widetilde{Q}^{(0)}_{\omega\gamma} \right) (3 - \beta \sqrt{4 - 2 n} - 2 \gamma \dw^{\alpha, \beta}  ) + 2 \omega \left(G_{2\omega\gamma} - G^{(0)}_{2\omega\gamma}\right) \left[ (2- \beta \sqrt{4 - 2 n}) (4 - \gamma \dw^{\alpha, \beta}  ) - 2 n   \right] \Big\} \notag .
\end{align}


\subsubsection{Two Remaining Constraints}

Two constraints that remain to be satisfied are given by Eqs.~\eqref{condition1b} and \eqref{conditions2b}. Solving these will allow us to obtain expressions for our final two remaining unknowns, the double-Laplace images $G_{2\omega\gamma}$ and $\widetilde{Q}_{\omega\gamma}$, and will complete our solution of the evolution equations \eqref{evoleqs_IR_1}. We rewrite the two constraints below:
\begin{subequations}\label{conditions_b}
\begin{align}
& 0 = e^{- \omega \soz} \gint  e^{\gamma\soz} \, \left[ \widetilde{G}^{(0)}_{\omega\gamma} + 2 \, G^{(0)}_{2\omega\gamma} \right] - \int \frac{d \omega'}{2\pi i } \sum_{\alpha,\beta = +,-} \frac{1}{\omega - \delta_{\omega'}^{\alpha\beta}} \,  \widetilde{\Gamma}_{\omega'}^{(\alpha\beta)}(\soz)  \\
    &\hspace{1cm} +  
   \frac{1}{\omega} \int \frac{d \omega'}{2\pi i } \sum_{\alpha,\beta = +,-} \widetilde{\Gamma}_{\omega'}^{(\alpha\beta)}(\soz) \frac{\omega'-\delta_{\omega'}^{\alpha\beta}}{\omega'}\, e^{\delta_{\omega'}^{\alpha\beta}\soz} , \notag \\
   & 0 = e^{- \omega \soz} \gint  e^{\gamma\soz} \, \left[ Q^{(0)}_{\omega\gamma} + 2 \, G^{(0)}_{2\omega\gamma} \right] - \int \frac{d \omega'}{2\pi i } \sum_{\alpha,\beta = +,-} \frac{1}{\omega - \delta_{\omega'}^{\alpha\beta}} \,  \widetilde{\Gamma}_{\omega'}^{(\alpha\beta)}(\soz) \, \frac{1 + \beta \, \sqrt{ 1 - \frac{n}{2}}}{n / 4 }   \\
    &\hspace{1cm} +  
   \frac{1}{\omega} \int \frac{d \omega'}{2\pi i } \sum_{\alpha,\beta = +,-} \widetilde{\Gamma}_{\omega'}^{(\alpha\beta)}(\soz) \frac{\omega'-\delta_{\omega'}^{\alpha\beta}}{\omega'}\, \frac{1 + \beta \, \sqrt{ 1 - \frac{n}{2} }}{n/ 4  }  \, e^{\delta_{\omega'}^{\alpha\beta}\soz} . \notag
\end{align}
\end{subequations}
It is straightforward to show using \eq{Gamma_final} that
\begin{align}\label{termscaling}
e^{\dw^{+\beta}\soz} \,  \widetilde{\Gamma}_{\omega}^{(+ \beta)}(\soz)  \to 0, \ \ \  e^{\dw^{-\beta}\soz} \, \omega \,  \widetilde{\Gamma}_{\omega}^{(- \beta)}(\soz)  \to 0, \ \ \ \textrm{when} \ \ \ \omega \to \infty.  
\end{align}
Then we can close the $\omega'$ contour to the right in the last term of each of the equations \eqref{conditions_b} and obtain zero. What remains of Eqs.~\eqref{conditions_b} can be rewritten using the identity in \eq{deltasumrule} along with another identity relating the $\dw^{\alpha\beta}$,
\begin{align}\label{deltaproductrule}
    \dw^{\alpha\beta}\dw^{-\alpha,\beta} = -\beta\sqrt{1-\frac{n}{2}} ,
\end{align}
which can be straightforwardly shown from the definition in \eq{deltapmpm}. We obtain
\begin{subequations}\label{conditions_b3}
\begin{align}
&  e^{- \omega \soz} \gint  e^{\gamma\soz} \, \left[ \widetilde{G}^{(0)}_{\omega\gamma} + 2 \, G^{(0)}_{2\omega\gamma} \right] = -  \frac{1}{\omega} \int \frac{d \omega'}{2\pi i } \sum_{\alpha,\beta = +,-}  \frac{\omega - \delta_{\omega'}^{-\alpha , \beta}}{\omega' - \left( \omega - \frac{\beta}{\omega} \, \sqrt{ 1 - \frac{n}{2}} \right) } \, \widetilde{\Gamma}_{\omega'}^{(\alpha\beta)}(\soz) ,  \\
   & e^{- \omega \soz} \gint  e^{\gamma\soz} \, \left[ Q^{(0)}_{\omega\gamma} + 2 \, G^{(0)}_{2\omega\gamma} \right] = -  \frac{1}{\omega} \int \frac{d \omega'}{2\pi i } \sum_{\alpha,\beta = +,-} \frac{\omega - \delta_{\omega'}^{-\alpha , \beta}}{\omega' - \left( \omega - \frac{\beta}{\omega} \, \sqrt{ 1 - \frac{n}{2}} \right) }  \,  \widetilde{\Gamma}_{\omega'}^{(\alpha\beta)}(\soz) \, \frac{1 + \beta \, \sqrt{ 1 - \frac{n}{2} }}{n/ 4 } .
\end{align}
\end{subequations}
On the right hand side of each of Eqs.~\eqref{conditions_b3}, we can close the contour to the right, picking up the poles at $\omega' = \omega - \tfrac{\beta}{\omega}\sqrt{1-\tfrac{n}{2}}$. One can show using \eq{deltapmpm} that
\begin{align}\label{deltasatpole}
\delta_{\omega - \frac{\beta}{\omega} \, \sqrt{ 1 - \frac{n}{2}} }^{\alpha , \beta} = \thalf \, \left[ \omega \, (1+\alpha) - (1-\alpha) \, \frac{\beta}{\omega} \, \sqrt{ 1 - \frac{n}{2}}  \right],
\end{align}
which subsequently gives
\begin{align}\label{deltasatpole2}
\delta_{\omega - \frac{\beta}{\omega} \, \sqrt{ 1 - \frac{n}{2}} }^{+ , \beta} = \omega, \ \ \ \delta_{\omega - \frac{\beta}{\omega} \, \sqrt{ 1 - \frac{n}{2 }} }^{- , \beta} = - \frac{\beta}{\omega} \, \sqrt{ 1 - \frac{n}{2}} . 
\end{align}
Then, in view of the factor $\omega - \delta_{\omega'}^{-\alpha,\beta}$ in both of Eqs.~\eqref{conditions_b3}, we conclude that the residues of the $\omega' = \omega - \tfrac{\beta}{\omega}\sqrt{1-\tfrac{n}{2}}$ poles for $\alpha = -$ are zero, leaving only the residues from the $\alpha = +$ contribution. Eqs.~\eqref{conditions_b3} thus become
\begin{subequations}\label{conditions_b4}
\begin{align}
&  e^{- \omega \soz} \gint  e^{\gamma\soz} \, \left[ \widetilde{G}^{(0)}_{\omega\gamma} + 2 \, G^{(0)}_{2\omega\gamma} \right] = \sum_{\beta = +,-} \left(1 +  \frac{\beta}{\omega^2} \, \sqrt{ 1 - \frac{n}{2}}  \right) \,   \widetilde{\Gamma}_{\omega - \frac{\beta}{\omega} \, \sqrt{ 1 - \frac{n}{2}}}^{(+\beta)}(\soz) ,  \\
   & e^{- \omega \soz} \gint  e^{\gamma\soz} \, \left[ Q^{(0)}_{\omega\gamma} + 2 \, G^{(0)}_{2\omega\gamma} \right] = \sum_{\beta = +,-} \left(1 +  \frac{\beta}{\omega^2} \, \sqrt{ 1 - \frac{n}{2}}  \right) \,   \widetilde{\Gamma}_{\omega - \frac{\beta}{\omega} \, \sqrt{ 1 - \frac{n}{2}}}^{(+\beta)}(\soz) \, \frac{1 + \beta \, \sqrt{ 1 - \frac{n}{2} }}{n / 4 } .
\end{align}
\end{subequations}

Next, we recall \eq{Gammatildestructure}, which, along with the first equality in \eq{deltasatpole2} tells us that
\begin{align}
    \widetilde{\Gamma}_{\omega-\tfrac{\beta}{\omega}\sqrt{1-\tfrac{n}{2}}}^{(+\beta)}(\soz) = e^{-\omega\soz}\gint \, e^{\gamma\soz} \, \widetilde{\Gamma}_{\omega-\tfrac{\beta}{\omega}\sqrt{1-\tfrac{n}{2}},\gamma}^{(+\beta)}.
\end{align}
Using this in Eqs.~\eqref{conditions_b4} and writing out the sum over $\beta$ explicitly, we have
\begin{subequations}\label{conditions_b6}
\begin{align}
&  \gint  e^{\gamma\soz} \, \left[ \widetilde{G}^{(0)}_{\omega\gamma} + 2 \, G^{(0)}_{2\omega\gamma} \right] = \left(1 +  \frac{1}{\omega^2} \, \sqrt{ 1 - \frac{n}{2}}  \right) \,   \gint e^{\gamma\soz} \widetilde{\Gamma}_{\omega - \frac{1}{\omega} \, \sqrt{ 1 - \frac{n}{2}},\gamma}^{(++)} \\
&\hspace{4.5cm} + \left(1 -  \frac{1}{\omega^2} \, \sqrt{ 1 - \frac{n}{2}}  \right) \,   \gint e^{\gamma\soz}\widetilde{\Gamma}_{\omega + \frac{1}{\omega} \, \sqrt{ 1 - \frac{n}{2}},\gamma}^{(+-)} \notag\,,  \\
   &\gint  e^{\gamma\soz} \, \left[ Q^{(0)}_{\omega\gamma} + 2 \, G^{(0)}_{2\omega\gamma} \right] =  \left(1 +  \frac{1}{\omega^2} \, \sqrt{ 1 - \frac{n}{2}}  \right) \,   \gint e^{\gamma\soz}\widetilde{\Gamma}_{\omega - \frac{1}{\omega} \, \sqrt{ 1 - \frac{n}{2}},\gamma}^{(++)} \, \frac{1 + \sqrt{ 1 - \frac{n}{2} }}{n / 4 } \\
   & \hspace{4.5cm} \notag +  \left(1 -  \frac{1}{\omega^2} \, \sqrt{ 1 - \frac{n}{2}}  \right) \,   \gint e^{\gamma\soz}\widetilde{\Gamma}_{\omega + \frac{1}{\omega} \, \sqrt{ 1 - \frac{n}{2}},\gamma}^{(+-)} \, \frac{1 - \sqrt{ 1 - \frac{n}{2} }}{n / 4 }.
\end{align}
\end{subequations}
Equations~\eqref{conditions_b6} can be written more compactly as a single equation,
\begin{align}\label{conditions_b8}
& \gint  e^{\gamma\soz} \, \left\{\frac{n}{4 } \left[ Q^{(0)}_{\omega\gamma} + 2 \, G^{(0)}_{2\omega\gamma} \right] - \left[1 - \beta \, \sqrt{ 1 - \frac{n}{2}}  \right]  \left[ \widetilde{G}^{(0)}_{\omega\gamma} + 2 \, G^{(0)}_{2\omega\gamma} \right] \right\} \\ 
& \hspace{5cm} \notag = \beta \, \left(1 +  \frac{\beta}{\omega^2} \, \sqrt{ 1 - \frac{n}{2}}  \right) \,   \gint e^{\gamma\soz}\widetilde{\Gamma}_{\omega - \frac{\beta}{\omega} \, \sqrt{ 1 - \frac{n}{2}},\gamma}^{(+\beta)} \, 2 \, \sqrt{ 1 - \frac{n}{2}} .
\end{align}

Recalling \eq{deltasatpole2}, we can rewrite this is as
\begin{align}\label{conditions_b9}
& \gint  e^{\gamma\soz} \, \left\{\frac{n}{4} \left[ Q^{(0)}_{\delta_{\omega - \frac{\beta}{\omega} \, \sqrt{ 1 - \frac{n}{2}} }^{+ , \beta} \ \gamma} + 2 \, G^{(0)}_{2 \ \delta_{\omega - \frac{\beta}{\omega} \, \sqrt{ 1 - \frac{n}{2}} }^{+ , \beta} \ \gamma} \right] \right. \\
&\hspace{2.5cm} \left. - \left[1 - \beta \, \sqrt{ 1 - \frac{n}{2}}  \right] \times  \left[ \widetilde{G}^{(0)}_{\delta_{\omega - \frac{\beta}{\omega} \, \sqrt{ 1 - \frac{n}{2}} }^{+ , \beta} \ \gamma} + 2 \, G^{(0)}_{2 \ \delta_{\omega - \frac{\beta}{\omega} \, \sqrt{ 1 - \frac{n}{2}} }^{+ , \beta} \ \gamma} \right] \right\} \notag \\
&\hspace{0.5cm}= \beta \, \left(1 +  \frac{\beta}{\left[ \delta_{\omega - \frac{\beta}{\omega} \, \sqrt{ 1 - \frac{n}{2}} }^{+ , \beta} \right]^2} \, \sqrt{ 1 - \frac{n}{2}}  \right) \,   \gint e^{\gamma\soz}\widetilde{\Gamma}_{\omega - \frac{\beta}{\omega} \, \sqrt{ 1 - \frac{n}{2}},\gamma}^{(+\beta)} \, 2 \, \sqrt{ 1 - \frac{n}{2 }} , \notag
\end{align}
which, upon the substitution 
\begin{align}
\omega - \frac{\beta}{\omega} \, \sqrt{ 1 - \frac{n}{2}} \to \omega \, ,
\end{align}
yields
\begin{align}\label{conditions_b11}
& \gint  e^{\gamma\soz} \, \left\{\frac{n}{4} \left[ Q^{(0)}_{\delta_{\omega }^{+ , \beta} \ \gamma} + 2 \, G^{(0)}_{2 \ \delta_{\omega  }^{+ , \beta} \ \gamma} \right] - \left[1 - \beta \, \sqrt{ 1 - \frac{n}{2}}  \right] \,  \left[ \widetilde{G}^{(0)}_{\delta_{\omega  }^{+ , \beta} \ \gamma} + 2 \, G^{(0)}_{2 \ \delta_{\omega }^{+ , \beta} \ \gamma} \right] \right\} \\
& \hspace{7cm}  = \beta \, \left(1 +  \frac{\beta}{\left[ \delta_{\omega }^{+ , \beta} \right]^2} \, \sqrt{ 1 - \frac{n}{2}}  \right) \, 2 \, \sqrt{ 1 - \frac{n}{2}} \,  \gint  e^{\gamma\soz} \, \widetilde{\Gamma}^{+ , \beta}_{\omega \gamma}  . \notag
\end{align}

We begin to invert the remaining inverse Laplace transform by writing
\begin{align}\label{conditions_b12}
& \frac{n}{4} \left[ Q^{(0)}_{\delta_{\omega }^{+ , \beta} \ \gamma} + 2 \, G^{(0)}_{2 \ \delta_{\omega  }^{+ , \beta} \ \gamma} \right] - \left[1 - \beta \, \sqrt{ 1 - \frac{n}{2}}  \right] \,  \left[ \widetilde{G}^{(0)}_{\delta_{\omega  }^{+ , \beta} \ \gamma} + 2 \, G^{(0)}_{2 \ \delta_{\omega }^{+ , \beta} \ \gamma} \right]  \\
& \hspace{5cm}  = \beta \, \left(1 +  \frac{\beta}{\left[ \delta_{\omega }^{+ , \beta} \right]^2} \, \sqrt{ 1 - \frac{n}{2}}  \right) \, 2 \, \sqrt{ 1 - \frac{n}{2}} \,  \int \frac{d \gamma'}{2 \pi i} \, \frac{1}{\gamma - \gamma'} \, \widetilde{\Gamma}^{+ , \beta}_{\omega \gamma'}  . \notag
\end{align}
Next, we would like to complete the inversion of the Laplace transform on the right-hand side of \eq{conditions_b12}. To do so, we must carefully recall the structure of $\widetilde{\Gamma}^{+,\beta}_{\omega\gamma}$. \eq{Gammatildescompact} tells us that $\widetilde{\Gamma}^{+,\beta}_{\omega\gamma}$ depends on the functions $A_{\omega\gamma}$ and $\overline{A}_{\omega\gamma}$. Defined in Eqs.~\eqref{Aeq} and \eqref{Abar_eq}, these functions have poles at $\gamma=\dw^{++}$ and $\gamma=\dw^{+-}$. As discussed in the text following Eqs.~\eqref{all}, although these poles lie to the left of the $\omega$-contour, they lie to the right of the $\gamma'$-contour. Then closing the $\gamma'$-contour to the right in \eq{conditions_b12} requires us to pick up three poles: $\gamma'=\gamma$, $\gamma' = \dw^{++}$, and $\gamma' = \dw^{+-}$. Doing so, we obtain 
\begin{align}\label{conditions_b13}
& \frac{n}{4} \left[ Q^{(0)}_{\delta_{\omega }^{+ , \beta} \ \gamma} + 2 \, G^{(0)}_{2 \ \delta_{\omega  }^{+ , \beta} \ \gamma} \right] - \left[1 - \beta \, \sqrt{ 1 - \frac{n}{2}}  \right] \,  \left[ \widetilde{G}^{(0)}_{\delta_{\omega  }^{+ , \beta} \ \gamma} + 2 \, G^{(0)}_{2 \ \delta_{\omega }^{+ , \beta} \ \gamma} \right]  \\
&  =  \frac{1}{8 (\dw^{+\beta} - \dw^{-, \beta}) } \,  \left(1 +  \frac{\beta}{\left[ \delta_{\omega }^{+ , \beta} \right]^2} \, \sqrt{ 1 - \frac{n}{2}}  \right) \,   \Bigg\{ \left[ \, {\overline A}_{\omega\gamma} 2 n - 4 A_{\omega\gamma} (2- \beta \sqrt{4 - 2 n}) \right] (\dw^{-, \beta} - \gamma) \notag  \\ 
& + n \omega  \left( \widetilde{Q}_{\omega\gamma} - \widetilde{Q}^{(0)}_{\omega\gamma} \right) (3 - \beta \sqrt{4 - 2 n} - 2 \gamma \dw^{+, \beta}  ) + 2 \omega \left(G_{2\omega\gamma} - G^{(0)}_{2\omega\gamma}\right) \left[ (2- \beta \sqrt{4 - 2 n}) (4 - \gamma \dw^{+, \beta}  ) - 2 n   \right] \notag \\ 
& - \frac{\dw^{-, \beta} - \dw^{++}}{\gamma - \dw^{++}} \lim_{\gamma' \to \dw^{++}} \Big[ (\gamma' -  \dw^{++}) \left[ \, {\overline A}_{\omega\gamma'} 2 n - 4 A_{\omega\gamma'} (2- \beta \sqrt{4 - 2 n}) \right]  \Big] \notag \\ 
& - \frac{\dw^{-, \beta} - \dw^{+-}}{\gamma - \dw^{+-}} \lim_{\gamma' \to \dw^{+-}} \Big[ (\gamma' -  \dw^{+-}) \left[ \, {\overline A}_{\omega\gamma'} 2 n - 4 A_{\omega\gamma'} (2- \beta \sqrt{4 - 2 n}) \right]  \Big]  \Bigg\} . \notag
\end{align}
Explicitly substituting $A_{\omega\gamma}$ and $\overline{A}_{\omega\gamma}$ from Eqs.~\eqref{Aeq} and \eqref{Abar_eq}, respectively, into \eq{conditions_b13}, after some lengthy but straightforward algebra, we can evaluate the residues (limits) in \eq{conditions_b13} and recast these two constraints (the first for $\beta = +$ and the second for $\beta = -$) as
\begin{subequations}\label{ppandpmcombo6}
\begin{align}
    \label{ppcombo6}
    &Q^{(0)}_{\dw^{++}\gamma} + 2G^{(0)}_{2\dw^{++}\gamma} - \frac{4}{n}\left(1-\sqrt{1-\frac{n}{2}}\right) \left( \widetilde{G}^{(0)}_{\dw^{++}\gamma} + 2G^{(0)}_{2\dw^{++}\gamma}\right) \\
    & = \frac{4}{n}\frac{1}{\gamma-\dw^{++}} \Bigg\{\frac{\omega}{4}\left(-2+\sqrt{4-2n}\right)\Bigg[G_{2\omega\gamma}\left(\gamma-r_1^{++}\right)\left(\gamma-r_1^{-+}\right) - G^{(0)}_{2\omega\gamma}\left(\left(\gamma-\dw^{++}\right)\left(\gamma-\dw^{-+}\right) + 2 - \frac{1}{2}\sqrt{4-2n}\right)\Bigg] \notag\\
    &\hspace{2.4cm} -\frac{n}{4} \, \omega \, \Bigg[ \widetilde{Q}_{\omega\gamma}\left(\gamma-r_2^{++}\right)\left(\gamma-r_2^{-+}\right) - \widetilde{Q}^{(0)}_{\omega\gamma}\left(\left(\gamma-\dw^{++}\right)\left(\gamma-\dw^{-+}\right) + \frac{3}{2}\right)\Bigg] - \left(\gamma\rightarrow\dw^{++}\right) \Bigg\} ,\notag \\  
    \label{pmcombo6}
    &Q^{(0)}_{\dw^{+-}\gamma} + 2G^{(0)}_{2\dw^{+-}\gamma} - \frac{4}{n}\left(1+\sqrt{1-\frac{n}{2}}\right) \left( \widetilde{G}^{(0)}_{\dw^{+-}\gamma} + 2G^{(0)}_{2\dw^{+-}\gamma}\right) \\
    & = -\frac{4}{n}\frac{1}{\gamma-\dw^{+-}} \Bigg\{\frac{\omega}{4}\left(2+\sqrt{4-2n}\right)\Bigg[G_{2\omega\gamma}\left(\gamma-r_1^{+-}\right)\left(\gamma-r_1^{--}\right) - G^{(0)}_{2\omega\gamma}\left(\left(\gamma-\dw^{+-}\right)\left(\gamma-\dw^{--}\right) + 2 + \frac{1}{2}\sqrt{4-2n}\right)\Bigg] \notag\\
    &\hspace{2.4cm} +\frac{n}{4} \, \omega \, \Bigg[ \widetilde{Q}_{\omega\gamma}\left(\gamma-r_2^{+-}\right)\left(\gamma-r_2^{--}\right) - \widetilde{Q}^{(0)}_{\omega\gamma}\left(\left(\gamma-\dw^{+-}\right)\left(\gamma-\dw^{--}\right) + \frac{3}{2}\right) \notag\Bigg]  -\left(\gamma\rightarrow\dw^{+-}\right) \Bigg\}\notag ,
\end{align}
\end{subequations}
where we have defined the roots of the second-order polynomials in $\gamma$ which multiply $G_{2\omega\gamma}$ and $\widetilde{Q}_{\omega\gamma}$ as
\begin{subequations}\label{rs}
\begin{align}
    \label{r1alphabeta}
    &r_1^{\alpha\beta} = \frac{1}{2}\left[\omega + \alpha \, \sqrt{\omega^2 - 8 \, \left(1- \beta \, \sqrt{1-  \frac{n}{2}}\right) \, \left( 1 - \frac{2}{\omega \, \delta^{+,\beta}_\omega} \right)  
    }     \,\,\right] ,\\
    \label{r2alphabeta}
    &r_2^{\alpha\beta} = \frac{1}{2}\left[\omega +\alpha \, \sqrt{\omega^2 - 2 - 4 \, \left(1- \beta \, \sqrt{1-  \frac{n}{2}}\right) \, \left( 1 - \frac{2}{\omega \, \delta^{+,\beta}_\omega} \right)  
    }    \,\,\right] \,.
\end{align}
\end{subequations}

Note that by properly accounting for the poles at $\gamma = \dw^{++}$ and $\gamma = \dw^{+-}$ which lie to the right of the $\gamma$-contour (that is, by picking up these poles when we inverted the inverse Laplace transform in \eq{conditions_b12}), we have ensured that there are no explicit poles at $\gamma = \dw^{++}$ and $\gamma = \dw^{+-}$ in Eqs.~\eqref{ppcombo6} and \eqref{pmcombo6}, respectively. 
This is consistent with the requirement that our inverse Laplace transform expressions \eqref{doubleLaplacestart} remain well defined with all the singularities of the integrand located to the left of the integration contours: for instance, requiring that $G_{2\omega\gamma}$ and $\widetilde{Q}_{\omega\gamma}$ have no singularities at $\gamma = \dw^{++}$ and $\gamma = \dw^{+-}$ appears to not lead to any contradictions in Eqs.~\eqref{ppandpmcombo6}.

We now have two equations \eqref{ppandpmcombo6} which we would like to solve for the double-Laplace images $G_{2\omega\gamma}$ and $\widetilde{Q}_{\omega\gamma}$: however, these functions appear in our equations with multiple different arguments, which complicates our task. The substitutions $\gamma\rightarrow \dw^{++}$ and $\gamma\rightarrow \dw^{+-}$ at the end of each of Eqs.~\eqref{ppandpmcombo6} also apply to the arguments of the double-Laplace images $G_{2\omega\gamma}$ and $Q_{\omega\gamma}$ as well, so that Eqs.~\eqref{ppandpmcombo6} contain terms proportional to $G_{2\omega\dw^{++}}$, $\widetilde{Q}_{\omega\dw^{++}}$, $G_{2\omega\dw^{+-}}$, and $\widetilde{Q}_{\omega\dw^{+-}}$. At first glance, in addition to $G_{2\omega\gamma}$ and $\widetilde{Q}_{\omega\gamma}$ we have these four additional functions to solve for: $G_{2\omega\dw^{++}}$, $\widetilde{Q}_{\omega\dw^{++}}$, $G_{2\omega\dw^{+-}}$, and $\widetilde{Q}_{\omega\dw^{+-}}$. However, only two specific linear combinations of these four unknowns enter Eqs.~\eqref{ppandpmcombo6}. They are 
\begin{subequations}
\begin{align}\label{Zpp}
    &Z^{(++)}(\omega) \equiv - \frac{\omega}{4}\left(-2+\sqrt{4-2n}\right) 
    \, G_{2\omega\dw^{++}}\left(\dw^{++} - r_1^{++}\right)\left(\dw^{++} - r_1^{-+}\right) + \frac{n}{4}\omega \widetilde{Q}_{\omega\dw^{++}} \left(\dw^{++} - r_2^{++}\right)\left(\dw^{++} - r_2^{-+}\right) ,\\
    \label{Zpm}
    &Z^{(+-)}(\omega) \equiv - \frac{\omega}{4}\left(2+\sqrt{4-2n}\right) \, G_{2\omega\dw^{+-}}\left(\dw^{+-} - r_1^{+-}\right)\left(\dw^{+-} - r_1^{--}\right) - \frac{n}{4}\omega \widetilde{Q}_{\omega\dw^{+-}} \left(\dw^{+-} - r_2^{+-}\right)\left(\dw^{+-} - r_2^{--}\right) .
\end{align}
\end{subequations}
In terms of these new objects, we re-write Eqs.~\eqref{ppandpmcombo6} as 
\begin{subequations}\label{ppandpmcombo6prime}
\begin{align}
    \label{ppcombo6prime}
    &Q^{(0)}_{\dw^{++}\gamma} + 2G^{(0)}_{2\dw^{++}\gamma} - \frac{4}{n}\left(1-\sqrt{1-\frac{n}{2}}\right) \left( \widetilde{G}^{(0)}_{\dw^{++}\gamma} + 2G^{(0)}_{2\dw^{++}\gamma}\right) \\
    & = \frac{4}{n}\frac{1}{\gamma-\dw^{++}} \Bigg\{\frac{\omega}{4}\left(-2+\sqrt{4-2n}\right)\Bigg[G_{2\omega\gamma}\left(\gamma-r_1^{++}\right)\left(\gamma-r_1^{-+}\right) - G^{(0)}_{2\omega\gamma}\left(\left(\gamma-\dw^{++}\right)\left(\gamma-\dw^{-+}\right) + 2 - \frac{1}{2}\sqrt{4-2n}\right) \notag \\
    &\hspace{5.5cm}+ G^{(0)}_{2\omega\dw^{++}}\left(2-\frac{1}{2}\sqrt{4-2n} \right) \Bigg] \notag\\
    &\hspace{2.4cm} -\frac{n}{4}\omega\Bigg[ \widetilde{Q}_{\omega\gamma}\left(\gamma-r_2^{++}\right)\left(\gamma-r_2^{-+}\right) - \widetilde{Q}^{(0)}_{\omega\gamma}\left(\left(\gamma-\dw^{++}\right)\left(\gamma-\dw^{-+}\right) + \frac{3}{2}\right) + \frac{3}{2}\widetilde{Q}^{(0)}_{\omega\dw^{++}}\Bigg] + Z^{(++)}(\omega) \Bigg\} ,\notag \\  
    \label{pmcombo6prime}
    &Q^{(0)}_{\dw^{+-}\gamma} + 2G^{(0)}_{2\dw^{+-}\gamma} - \frac{4}{n}\left(1+\sqrt{1-\frac{n}{2}}\right) \left( \widetilde{G}^{(0)}_{\dw^{+-}\gamma} + 2G^{(0)}_{2\dw^{+-}\gamma}\right) \\
    & = -\frac{4}{n}\frac{1}{\gamma-\dw^{+-}} \Bigg\{\frac{\omega}{4}\left(2+\sqrt{4-2n}\right)\Bigg[G_{2\omega\gamma}\left(\gamma-r_1^{+-}\right)\left(\gamma-r_1^{--}\right) - G^{(0)}_{2\omega\gamma}\left(\left(\gamma-\dw^{+-}\right)\left(\gamma-\dw^{--}\right) + 2 + \frac{1}{2}\sqrt{4-2n}\right) \notag \\
    &\hspace{5.5cm}+ G^{(0)}_{2\omega\dw^{+-}}\left(2+\frac{1}{2}\sqrt{4-2n}\right)  \Bigg] \notag\\
    &\hspace{2.4cm} +\frac{n}{4}\omega\Bigg[ \widetilde{Q}_{\omega\gamma}\left(\gamma-r_2^{+-}\right)\left(\gamma-r_2^{--}\right) - \widetilde{Q}^{(0)}_{\omega\gamma}\left(\left(\gamma-\dw^{+-}\right)\left(\gamma-\dw^{--}\right) + \frac{3}{2}\right) + \frac{3}{2}\widetilde{Q}^{(0)}_{\omega\dw^{+-}}\Bigg] + Z^{(+-)}(\omega) \Bigg\} .\notag 
\end{align}
\end{subequations}
Solving \eq{ppcombo6prime} algebraically for $G_{2\omega\gamma}$ (still in terms of the unknowns $\widetilde{Q}_{\omega\gamma}$ and $Z^{(++)}(\omega)$), and substituting the result into \eq{pmcombo6prime}, we obtain
\begin{align}\label{pmcombo7}
    &\frac{2+\sqrt{4-2n}}{-2+\sqrt{4-2n}}\frac{\left(\gamma-r_1^{+-}\right)\left(\gamma-r_1^{--}\right)}{\left(\gamma-r_1^{++}\right)\left(\gamma-r_1^{-+}\right)}Z^{(++)}(\omega) - Z^{(+-)}(\omega) \\
    & = \frac{\omega}{4}\left(2+\sqrt{4-2n}\right)\Bigg\{ \frac{\left(\gamma-r_1^{+-}\right)\left(\gamma-r_1^{--}\right)}{\left(\gamma-r_1^{++}\right)\left(\gamma-r_1^{-+}\right)} \bigg[G^{(0)}_{2\omega\gamma}\left(\left(\gamma-\dw^{++}\right)\left(\gamma-\dw^{-+}\right) + 2 - \frac{1}{2}\sqrt{4-2n}  \right) \notag \\
    & \hspace{6.8cm}- G^{(0)}_{2\omega\dw^{++}}\left(2-\frac{1}{2}\sqrt{4-2n}\right) \bigg] \notag \\
    &\hspace{3.9cm}- G^{(0)}_{2\omega\gamma}\left(\left(\gamma-\dw^{+-}\right)\left(\gamma-\dw^{--}\right) + 2 + \frac{1}{2}\sqrt{4-2n}  \right) + G^{(0)}_{2\omega\dw^{+-}}\left(2+\frac{1}{2}\sqrt{4-2n}\right)\Bigg\} \notag \\
    & + \frac{n}{4}\omega\Bigg\{ \frac{2+\sqrt{4-2n}}{-2+\sqrt{4-2n}}\frac{\left(\gamma-r_1^{+-}\right)\left(\gamma-r_1^{--}\right)}{\left(\gamma-r_1^{++}\right)\left(\gamma-r_1^{-+}\right)} \bigg[ -\widetilde{Q}^{(0)}_{\omega\gamma}\left(\left(\gamma-\dw^{++}\right)\left(\gamma-\dw^{-+}\right) + \frac{3}{2}\right) + \widetilde{Q}^{(0)}_{\omega\dw^{++}}\left(\frac{3}{2}\right) \notag \\
    &\hspace{5.8cm} + \frac{\gamma-\dw^{++}}{\omega} \left(Q^{(0)}_{\dw^{++}\gamma} + 2G^{(0)}_{2\dw^{++}\gamma} - \frac{4}{n}\left(1-\sqrt{1-\frac{n}{4}}\right)\left(\widetilde{G}^{(0)}_{\dw^{++}\gamma} + 2G^{(0)}_{2\dw^{++}\gamma}\right) \right) \bigg] \notag \\
    &\hspace{1.3cm} - \widetilde{Q}^{(0)}_{\omega\gamma}\left(\left(\gamma-\dw^{+-}\right)\left(\gamma-\dw^{--}\right) + \frac{3}{2}  \right) + \widetilde{Q}^{(0)}_{\omega\dw^{+-}}\left(\frac{3}{2}\right) \notag \\
    &\hspace{1.3cm}+  \frac{\gamma-\dw^{+-}}{\omega} \left(Q^{(0)}_{\dw^{+-}\gamma} + 2G^{(0)}_{2\dw^{+-}\gamma} - \frac{4}{n}\left(1+\sqrt{1-\frac{n}{4}}\right)\left(\widetilde{G}^{(0)}_{\dw^{+-}\gamma} + 2G^{(0)}_{2\dw^{+-}\gamma}\right) \right)
    \Bigg\} \notag \\
    & + \frac{n}{4}\omega \widetilde{Q}_{\omega\gamma} \left\{ \frac{2\sqrt{4-2n}}{\left(-2+\sqrt{4-2n}\right)\left(\gamma-r_1^{++}\right)\left(\gamma-r_1^{-+}\right)}\left(\gamma-\gw^{++}\right)\left(\gamma-\gw^{+-}\right)\left(\gamma-\gw^{-+}\right)\left(\gamma-\gw^{--}\right) \right\} \notag.
\end{align}
In writing the last line of \eq{pmcombo7} we have defined the following:
\begin{align}
    & 2\sqrt{4-2n}\left(\gamma - \gw^{++}\right)\left(\gamma - \gw^{+-}\right)\left(\gamma - \gw^{-+}\right)\left(\gamma - \gw^{--}\right) 
    \equiv \left(2+\sqrt{4-2n}\right)\left(\gamma-r_1^{+-}\right)\left(\gamma-r_1^{--}\right)\left(\gamma-r_2^{++}\right)\left(\gamma-r_2^{-+}\right) 
    \notag \\
    &\hspace{7cm}
    - \left(2-\sqrt{4-2n}\right)\left(\gamma-r_1^{++}\right)\left(\gamma-r_1^{-+}\right)\left(\gamma-r_2^{+-}\right)\left(\gamma-r_2^{--}\right), 
\end{align}
where the functions $\gw^{\pm\pm}$ are given by 

\begin{align}\label{gammapmpmfull}
    &\gamma_\omega^{\pm\pm} \equiv \frac{1}{2}\left[\omega \pm \sqrt{\omega^2 + s_1(\omega) \pm \sqrt{s_2(\omega)}}  \right] 
\end{align}
with
\begin{subequations}\label{gammapmpmfull12}
\begin{align} 
    \label{news1}
    &s_1(\omega) = -9 + \frac{2\left(8-3n \right)\left(\dw^{--} + \dw^{-+}\right) + 8\sqrt{4-2n}\left(\dw^{--} - \dw^{-+}\right)}{\omega \left(2-n\right)} , \\
    \label{news2}
    &s_2(\omega) = \frac{1}{\left(2-n\right)^2}\frac{1}{\omega^2} \Bigg\{\omega^2\left(2-n\right)^2\left(49-16n\right) - 64\left(2-n\right)\left(8-3n\right) + 8\omega\left(\dw^{--}+\dw^{-+} \right)\left(8 + 11n - 7n^2 \right) \\
    &\hspace{3.5cm} + 16\omega\left(\dw^{--}-\dw^{-+}\right) \sqrt{4 - 2n}\left(2 + 5n -2n^2\right) + 8n\dw^{--}\dw^{-+}\left(16 - 7n\right)\Bigg\} .\notag
\end{align}
\end{subequations}
Note that in obtaining Eqs.~\eqref{gammapmpmfull} and \eqref{gammapmpmfull12} we have extensively used the identities involving the functions $\dw^{\alpha\beta}$ in Eqs.~\eqref{deltasumrule} and \eqref{deltaproductrule}.

Equation~\eqref{pmcombo7} contains three unknowns: $\widetilde{Q}_{\omega\gamma}$, $Z^{(++)}(\omega)$, and $Z^{(+-)}(\omega)$. However, by evaluating \eq{pmcombo7} at particular values of $\gamma$, we could completely eliminate the term containing $\widetilde{Q}_{\omega\gamma}$ (the last line), as long as $\widetilde{Q}_{\omega\gamma}$ has no pole at our choice of $\gamma$.  As can be seen from Eqs.~\eqref{gammapmpmfull} and \eqref{gammapmpmfull12}, we have the scalings
\begin{align}\label{gammascaling}
    \gw^{++} \sim \gw^{+-} \sim \omega \qquad \text{when} \qquad \omega \rightarrow \infty ,
\end{align}
and so we conclude that $\widetilde{Q}_{\omega\gamma}$ cannot have a singularity at either $\gamma = \gw^{++}$ or $\gamma = \gw^{+-}$, for the same reasoning discussed in the text after \eq{polescalingslargew}, that is, to avoid having singularities to the right of the $\gamma$ integration contour. Thus, if we evaluate \eq{pmcombo7} first for $\gamma = \gw^{++}$ then again for $\gamma = \gw^{+-}$, while requiring that $\widetilde{Q}_{\omega\gamma}$ is finite at these values of $\gamma$, we will obtain two equations for the two unknowns $Z^{(++)}(\omega)$ and $Z^{(+-)}(\omega)$, which we can solve. Then, having explicit expressions for the functions $Z^{(++)}(\omega)$ and $Z^{(+-)}(\omega)$, we can construct $G_{2\omega\gamma}$ and $\widetilde{Q}_{\omega\gamma}$ by using Eqs.~\eqref{ppandpmcombo6prime}.
The system of two equations we can solve for $Z^{(++)}(\omega)$ and $Z^{(+-)}(\omega)$ is then
\begin{align}\label{ZppZmmsystem}
    &\Bigg[ - \frac{2+\sqrt{4-2n}}{-2+\sqrt{4-2n}}\frac{\left(\gamma-r_1^{+-}\right)\left(\gamma-r_1^{--}\right)}{\left(\gamma-r_1^{++}\right)\left(\gamma-r_1^{-+}\right)}Z^{(++)}(\omega) + Z^{(+-)}(\omega) \\
    & + \frac{\omega}{4}\left(2+\sqrt{4-2n}\right)\Bigg\{ \frac{\left(\gamma-r_1^{+-}\right)\left(\gamma-r_1^{--}\right)}{\left(\gamma-r_1^{++}\right)\left(\gamma-r_1^{-+}\right)} \bigg[G^{(0)}_{2\omega\gamma}\left(\left(\gamma-\dw^{++}\right)\left(\gamma-\dw^{-+}\right) + 2 - \frac{1}{2}\sqrt{4-2n}  \right) \notag \\
    & \hspace{6.8cm}- G^{(0)}_{2\omega\dw^{++}}\left(2-\frac{1}{2}\sqrt{4-2n}\right) \bigg] \notag \\
    &\hspace{3.9cm}- G^{(0)}_{2\omega\gamma}\left(\left(\gamma-\dw^{+-}\right)\left(\gamma-\dw^{--}\right) + 2 + \frac{1}{2}\sqrt{4-2n}  \right) + G^{(0)}_{2\omega\dw^{+-}}\left(2+\frac{1}{2}\sqrt{4-2n}\right)\Bigg\} \notag \\
    & + \frac{n}{4}\omega\Bigg\{ \frac{2+\sqrt{4-2n}}{-2+\sqrt{4-2n}}\frac{\left(\gamma-r_1^{+-}\right)\left(\gamma-r_1^{--}\right)}{\left(\gamma-r_1^{++}\right)\left(\gamma-r_1^{-+}\right)} \bigg[ -\widetilde{Q}^{(0)}_{\omega\gamma}\left(\left(\gamma-\dw^{++}\right)\left(\gamma-\dw^{-+}\right) + \frac{3}{2}\right) + \widetilde{Q}^{(0)}_{\omega\dw^{++}}\left(\frac{3}{2}\right) \notag \\
    &\hspace{5.8cm} + \frac{\gamma-\dw^{++}}{\omega} \left(Q^{(0)}_{\dw^{++}\gamma} + 2G^{(0)}_{2\dw^{++}\gamma} - \frac{4}{n}\left(1-\sqrt{1-\frac{n}{4}}\right)\left(\widetilde{G}^{(0)}_{\dw^{++}\gamma} + 2G^{(0)}_{2\dw^{++}\gamma}\right) \right) \bigg] \notag \\
    &\hspace{1.3cm} - \widetilde{Q}^{(0)}_{\omega\gamma}\left(\left(\gamma-\dw^{+-}\right)\left(\gamma-\dw^{--}\right) + \frac{3}{2}  \right) + \widetilde{Q}^{(0)}_{\omega\dw^{+-}}\left(\frac{3}{2}\right) \notag \\
    &\hspace{1.3cm}+  \frac{\gamma-\dw^{+-}}{\omega} \left(Q^{(0)}_{\dw^{+-}\gamma} + 2G^{(0)}_{2\dw^{+-}\gamma} - \frac{4}{n}\left(1+\sqrt{1-\frac{n}{4}}\right)\left(\widetilde{G}^{(0)}_{\dw^{+-}\gamma} + 2G^{(0)}_{2\dw^{+-}\gamma}\right) \right)
    \Bigg\}\Bigg]_{\gamma=\gw^{++}, \gamma=\gw^{+-}} = 0 \notag,
\end{align}
where the subscripts after the square bracket in the last line denote the fact that \eq{ZppZmmsystem} contains two separate equations, one with the left-hand side evaluated at $\gamma = \gw^{++}$ and another one with the left-hand side evaluated at $\gamma = \gw^{+-}$.

Solving this system of equations and substituting the results for $Z^{(++)}(\omega)$ and $Z^{(+-)}(\omega)$ back into Eqs.~\eqref{ppandpmcombo6prime}, we explicitly construct $G_{2\omega\gamma}$ and $\widetilde{Q}_{\omega\gamma}$. After some significant algebra, the results can be written as
\begin{subequations}\label{QtildeandG2full}
\begin{align}
    \label{newQomegagamma}
    &\widetilde{Q}_{\omega\gamma} = \frac{f^{(\widetilde{Q})}(\omega,\gamma) - \frac{2\left(\gamma-\gw^{+-}\right)\left(\gamma-\gw^{--}\right)}{\sqrt{s_2(\omega)}}f^{(\widetilde{Q})}(\omega,\gamma=\gw^{++}) + \frac{2\left(\gamma-\gw^{++}\right)\left(\gamma-\gw^{-+}\right)}{\sqrt{s_2(\omega)}}f^{(\widetilde{Q})}(\omega,\gamma=\gw^{+-}) }{2\sqrt{4-2n}\left(\gamma-\gw^{++}\right)\left(\gamma-\gw^{+-}\right)\left(\gamma-\gw^{-+}\right)\left(\gamma-\gw^{--}\right) } \, , \\
    \label{newG2omegagamma}
    &G_{2\omega\gamma} = -\,\frac{f^{(G_2)}(\omega,\gamma) - \frac{2\left(\gamma-\gw^{+-}\right)\left(\gamma-\gw^{--}\right)}{\sqrt{s_2(\omega)}}f^{(G_2)}(\omega,\gamma=\gw^{++}) + \frac{2\left(\gamma-\gw^{++}\right)\left(\gamma-\gw^{-+}\right)}{\sqrt{s_2(\omega)}}f^{(G_2)}(\omega,\gamma=\gw^{+-}) }{2\sqrt{4-2n}\left(\gamma-\gw^{++}\right)\left(\gamma-\gw^{+-}\right)\left(\gamma-\gw^{-+}\right)\left(\gamma-\gw^{--}\right) }  
\end{align}
\end{subequations}
with $\gw^{\alpha\beta}$ and $s_2(\omega)$ as defined in Eqs.~\eqref{gammapmpmfull} and \eqref{gammapmpmfull12}. We have defined two new quantities,
\begin{subequations}\label{fQtandfG2}
\begin{align}
    \label{fQt}
    &f^{(\widetilde{Q})}(\omega,\gamma) \equiv \left(\gamma-r_1^{+-}\right)\left(\gamma-r_1^{--}\right)f^{(+)}(\omega,\gamma) - \left(\gamma-r_1^{++}\right)\left(\gamma-r_1^{-+}\right)f^{(-)}(\omega,\gamma) \, , \\
    \label{fG2}
    &f^{(G_2)}(\omega,\gamma) \equiv n \frac{\left(\gamma-r_2^{+-}\right)\left(\gamma-r_2^{--}\right)  }{2+\sqrt{4-2n}} f^{(+)}(\omega,\gamma) - n \frac{\left(\gamma-r_2^{++}\right)\left(\gamma-r_2^{-+}\right)  }{2-\sqrt{4-2n}} f^{(-)}(\omega,\gamma) ,
\end{align}
\end{subequations}
where $r_1^{\alpha\beta}$ and $r_2^{\alpha\beta}$ are defined in Eqs.~\eqref{rs}, while
$f^{(+)}(\omega,\gamma)$ and $f^{(-)}(\omega,\gamma)$ are given by
\begin{subequations}
\begin{align}
    \label{fplus}
    &f^{(+)}(\omega,\gamma) \\
    &\hspace{.5cm} \equiv \bigg\{ 2G_{2\omega\gamma}^{(0)}\left[\left(\gamma-\dw^{++}\right)\left(\gamma-\dw^{-+}\right) + 2-\tfrac{1}{2}\sqrt{4-2n}  \right] + \widetilde{Q}^{(0)}_{\omega\gamma} \left(2+\sqrt{4-2n}\right)\left[\left(\gamma-\dw^{++}\right)\left(\gamma-\dw^{-+}\right) + \frac{3}{2}\right] \notag\\
    &\hspace{1cm} - \left(2+\sqrt{4-2n}\right)\frac{\gamma-\dw^{++}}{\omega}\left[Q^{(0)}_{\dw^{++}\gamma} + 2G^{(0)}_{2\dw^{++}\gamma} - \tfrac{4}{n}\left(1-\sqrt{1-\tfrac{n}{2}}\right)\left(\widetilde{G}^{(0)}_{\dw^{++}\gamma} + 2G^{(0)}_{2\dw^{++}\gamma}  \right)    \right]   \bigg\} - \left\{\gamma \rightarrow \dw^{++} \right\} \, ,  \notag\\
    \label{fminus}
    &f^{(-)}(\omega,\gamma) \\
    & \hspace{.5cm} \equiv \bigg\{ 2G_{2\omega\gamma}^{(0)}\left[\left(\gamma-\dw^{+-}\right)\left(\gamma-\dw^{--}\right) + 2+\tfrac{1}{2}\sqrt{4-2n}  \right] + \widetilde{Q}^{(0)}_{\omega\gamma} \left(2-\sqrt{4-2n}\right)\left[\left(\gamma-\dw^{+-}\right)\left(\gamma-\dw^{--}\right) + \frac{3}{2}\right] \notag \\
    &\hspace{1cm} - \left(2-\sqrt{4-2n}\right)\frac{\gamma-\dw^{+-}}{\omega}\left[Q^{(0)}_{\dw^{+-}\gamma} + 2G^{(0)}_{2\dw^{+-}\gamma} - \tfrac{4}{n}\left(1+\sqrt{1-\tfrac{n}{2}}\right)\left(\widetilde{G}^{(0)}_{\dw^{+-}\gamma} + 2G^{(0)}_{2\dw^{+-}\gamma}  \right)    \right]   \bigg\} - \left\{\gamma \rightarrow \dw^{+-} \right\} .\notag
\end{align}
\end{subequations}
All four new objects, $f^{(\widetilde{Q})}(\omega,\gamma)$, $f^{(G_2)}(\omega,\gamma)$, $f^{(+)}(\omega,\gamma)$, and $f^{(-)}(\omega,\gamma)$ are determined by the initial conditions/inhomogeneous terms in our evolution equations \eqref{evoleqs_IR_1}. 

One can easily show using Eqs.~\eqref{gammapmpmfull} and \eqref{gammapmpmfull12} that 
\begin{subequations}
\begin{align}
    &\frac{2\left(\gamma - \gw^{+-}\right)\left(\gamma - \gw^{--}\right)}{\sqrt{s_2(\omega)}}\bigg|_{\gamma = \gw^{++}} = 1 \qquad \text{and} \\
    &\frac{2\left(\gamma - \gw^{++}\right)\left(\gamma - \gw^{-+}\right)}{\sqrt{s_2(\omega)}}\bigg|_{\gamma = \gw^{+-}} = -1.
\end{align}
\end{subequations}
This makes it clear that both $\widetilde{Q}_{\omega\gamma}$ and $G_{2\omega\gamma}$ as expressed in Eqs.~\eqref{QtildeandG2full} have vanishing residues at both $\gamma = \gw^{++}$ and $\gamma = \gw^{+-}$, as indeed they must in order for the inverse Laplace transforms in Eqs.~\eqref{doubleLaplaceG2} and \eqref{doubleLaplaceQtilde} to be well defined, with the integrands having no singularities to the right of the integration contours. 

Having explicitly constructed $\widetilde{Q}_{\omega\gamma}$ and $G_{2\omega\gamma}$, the full solution of the evolution Eqs.~\eqref{evoleqs_IR_1} is now formally complete. In the next Section, we summarize the results of our calculation.


\section{Summary of the Solution}\label{sec:solnsummary}

Returning to the original variables $zs$, $z's$, $\xoz^2$, $\xto^2$ (see \eq{rescaledvars}) and employing the definition in \eq{alphabar}
we collect all the pieces here which form the complete analytic solution to the evolution equations \eqref{evoleqs_IR}. In addition, we undo the rescaling introduced in defining the variables in \eq{rescaledvars} by replacing
\begin{align}\label{resc}
 & \omega \to \frac{\omega}{\sqrt{\bas}}, \ \gamma \to \frac{\gamma}{\sqrt{\bas}}, \ \widetilde{Q}_{\omega\gamma} \to \bas \, \widetilde{Q}_{\omega\gamma}, \ G_{2\omega\gamma} \to \bas \, G_{2\omega\gamma}, \ G^{(0)}_{2\omega\gamma} \to \bas \, G^{(0)}_{2\omega\gamma}, \ \widetilde{G}^{(0)}_{\omega\gamma} \to \bas \, \widetilde{G}^{(0)}_{\omega\gamma}, \ Q^{(0)}_{\omega\gamma} \to \bas \, Q^{(0)}_{\omega\gamma}, \\ & \widetilde{Q}^{(0)}_{\omega\gamma} \to \bas \, \widetilde{Q}^{(0)}_{\omega\gamma}, \ r_1^{\alpha\beta} \to \frac{r_1^{\alpha\beta}}{\sqrt{\bas}}, \ r_2^{\alpha\beta} \to \frac{r_2^{\alpha\beta}}{\sqrt{\bas}}, \ f^{(\widetilde{Q})}(\omega,\gamma) \to \frac{f^{(\widetilde{Q})}(\omega,\gamma)}{\bas}, \  f^{(G_2)}(\omega,\gamma) \to \frac{f^{G_2}(\omega,\gamma)}{\bas}, \notag \\ & A_{\omega\gamma} \to \bas \, A_{\omega\gamma}, \  {\overline A}_{\omega\gamma} \to \bas \, {\overline A}_{\omega\gamma}, \ \widetilde{\Gamma}_{\omega}^{(\alpha\beta)}(\xoz^2) \to \sqrt{\bas} \, \widetilde{\Gamma}_{\omega}^{(\alpha\beta)}(\xoz^2), \ \delta_{\omega}^{\pm\pm} \to \frac{\delta_{\omega}^{\pm\pm}}{\sqrt{\bas}}, \ \gamma_{\omega}^{\pm\pm} \to \frac{\gamma_{\omega}^{\pm\pm}}{\sqrt{\bas}} ,\notag
\end{align}
with the arguments $\omega, \gamma$ of the double inverse Laplace transform images of the dipole amplitudes not reflecting the rescaling of $\omega, \gamma$. 
Recalling that $n = N_f/N_c$ we write
\begin{subequations}\label{fullsoln}\allowdisplaybreaks
\begin{align}
    \label{fullsolnQtilde}
    &\widetilde{Q}(\xoz^2,zs) = \wint \gint e^{\omega \, \ln(zs\xoz^2) + \gamma \, \ln\left(\tfrac{1}{\xoz^2\Lambda^2} \right)}\widetilde{Q}_{\omega\gamma}\,,\\
    \label{fullsolnG2}
    &G_2(\xoz^2,zs) = \wint \gint e^{\omega\, \ln(zs\xoz^2) + \gamma\, \ln\left(\tfrac{1}{\xoz^2\Lambda^2} \right)}G_{2\omega\gamma}\,, \\
    \label{fullsolnGtilde}
    &\widetilde{G}(\xoz^2,zs) = \wint \gint e^{\omega\, \ln(zs\xoz^2) + \gamma\, \ln\left(\tfrac{1}{\xoz^2\Lambda^2} \right)}\left[\frac{\omega\gamma}{2 \, \bas}\left(G_{2\omega\gamma} - G^{(0)}_{2\omega\gamma} \right) - 2 \, G_{2\omega\gamma}\right]\,, \\
    \label{fullsolnQ}
    &Q(\xoz^2,zs) = \wint \gint e^{\omega\, \ln(zs\xoz^2) + \gamma\, \ln\left(\tfrac{1}{\xoz^2\Lambda^2} \right)}\left[-\frac{\omega\gamma}{\bas} \, \left(\widetilde{Q}_{\omega\gamma} - \widetilde{Q}^{(0)}_{\omega\gamma}\right) - 2 \, G_{2\omega\gamma}\right]\,, \\
    \label{fullsolnGamma2}
    &\Gamma_2(\xoz^2,\xto^2,z's) = \wint \gint \bigg[e^{\omega\, \ln(z's\xto^2)+\gamma\, \ln\left(\tfrac{1}{\xoz^2\Lambda^2} \right)}\left(G_{2\omega\gamma} - G^{(0)}_{2\omega\gamma}  \right)  \\
    &\hspace{4.8cm}+ e^{\omega\, \ln(z's\xoz^2)+\gamma\, \ln\left(\tfrac{1}{\xoz^2\Lambda^2}\right)} G^{(0)}_{2\omega\gamma}  \bigg] \,, \notag\\
    \label{fullsolnGammatilde}
    &\widetilde{\Gamma}(\xoz^2,\xto^2,z's) =  \wint \, e^{\omega\, \ln(z's\xto^2)}\sum_{\alpha,\beta = +,-} e^{\dw^{\alpha\beta}\, \ln\left(\tfrac{1}{\xoz^2\Lambda^2}\right)} \, \widetilde{\Gamma}_\omega^{(\alpha\beta)}(\xoz^2) \\
    &\hspace{2cm} + \wint\gint \bigg[e^{\omega\, \ln(z's\xto^2)+\gamma\, \ln\left(\tfrac{1}{\xto^2\Lambda^2}\right)}A_{\omega\gamma} -2e^{\omega\, \ln(z's\xto^2)+\gamma\, \ln\tfrac{1}{\xoz^2\Lambda^2}}\left(G_{2\omega\gamma} - G^{(0)}_{2\omega\gamma}\right) \notag \\
    &\hspace{4.8cm}-2e^{\omega\, \ln(z's\xoz^2)+\gamma\, \ln\left(\tfrac{1}{\xoz^2\Lambda^2}\right)}G^{(0)}_{2\omega\gamma} \bigg] , \notag \\
    \label{fullsolnGammabar}
    &\overline{\Gamma}(\xoz^2,\xto^2,z's) = \wint \, e^{\omega\, \ln(z's\xto^2)}\sum_{\alpha,\beta = +,-} e^{\dw^{\alpha\beta}\, \ln\left(\tfrac{1}{\xto^2\Lambda^2}\right)} \frac{1 + \beta \, \sqrt{ 1 - \frac{N_f}{2 \, N_c} }}{N_f / (4 N_c) }  \,  \widetilde{\Gamma}_\omega^{(\alpha\beta)}(\xoz^2) \\
    &\hspace{2cm} + \wint\gint \bigg[ e^{\omega\, \ln(z's\xto^2)+\gamma\, \ln\left(\tfrac{1}{\xto^2\Lambda^2}\right)} \, {\overline A}_{\omega\gamma} - 2e^{\omega\, \ln(z's\xto^2)+\gamma\, \ln\left(\tfrac{1}{\xoz^2\Lambda^2}\right)}\left(G_{2\omega\gamma} - G^{(0)}_{2\omega\gamma}\right) \notag \\
    &\hspace{4.8cm}- 2e^{\omega\, \ln(z's\xoz^2)+\gamma\, \ln\left(\tfrac{1}{\xoz^2\Lambda^2}\right)}G^{(0)}_{2\omega\gamma} \bigg] , \notag
    \end{align}
\end{subequations}
\vspace{.5cm}
with 
\vspace{.5cm}
\begin{subequations}\label{fullsoln1}\allowdisplaybreaks
\begin{align}
    \label{fullsoln1Qomegagamma}
    &\widetilde{Q}_{\omega\gamma} = \frac{f^{(\widetilde{Q})}(\omega,\gamma) - \frac{2\left(\gamma-\gw^{+-}\right)\left(\gamma-\gw^{--}\right)}{\bas \, \sqrt{s_2(\omega)}}f^{(\widetilde{Q})}(\omega,\gamma=\gw^{++}) + \frac{2\left(\gamma-\gw^{++}\right)\left(\gamma-\gw^{-+}\right)}{\bas \, \sqrt{s_2(\omega)}}f^{(\widetilde{Q})}(\omega,\gamma=\gw^{+-}) }{2\sqrt{4-\frac{2N_f}{N_c}}\left(\gamma-\gw^{++}\right)\left(\gamma-\gw^{+-}\right)\left(\gamma-\gw^{-+}\right)\left(\gamma-\gw^{--}\right) }\,, \\
    \label{fullsoln1G2omegagamma}
    &G_{2\omega\gamma} = -\,\frac{f^{(G_2)}(\omega,\gamma) - \frac{2\left(\gamma-\gw^{+-}\right)\left(\gamma-\gw^{--}\right)}{\bas \, \sqrt{s_2(\omega)}}f^{(G_2)}(\omega,\gamma=\gw^{++}) + \frac{2\left(\gamma-\gw^{++}\right)\left(\gamma-\gw^{-+}\right)}{\bas \, \sqrt{s_2(\omega)}}f^{(G_2)}(\omega,\gamma=\gw^{+-}) }{2\sqrt{4-\frac{2N_f}{N_c}}\left(\gamma-\gw^{++}\right)\left(\gamma-\gw^{+-}\right)\left(\gamma-\gw^{-+}\right)\left(\gamma-\gw^{--}\right) } \,, \\
    \label{fullsoln1fQt}
    &f^{(\widetilde{Q})}(\omega,\gamma) = \left(\gamma-r_1^{+-}\right)\left(\gamma-r_1^{--}\right)f^{(+)}(\omega,\gamma) - \left(\gamma-r_1^{++}\right)\left(\gamma-r_1^{-+}\right)f^{(-)}(\omega,\gamma) \,,\\
    \label{fullsoln1fG2}
    &f^{(G_2)}(\omega,\gamma) = \frac{N_f}{N_c} \frac{\left(\gamma-r_2^{+-}\right)\left(\gamma-r_2^{--}\right)  }{2+\sqrt{4-\frac{2N_f}{N_c}}} f^{(+)}(\omega,\gamma) - \frac{N_f}{N_c} \frac{\left(\gamma-r_2^{++}\right)\left(\gamma-r_2^{-+}\right)  }{2-\sqrt{4-\frac{2N_f}{N_c}}} f^{(-)}(\omega,\gamma) \,, \\
    \label{fullsolnfplus}
    &f^{(+)}(\omega,\gamma) \\
    & = \bigg\{ 2G_{2\omega\gamma}^{(0)}\left[\left(\gamma-\dw^{++}\right)\left(\gamma-\dw^{-+}\right) + \bas \, \left( 2-\tfrac{1}{2}\sqrt{4-\tfrac{2N_f}{N_c}} \right)  \right] + \widetilde{Q}^{(0)}_{\omega\gamma} \left(2+\sqrt{4-\tfrac{2N_f}{N_c}}\right)\left[\left(\gamma-\dw^{++}\right)\left(\gamma-\dw^{-+}\right) + \frac{3}{2} \, \bas \right] \notag\\
    & - \bas \left(2+\sqrt{4-\tfrac{2N_f}{N_c}}\right)\frac{\gamma-\dw^{++}}{\omega}\left[Q^{(0)}_{\dw^{++}\gamma} + 2G^{(0)}_{2\dw^{++}\gamma} - \tfrac{4N_c}{N_f}\left(1-\sqrt{1-\tfrac{N_f}{2N_c}}\right)\left(\widetilde{G}^{(0)}_{\dw^{++}\gamma} + 2G^{(0)}_{2\dw^{++}\gamma}  \right)    \right]   \bigg\} - \left\{\gamma \rightarrow \dw^{++} \right\}\,, \notag\\
    \label{fullsolnfminus}
    &f^{(-)}(\omega,\gamma) \\
    & = \bigg\{ 2G_{2\omega\gamma}^{(0)}\left[\left(\gamma-\dw^{+-}\right)\left(\gamma-\dw^{--}\right) + \bas \, \left( 2+\tfrac{1}{2}\sqrt{4-\tfrac{2N_f}{N_c}} \right)  \right] + \widetilde{Q}^{(0)}_{\omega\gamma} \left(2-\sqrt{4-\tfrac{2N_f}{N_c}}\right)\left[\left(\gamma-\dw^{+-}\right)\left(\gamma-\dw^{--}\right) + \frac{3}{2} \, \bas \right] \notag \\
    & - \bas \left(2-\sqrt{4-\tfrac{2N_f}{N_c}}\right)\frac{\gamma-\dw^{+-}}{\omega}\left[Q^{(0)}_{\dw^{+-}\gamma} + 2G^{(0)}_{2\dw^{+-}\gamma} - \tfrac{4N_c}{N_f}\left(1+\sqrt{1-\tfrac{N_f}{2N_c}}\right)\left(\widetilde{G}^{(0)}_{\dw^{+-}\gamma} + 2G^{(0)}_{2\dw^{+-}\gamma}  \right)    \right]   \bigg\} - \left\{\gamma \rightarrow \dw^{+-} \right\} \,,\notag \\
    \label{fullsoln1r1alphabeta}
    & r_1^{\alpha\beta} = \frac{\omega}{2} \left[1 + \alpha \, \sqrt{1 - \frac{8 \, \bas}{\omega^2} \, \left(1- \beta \, \sqrt{1-  \frac{N_f}{2 \, N_c}}\right) \, \left( 1 - \frac{2 \, \bas}{\omega \, \delta^{+,\beta}_\omega} \right)  
    }     \,\,\right] \, ,  \\
    \label{fullsoln1r2alphabeta}
     & r_2^{\alpha\beta} = \frac{\omega}{2} \left[1 + \alpha \, \sqrt{1 - \frac{2 \, \bas}{\omega^2} - \frac{4 \, \bas}{\omega^2} \, \left(1- \beta \, \sqrt{1-  \frac{N_f}{2 \, N_c}}\right) \, \left( 1 - \frac{2 \, \bas}{\omega \, \delta^{+,\beta}_\omega} \right)  
    }     \,\,\right] \, , \\
    \label{fullsoln1doubleLaplaceG20}
    &G_2^{(0)}(\xoz^2,zs) = \wint \gint e^{\omega\, \ln(zs\xoz^2)+\gamma\, \ln\left(\tfrac{1}{\xoz^2\Lambda^2}\right)}G^{(0)}_{2\omega\gamma}\,, \\
    \label{fullsoln1doubleLaplaceGtilde0}
    &\widetilde{G}^{(0)}(\xoz^2,zs) = \wint \gint e^{\omega\, \ln(zs\xoz^2)+\gamma\, \ln\left(\tfrac{1}{\xoz^2\Lambda^2}\right)}\widetilde{G}^{(0)}_{\omega\gamma}\,, \\
    \label{fullsoln1doubleLaplaceQ0}
    &Q^{(0)}(\xoz^2,zs) = \wint \gint e^{\omega\, \ln(zs\xoz^2)+\gamma\, \ln\left(\tfrac{1}{\xoz^2\Lambda^2}\right)}Q^{(0)}_{\omega\gamma}\,, \\
    \label{fullsoln1doubleLaplaceQtilde0}
    &\widetilde{Q}^{(0)}(\xoz^2,zs) = \wint \gint e^{\omega\, \ln(zs\xoz^2)+\gamma\, \ln\left(\tfrac{1}{\xoz^2\Lambda^2}\right)}\widetilde{Q}^{(0)}_{\omega\gamma}\,,\\
    \label{fullsolnGammatildes}
    &  \widetilde{\Gamma}_{\omega}^{(\alpha\beta)}(\xoz^2) = e^{-\dw^{\alpha\beta}\, \ln\left(\tfrac{1}{\xoz^2\Lambda^2} \right) }\gint \, e^{\gamma \, \ln\left(\tfrac{1}{\xoz^2\Lambda^2} \right) } \frac{\beta}{8 (\dw^{\alpha\beta} - \dw^{-\alpha, \beta}) \, \sqrt{4 - \tfrac{2N_f}{N_c}}} \\ 
    & \times \, \Big\{ \left[ \, {\overline A}_{\omega\gamma} \tfrac{2N_f}{N_c} - 4 A_{\omega\gamma} \left(2- \beta \sqrt{4 - \tfrac{2N_f}{N_c}})\right) \right] (\dw^{-\alpha, \beta} - \gamma) + \tfrac{N_f}{N_c} \, \omega  \left( \widetilde{Q}_{\omega\gamma} - \widetilde{Q}^{(0)}_{\omega\gamma} \right) \left(3 - \beta \sqrt{4 - \tfrac{2N_f}{N_c}} - \frac{2}{\bas} \, \gamma \, \dw^{\alpha, \beta}  \right) \notag \\ 
    &\hspace{1cm} + 2 \, \omega \left(G_{2\omega\gamma} - G^{(0)}_{2\omega\gamma}\right) \left[ \left(2- \beta \sqrt{4 - \tfrac{2N_f}{N_c}}\right) \left(4 - \frac{1}{\bas} \, \gamma \, \dw^{\alpha, \beta}  \right) - \tfrac{2N_f}{N_c}   \right] \Big\} \notag \,,\\
    \label{fullsoln1Aeq}
    & A_{\omega\gamma} =  \frac{1}{\left(\gamma-\dw^{++} \right)\left(\gamma-\dw^{+-} \right)\left(\gamma-\dw^{-+} \right)\left(\gamma-\dw^{--} \right)} \, \Bigg\{  \frac{\omega\gamma}{2} \, \left[ \bas \, \left( 3 - \tfrac{N_f}{2 N_c} \right) - 3 \,  \gamma \, (\gamma - \omega) \right] \, \left(G_{2\omega\gamma} - G^{(0)}_{2\omega\gamma} \right) \notag \\
    &  
    + \bas \, \tfrac{3 N_f}{8 N_c} \, \omega \, \gamma \, \left( \widetilde{Q}_{\omega\gamma} - \widetilde{Q}^{(0)}_{\omega\gamma} \right)  
    + \bas \left[ 4 \, \gamma \, (\gamma - \omega) - \bas \, \left( 4 - \tfrac{N_f}{N_c} \right) \right] \, G_{2\omega\gamma} + \bas \, \tfrac{N_f}{2 N_c} \, \left[ \gamma \, (\gamma - \omega) - \bas \right] \, \widetilde{Q}_{\omega\gamma} \Bigg\},  \\
    \label{fullsoln1Abar_eq}
    & \overline{A}_{\omega\gamma} =  \frac{1}{\left(\gamma-\dw^{++} \right)\left(\gamma-\dw^{+-} \right)\left(\gamma-\dw^{-+} \right)\left(\gamma-\dw^{--} \right)} \, \Bigg\{ \omega\gamma \, [2 \, \bas - \gamma \, (\gamma - \omega)] \, \left(G_{2\omega\gamma} - G^{(0)}_{2\omega\gamma} \right) \\
    & \hspace{2cm} + 4 \, \bas \, [\gamma \, (\gamma - \omega) - \bas ] \, G_{2\omega\gamma} + \tfrac{3}{2} \, \omega\gamma \, [\gamma \, (\gamma - \omega) + \bas] \, \left( \widetilde{Q}_{\omega\gamma} - \widetilde{Q}^{(0)}_{\omega\gamma} \right) - \bas^2 \, \tfrac{N_f}{N_c} \, \widetilde{Q}_{\omega\gamma}  \Bigg\} \notag  \,,\\
    \label{fullsoln1deltapms}
    &\delta_{\omega}^{\pm\pm} = \frac{\omega}{2}\left[1 \pm \sqrt{1 \pm \frac{4 \, \bas}{\omega^2} \, \sqrt{1-\frac{N_f}{2N_c} } } \right] \,, \\
    \label{fullsoln1neweigs}
    &\gamma_\omega^{\pm\pm} = \frac{\omega}{2}\left[1 \pm \sqrt{1 + \frac{\bas}{\omega^2} \, \left[ s_1(\omega) \pm \sqrt{s_2(\omega)}\right] }  \, \right] \,,\\
    \label{fullsoln1news1}
    &s_1(\omega) = -9 + \frac{2\left( 8-3 \, \frac{N_f}{N_c} \right)\left(\dw^{--} + \dw^{-+}\right) + 8\sqrt{4-2 \, \frac{N_f}{N_c}}\left(\dw^{--} - \dw^{-+}\right)}{\omega \left(2-\frac{N_f}{N_c}\right)} \,,\\
    \label{fullsoln1news2}
    & s_2(\omega) = \frac{1}{\left(2-\frac{N_f}{N_c}\right)^2} \, \frac{1}{\omega^2} \, \Bigg\{\omega^2\left(2-\frac{N_f}{N_c}\right)^2\left(49-16 \, \frac{N_f}{N_c}\right) - 64 \, \bas \, \left(2-\frac{N_f}{N_c} \right)\left(8-3 \, \frac{N_f}{N_c} \right) \\
    & + 8 \, \omega \, \left(\dw^{--}+\dw^{-+} \right)\left[ 8 + 11 \, \frac{N_f}{N_c} - 7 \, \left( \frac{N_f}{N_c}\right)^2 \right]  
    + 16 \, \omega \, \left(\dw^{--}-\dw^{-+}\right) \sqrt{4 - 2\, \frac{N_f}{N_c}} \left[ 2 + 5\, \frac{N_f}{N_c} -2\, \left( \frac{N_f}{N_c} \right)^2\right] \notag \\ 
    & + 8 \, \frac{N_f}{N_c} \, \dw^{--} \, \dw^{-+} \, \left(16 - 7 \, \frac{N_f}{N_c} \right)\Bigg\} .\notag
\end{align}
\end{subequations}
The solution contained in Eqs.~\eqref{fullsoln} and \eqref{fullsoln1} represents a fully analytic solution to the small-$x$, large-$N_c\&N_f$ evolution equations~\eqref{evoleqs_IR}. It is valid for any initial conditions $G^{(0)}_{2}(\xoz^2,zs)$, $\widetilde{G}^{(0)}(\xoz^2,zs)$, $Q^{(0)}(\xoz^2,zs)$, $\widetilde{Q}^{(0)}(\xoz^2,zs)$. This is the main result of this work.

With the solution we have constructed here, we can straightforwardly express the gluon and flavor-singlet quark helicity PDFs within the DLA using Eqs.~\eqref{pdfsfromdipoles}. We immediately obtain
\begin{subequations}\label{fullsolnpdfs}
\begin{align}
    \label{fullsolnDeltaG}
    &\Delta G(x,Q^2) = \frac{2N_c}{\as \pi^2}\wint\gint e^{\omega\, \ln(1/x) + \gamma\, \ln(Q^2/\Lambda^2)}G_{2\omega\gamma}\,, \\
    \label{fullsolnDeltaSigma}
    &\Delta\Sigma(x,Q^2) = \frac{N_f}{\as \pi^2}\wint\gint e^{\omega \, \ln(1/x) + \gamma \, \ln(Q^2/\Lambda^2)}\widetilde{Q}_{\omega\gamma}\,.
\end{align}
\end{subequations}
Next, we employ \eq{g1_DLA} with our double-Laplace solution and carry out the integrals over $\xoz^2$ and $z$ to obtain
\begin{align}\label{fullsolng1}
    g_1(x,Q^2) = \sum_{f} \frac{Z_f^2}{\as 2\pi^2}\wint\gint \frac{\omega}{\omega-\gamma} \left(\widetilde{Q}_{\omega\gamma} - \widetilde{Q}^{(0)}_{\omega\gamma} \right)e^{\omega\, \ln(1/x) + \gamma\, \ln(Q^2/\Lambda^2)}.
\end{align}

We specify once again that $\text{Re}\,\omega > \text{Re}\,\gamma$ along the integration contours in all the double-inverse Laplace transforms in this Section.


\section{Connecting to DGLAP}\label{sec:DGLAP}

As the small-$x$ evolution studied here contains the resummation parameter $\as\ln(1/x)\ln(Q^2/\Lambda^2)$ (along with $\as \, \ln^2 (1/x)$, another resummation parameter), the solution constructed herein should contain the solution to the small-$x$ limit of the polarized DGLAP evolution equations \cite{Altarelli:1977zs,Dokshitzer:1977sg,Gribov:1972ri}. We can compare the DGLAP part of our results both to the predictions of BER \cite{Bartels:1996wc} and to the small-$x$, large-$N_c\&N_f$ limit of the existing finite order calculations \cite{Altarelli:1977zs,Dokshitzer:1977sg,Zijlstra:1993sh,Mertig:1995ny,Moch:1999eb,vanNeerven:2000uj,Vermaseren:2005qc,Moch:2014sna,Blumlein:2021ryt,Blumlein:2021lmf,Davies:2022ofz,Blumlein:2022gpp}. 

The analytic solution to the large-$N_c$ version of the small-$x$ helicity evolution constructed in \cite{Borden:2023ugd} allowed the authors to extract an analytic expression for the (small-$x$, large-$N_c$) $GG$ polarized anomalous dimension $\Delta \gamma_{GG} (\omega)$. This expression, when expanded in powers of $\alpha_s$, completely agreed with all three existing loops of the (small-$x$ and large-$N_c$ limit of the) finite-order calculations \cite{Altarelli:1977zs,Dokshitzer:1977sg,Zijlstra:1993sh,Mertig:1995ny,Moch:1999eb,vanNeerven:2000uj,Vermaseren:2005qc,Moch:2014sna,Blumlein:2021ryt,Blumlein:2021lmf,Davies:2022ofz,Blumlein:2022gpp} and agreed with the first three loops of the perturbative expansion of the BER $GG$ anomalous dimension,  derived in \cite{Borden:2023ugd} using the BER IREE technique. However, $\Delta \gamma_{GG} (\omega)$ from \cite{Borden:2023ugd} disagreed in the overall functional shape with that of BER, which led to a numerically minor disagreement at four loops and beyond in the perturbative expansion of the anomalous dimensions. Here, in the large-$N_c\&N_f$ limit of the evolution, we now have access to all four polarized anomalous dimensions: $\Delta\gamma_{GG}(\omega)$, $\Delta\gamma_{qq}(\omega)$, $\Delta\gamma_{qG}(\omega)$, and $\Delta\gamma_{Gq}(\omega)$. The goal of this Section is to extract an analytic expression (at small-$x$, in the large-$N_c\&N_f$ limit) for each of these polarized anomalous dimensions. Since the large-$N_c$ limit can be taken as a $N_f/N_c \to 0$ sub-limit of the large-$N_c \& N_f$ calculation at hand, we see right away that the disagreement between BER and our $\Delta \gamma_{GG} (\omega)$ anomalous dimensions established in \cite{Borden:2023ugd} should also be contained in the present calculation.

The solution to the spin-dependent DGLAP equations at fixed coupling (e.g., in the DLA) can be written as
\begin{align}\label{DGLAP}
    \begin{pmatrix}
        \Delta \Sigma(x,Q^2) \\
        \Delta G(x,Q^2)
    \end{pmatrix}
    = \wint e^{\omega\ln(1/x)}
    \begin{pmatrix}
        \Delta\Sigma_{\omega}(Q^2) \\
        \Delta G_{\omega}(Q^2)
    \end{pmatrix}
    = \wint e^{\omega\ln(1/x)} \text{exp}\left\{\begin{pmatrix}
        \Delta\gamma_{qq}(\omega) & \Delta\gamma{qG}(\omega) \\
        \Delta\gamma_{Gq}(\omega) & \Delta\gamma{GG}(\omega)
    \end{pmatrix} \ln\frac{Q^2}{\Lambda^2}  \right\}
    \begin{pmatrix}
        \Delta\Sigma_{\omega}(\Lambda^2) \\
        \Delta G_{\omega}(\Lambda^2)
    \end{pmatrix} \,,
\end{align}
with $\Delta\Sigma_{\omega}(\Lambda^2)$ and $\Delta G_{\omega}(\Lambda^2)$ specifying the initial conditions of the helicity PDFs for the evolution at the input scale $\Lambda^2$. As in \cite{Adamiak:2023okq}, let us define the eigenvalues of the anomalous dimension matrix (multiplied by $\ln (Q^2/\Lambda^2)$), 
\begin{subequations}\label{anomdim_eigenvalues}
\begin{align}
    \label{lambda1}
    &\lambda_1 \equiv \frac{1}{2}\left[\Delta\gamma_{qq} + \Delta\gamma_{GG} + \sqrt{\left(\Delta\gamma_{qq} - \Delta\gamma_{GG} \right)^2 + 4 \, \Delta\gamma_{qG} \, \Delta\gamma_{Gq} } \, \right] \, \ln \frac{Q^2}{\Lambda^2} \,,\\
    \label{lambda2}
    &\lambda_2 \equiv \frac{1}{2}\left[\Delta\gamma_{qq} + \Delta\gamma_{GG} -\sqrt{\left(\Delta\gamma_{qq} - \Delta\gamma_{GG} \right)^2 + 4 \, \Delta\gamma_{qG} \, \Delta\gamma_{Gq} } \, \right] \, \ln \frac{Q^2}{\Lambda^2}, 
\end{align}
\end{subequations}
which we can use to exponentiate the matrix of anomalous dimensions in \eq{DGLAP}. Employing these eigenvalues, we write the DGLAP equation's solution as
\begin{align}\label{DGLAPineigs}
    &\begin{pmatrix}
        \Delta\Sigma(x,Q^2) \\
        \Delta G(x,Q^2)
    \end{pmatrix}
    = \wint \, e^{\omega\ln\tfrac{1}{x}} \\
    &\times\begin{pmatrix}
        \frac{e^{\lambda_1}+e^{\lambda_2}}{2} + \left(\Delta\gamma_{qq}(\omega) - \Delta\gamma_{GG}(\omega)\right)\frac{e^{\lambda_1}-e^{\lambda_2}}{2\left(\lambda_1-\lambda_2 \right)}\ln\tfrac{Q^2}{\Lambda^2} & \Delta\gamma_{qG}(\omega) \frac{e^{\lambda_1}-e^{\lambda_2}}{\lambda_1-\lambda_2}\ln\tfrac{Q^2}{\Lambda^2} \\
        \Delta\gamma_{Gq}(\omega) \frac{e^{\lambda_1}-e^{\lambda_2}}{\lambda_1-\lambda_2}\ln\tfrac{Q^2}{\Lambda^2} & \frac{e^{\lambda_1}+e^{\lambda_2}}{2} - \left(\Delta\gamma_{qq}(\omega) - \Delta\gamma_{GG}(\omega) \right) \frac{e^{\lambda_1}-e^{\lambda_2}}{2\left(\lambda_1-\lambda_2 \right)}\ln\tfrac{Q^2}{\Lambda^2}
    \end{pmatrix}
    \begin{pmatrix}
        \Delta\Sigma_{\omega}(\Lambda^2) \\
        \Delta G_{\omega}(\Lambda^2)
    \end{pmatrix} \,. \notag
\end{align}

Let us choose the simple initial conditions 
\begin{align}\label{DGLAPics}
    \Delta G(x,\Lambda^2) = \frac{2N_c}{\as \pi^2} \quad \text{and} \quad \Delta \Sigma(x,\Lambda^2) = 0\, ,
\end{align}
which correspond in Mellin space to
\begin{align}\label{DGLAPicsmellin}
    \Delta G_{\omega}(\Lambda^2) = \frac{2N_c}{\as\pi^2}\frac{1}{\omega} \quad \text{and}\quad \Delta\Sigma_{\omega}(\Lambda^2) = 0\,.
\end{align}
With these we obtain from \eq{DGLAPineigs} the following expressions for $\Delta\Sigma(x,Q^2)$ and $\Delta G(x,Q^2)$:
\begin{subequations}\label{pdfsfromdglap}
\begin{align}
    \label{DeltaSigmaics1}
    &\Delta\Sigma(x,Q^2) = \frac{2N_c}{\as \pi^2}\wint e^{\omega\ln\tfrac{1}{x}} \frac{1}{\omega}\Delta\gamma_{qG}(\omega)\frac{e^{\lambda_1}-e^{\lambda_2}}{\lambda_1-\lambda_2}\ln\tfrac{Q^2}{\Lambda^2}\,,\\
    \label{DeltaGics1}
    &\Delta G(x,Q^2) = \frac{2N_c}{\as \pi^2}\wint e^{\omega\ln\tfrac{1}{x}} \frac{1}{\omega}\left[\frac{e^{\lambda_1} + e^{\lambda_2}}{2} - \left(\Delta\gamma_{qq}(\omega)-\Delta\gamma_{GG}(\omega)\right)\frac{e^{\lambda_1}-e^{\lambda_2}}{2\left(\lambda_1-\lambda_2\right)}\ln\tfrac{Q^2}{\Lambda^2} \right] \,.
\end{align}
\end{subequations}

In order to extract the anomalous dimensions, we would like to compare Eqs.~\eqref{pdfsfromdglap} to the predictions for $\Delta\Sigma$ and $\Delta G$ obtained in the solution to the small-$x$ helicity evolution we constructed in the previous Sections. To match the initial conditions \eqref{DGLAPics} we take the inhomogeneous terms for our helicity evolution to be (cf. Eqs.~\eqref{pdfsfromdipoles})
\begin{align}\label{DGLAPicsfordipoles}
    G^{(0)}_{2}(\xoz^2, zs) = 1\,,\quad\quad Q^{(0)}(\xoz^2,zs) = \widetilde{Q}^{(0)}(\xoz^2,zs) = \widetilde{G}^{(0)}(\xoz^2,zs) = 0\,,
\end{align}
which straightforwardly give (see Eqs.~\eqref{fullsoln1doubleLaplaceG20}, \eqref{fullsoln1doubleLaplaceGtilde0}, \eqref{fullsoln1doubleLaplaceQ0}, and \eqref{fullsoln1doubleLaplaceQtilde0})
\begin{align}\label{DGLAPicsdoublelaplace}
    G_{2\omega\gamma}^{(0)} = \frac{1}{\omega\gamma}\,,\quad\quad Q^{(0)}_{\omega\gamma} = \widetilde{Q}^{(0)}_{\omega\gamma} = \widetilde{G}^{(0)}_{\omega\gamma} = 0\,.
\end{align}
Using Eqs.~\eqref{DGLAPicsdoublelaplace}, we can construct expressions for the double-Laplace images $G_{2\omega\gamma}$ and $\widetilde{Q}_{\omega\gamma}$ by using Eqs.~\eqref{fullsoln1}. The results are
\begin{subequations}\label{laplaceimagesfromics}
\begin{align}
    \label{G2omegagammafromics}
    &G_{2\omega\gamma} = \frac{-1}{2\sqrt{4-\tfrac{2N_f}{N_c}} \left(\gamma-\gw^{++}\right)\left(\gamma-\gw^{+-}\right)\left(\gamma-\gw^{-+}\right)\left(\gamma-\gw^{--}\right)} \\
    &\times\left\{ \frac{2N_f}{N_c}\frac{1}{\omega}\frac{\left(\gamma-r_2^{+-}\right)\left(\gamma-r_2^{--}\right)}{2+\sqrt{4-\tfrac{2N_f}{N_c}}}\left(\gamma-\dw^{++}\right) - \frac{2N_f}{N_c}\frac{1}{\omega}\frac{\left(\gamma-r_2^{++}\right)\left(\gamma-r_2^{-+}\right)}{2-\sqrt{4-\tfrac{2N_f}{N_c}}}\left(\gamma-\dw^{+-}\right) \right. \notag\\
    &- \frac{2\left(\gamma-\gw^{+-}\right)\left(\gamma-\gw^{--}\right)}{\bas\sqrt{s_2(\omega)}}
    \left[\frac{2N_f}{N_c}\frac{1}{\omega}\frac{\left(\gw^{++}-r_2^{+-}\right)\left(\gw^{++}-r_2^{--}\right)}{2+\sqrt{4-\tfrac{2N_f}{N_c}}}\left(\gw^{++}-\dw^{++}\right) \right. \notag\\
    & \left. \hspace{7cm}- \frac{2N_f}{N_c}\frac{1}{\omega}\frac{\left(\gw^{++}-r_2^{++}\right)\left(\gw^{++}-r_2^{-+}\right)}{2-\sqrt{4-\tfrac{2N_f}{N_c}}}\left(\gw^{++}-\dw^{+-}\right) \right] \notag \\
    &+ \frac{2\left(\gamma-\gw^{++}\right)\left(\gamma-\gw^{-+}\right)}{\bas\sqrt{s_2(\omega)}} \left[ \frac{2N_f}{N_c}\frac{1}{\omega}\frac{\left(\gw^{+-}-r_2^{+-}\right)\left(\gw^{+-}-r_2^{--}\right)}{2+\sqrt{4-\tfrac{2N_f}{N_c}}}\left(\gw^{+-}-\dw^{++}\right) \right. \notag\\
    & \left. \left. \hspace{7cm}- \frac{2N_f}{N_c}\frac{1}{\omega}\frac{\left(\gw^{+-}-r_2^{++}\right)\left(\gw^{+-}-r_2^{-+}\right)}{2-\sqrt{4-\tfrac{2N_f}{N_c}}}\left(\gw^{+-}-\dw^{+-}\right) \right]     \right\} \, ,  \notag \\
    \label{Qomegagammafromics}
    &\widetilde{Q}_{\omega\gamma} = \frac{1}{2\sqrt{4-\tfrac{2N_f}{N_c}} \left(\gamma-\gw^{++}\right)\left(\gamma-\gw^{+-}\right)\left(\gamma-\gw^{-+}\right)\left(\gamma-\gw^{--}\right)} \\
    &\times \left\{ \frac{2}{\omega}\left(\gamma-r_1^{+-}\right)\left(\gamma-r_1^{--}\right)\left(\gamma-\dw^{++}\right) - \frac{2}{\omega}\left(\gamma-r_1^{++}\right)\left(\gamma-r_1^{-+}\right)\left(\gamma-\dw^{+-}\right) \right. \notag\\
    &- \frac{2\left(\gamma-\gw^{+-}\right)\left(\gamma-\gw^{--}\right)}{\bas\sqrt{s_2(\omega)}} \left[ \frac{2}{\omega}\left(\gw^{++}-r_1^{+-}\right)\left(\gw^{++}-r_1^{--}\right)\left(\gw^{++}-\dw^{++}\right) \right. \notag \\
    & \left. \hspace{7cm}- \frac{2}{\omega}\left(\gw^{++}-r_1^{++}\right)\left(\gw^{++}-r_1^{-+}\right)\left(\gw^{++}-\dw^{+-}\right) \right] \notag \\
    &+ \frac{2\left(\gamma-\gw^{++}\right)\left(\gamma-\gw^{-+}\right)}{\bas\sqrt{s_2(\omega)}} \left[ \frac{2}{\omega}\left(\gw^{+-}-r_1^{+-}\right)\left(\gw^{+-}-r_1^{--}\right)\left(\gw^{+-}-\dw^{++}\right) \right. \notag \\
    &\hspace{7cm} \left. \left. - \frac{2}{\omega}\left(\gw^{+-}-r_1^{++}\right)\left(\gw^{+-}-r_1^{-+}\right)\left(\gw^{+-}-\dw^{+-}\right) \right] \right\} \notag \,.
\end{align}
\end{subequations}

Now with $\Delta G(x,Q^2)$ and $\Delta\Sigma(x,Q^2)$ given in terms of $G_{2\omega\gamma}$ and $\widetilde{Q}_{\omega\gamma}$ in Eqs.~\eqref{fullsolnpdfs}, we would first like to carry out the integrals over $\gamma$. The only non-vanishing poles in the $\gamma$-plane contained in Eqs.~\eqref{laplaceimagesfromics} are those at $\gamma = \gw^{--}$ and $\gamma = \gw^{-+}$. So we carry out the $\gamma$-integrals in Eqs.~\eqref{fullsolnpdfs} by closing the contour to the left and picking up these two simple poles. Schematically,
\begin{subequations}\label{DGLAPgammaintegrals}
\begin{align}
    \label{DeltaGgammaintegral}
    &\Delta G(x,Q^2) = \frac{2N_c}{\as \pi^2}\wint e^{\omega\ln(1/x)} \bigg[e^{\gw^{--}\ln(Q^2/\Lambda^2)} \lim_{\gamma\rightarrow\gw^{--}}\left(\gamma-\gw^{--}\right)G_{2\omega\gamma} \\
    &\hspace{5.1cm}+ e^{\gw^{-+}\ln(Q^2/\Lambda^2)} \lim_{\gamma\rightarrow\gw^{-+}}\left(\gamma-\gw^{-+}\right)G_{2\omega\gamma} \bigg] \,,\notag \\
    \label{DeltaSigmagammaintegral}
    &\Delta\Sigma(x,Q^2) = \frac{N_f}{\as \pi^2}\wint e^{\omega\ln(1/x)} \bigg[e^{\gw^{--}\ln(Q^2/\Lambda^2)} \lim_{\gamma\rightarrow\gw^{--}}\left(\gamma-\gw^{--}\right)\widetilde{Q}_{\omega\gamma} \\
    &\hspace{5.1cm}+ e^{\gw^{-+}\ln(Q^2/\Lambda^2)} \lim_{\gamma\rightarrow\gw^{-+}}\left(\gamma-\gw^{-+}\right)\widetilde{Q}_{\omega\gamma}\bigg] \,,\notag
\end{align}
\end{subequations}
with $G_{2\omega\gamma}$ and $\widetilde{Q}_{\omega\gamma}$ as written in Eqs.~ \eqref{laplaceimagesfromics}. We can rewrite each of Eqs.~\eqref{DGLAPgammaintegrals} in terms of the sum and difference of the two exponential structures:
\begin{subequations}\label{DGLAPgammaintegrals1}
\begin{align}
    \label{DeltaGgammaintegral1}
    &\Delta G(x,Q^2) = \frac{2N_c}{\as \pi^2}\wint e^{\omega\ln(1/x)} \\
    &\hspace{.2cm}\times\bigg[\left(e^{\gw^{--}\ln(Q^2/\Lambda^2)} + e^{\gw^{-+}\ln(Q^2/\Lambda^2)}\right)\frac{1}{2}\left(\lim_{\gamma\rightarrow\gw^{--}}\left(\gamma-\gw^{--}\right)G_{2\omega\gamma} + \lim_{\gamma\rightarrow\gw^{-+}}\left(\gamma-\gw^{-+}\right)G_{2\omega\gamma} \right) \notag\\
    &\hspace{.4cm} + \left(e^{\gw^{--}\ln(Q^2/\Lambda^2)} - e^{\gw^{-+}\ln(Q^2/\Lambda^2)}\right)\frac{1}{2}\left(\lim_{\gamma\rightarrow\gw^{--}}\left(\gamma-\gw^{--}\right)G_{2\omega\gamma} - \lim_{\gamma\rightarrow\gw^{-+}}\left(\gamma-\gw^{-+}\right)G_{2\omega\gamma} \right)\bigg] \,,\notag  \\
    \label{DeltaSigmagammaintegral1}
    &\Delta\Sigma(x,Q^2) = \frac{N_f}{\as \pi^2}\wint e^{\omega\ln(1/x)} \\
    &\hspace{.4cm}\times\bigg[\left(e^{\gw^{--}\ln(Q^2/\Lambda^2)} + e^{\gw^{-+}\ln(Q^2/\Lambda^2)}\right)\frac{1}{2}\left(\lim_{\gamma\rightarrow\gw^{--}}\left(\gamma-\gw^{--}\right)\widetilde{Q}_{\omega\gamma} + \lim_{\gamma\rightarrow\gw^{-+}}\left(\gamma-\gw^{-+}\right)\widetilde{Q}_{\omega\gamma} \right) \notag\\
    &\hspace{.4cm} + \left(e^{\gw^{--}\ln(Q^2/\Lambda^2)} - e^{\gw^{-+}\ln(Q^2/\Lambda^2)}\right)\frac{1}{2}\left(\lim_{\gamma\rightarrow\gw^{--}}\left(\gamma-\gw^{--}\right)\widetilde{Q}_{\omega\gamma} - \lim_{\gamma\rightarrow\gw^{-+}}\left(\gamma-\gw^{-+}\right)\widetilde{Q}_{\omega\gamma} \right)\bigg] \,.\notag
\end{align}
\end{subequations}

Comparing Eqs.~\eqref{DGLAPgammaintegrals1} to Eqs.~\eqref{pdfsfromdglap}, we make several identifications. First, from the exponentials themselves, we conclude that
\begin{subequations}\label{gammasareeigs}
\begin{align}
    &\frac{\lambda_1(\omega)}{\ln\tfrac{Q^2}{\Lambda^2}} = \gw^{--} \,,\\
    &\frac{\lambda_2(\omega)}{\ln\tfrac{Q^2}{\Lambda^2}} = \gw^{-+} \,.
\end{align}
\end{subequations}
That is, the functions $\gw^{--}$ and $\gw^{-+}$ (which correspond to the two pole structures that survive in the double-Laplace images $G_{2\omega\gamma}$ and $\widetilde{Q}_{\omega\gamma}$) are the eigenvalues of the anomalous dimension matrix. To cross-check this result against the finite-order calculations \cite{Altarelli:1977zs,Dokshitzer:1977sg,Zijlstra:1993sh,Mertig:1995ny,Moch:1999eb,vanNeerven:2000uj,Vermaseren:2005qc,Moch:2014sna,Blumlein:2021ryt,Blumlein:2021lmf,Davies:2022ofz,Blumlein:2022gpp} we can expand the quantities in Eqs.~\eqref{gammasareeigs} (or, equivalently, in \eq{fullsoln1neweigs}) in powers of $\as$ (while employing Eqs.~\eqref{fullsoln1news1} and \eqref{fullsoln1news2}). For our functions $\gw^{--}$ and $\gw^{-+}$, we find ($\beta = \pm$)
\begin{align}\label{eigenvalueexpansions}
    &\gamma^{-,\beta}_{\omega} =
    \left(\frac{\as N_c}{4\pi} \right)\frac{1}{2}\left[9 + \sqrt{49 - 16\tfrac{N_f}{N_c}}  \right] \frac{1}{\omega} \\
    &\hspace{1.8cm}+ \left(\frac{\as N_c}{4\pi} \right)^2 \frac{1}{2}\frac{1}{\left(49 - 16\tfrac{N_f}{N_c}\right)} \left[\left(49-16\tfrac{N_f}{N_c}\right)\left(33-8\tfrac{N_f}{N_c}\right) -\beta \sqrt{49 - 16\tfrac{N_f}{N_c}}\left(217 - 80\tfrac{N_f}{N_c} \right) \right] \frac{1}{\omega^3} \notag \\ 
    &\hspace{1.8cm} + \left(\frac{\as N_c}{4\pi} \right)^3 \frac{1}{\left(49 - 16\tfrac{N_f}{N_c}\right)^2} \bigg[\left(49 - 16\tfrac{N_f}{N_c}\right)^2 \left(225 - 64\tfrac{N_f}{N_c} \right) \notag \\
    &\hspace{5.6cm} -\beta \sqrt{49 - 16\tfrac{N_f}{N_c}} \left(76489 - 60712\tfrac{N_f}{N_c} + 14784\left(\tfrac{N_f}{N_c}\right)^2 -1024\left(\tfrac{N_f}{N_c}\right)^3 \right) \bigg] \frac{1}{\omega^5} \notag \\
    &\hspace{1.8cm} + \mathcal{O}\left(\as^4\right) \notag .
\end{align}
Meanwhile, the small-$x$ large-$N_c\&N_f$ limit of the polarized splitting functions calculated to three loops is \cite{Altarelli:1977zs,Dokshitzer:1977sg,Mertig:1995ny,Moch:2014sna} (with the bar over each splitting function denoting that it was calculated in the $\overline{\text{MS}}$ scheme)
\begin{subequations}\label{msbarsplittingfuncs}
    \begin{align}
    &\Delta \overline{P}_{qq}(x) = \left(\frac{\alpha_sN_c}{4\pi}\right) + \left(\frac{\alpha_s N_c}{4\pi}\right)^2 \left( \frac{1}{2}-2\frac{N_f}{N_c} \right)\ln^2\frac{1}{x} + \left(\frac{\alpha_sN_c}{4\pi}\right)^3\frac{1}{12} \left( 1-20\frac{N_f}{N_c} \right) \ln^4\frac{1}{x} + {\cal O} (\alpha_s^4) \,,     \label{Pqq} \\
    &\Delta \overline{P}_{qG}(x) =  - \left(\frac{\alpha_sN_c}{4\pi}\right)\frac{2N_f}{N_c} - \left(\frac{\alpha_sN_c}{4\pi}\right)^2 5\frac{N_f}{N_c}\ln^2\frac{1}{x} - \left(\frac{\alpha_sN_c}{4\pi}\right)^3\frac{1}{6}\frac{N_f}{N_c} \left( 34-4\frac{N_f}{N_c} \right) \ln^4\frac{1}{x} + {\cal O} (\alpha_s^4) \,,     \label{eq:PqG} \\
    &\Delta \overline{P}_{Gq}(x) =   2\left(\frac{\alpha_sN_c}{4\pi}\right) + 5\left(\frac{\alpha_s N_c}{4\pi}\right)^2 \ln^2\frac{1}{x} + \left(\frac{\alpha_s N_c}{4\pi}\right)^3\frac{1}{6} \left( 36-4\frac{N_f}{N_c} \right) \ln^4\frac{1}{x} + {\cal O} (\alpha_s^4)  \,,   \label{eq:PGq} \\
    &\Delta \overline{P}_{GG}(x) =   8 \left(\frac{\alpha_sN_c}{4\pi}\right) + \left(\frac{\alpha_s N_c}{4\pi}\right)^2 \left( 16-2\frac{N_f}{N_c} \right) \ln^2\frac{1}{x} + \left(\frac{\alpha_s N_c}{4\pi}\right)^3\frac{1}{3} \left( 56-11\frac{N_f}{N_c} \right) \ln^4\frac{1}{x} + {\cal O} (\alpha_s^4) \,.   \label{PGG} 
\end{align}
\end{subequations}
We have explicitly checked that if one converts these finite-order polarized splitting functions $\Delta\overline{P}_{ij}(x)$ to the corresponding polarized anomalous dimensions $\Delta\overline{\gamma}_{ij}(\omega)$ (with the bar denoting the $\overline{\text{MS}}$ scheme again) defined by
\begin{align}\label{anomdimdef}
    \Delta\overline{\gamma}_{ij}(\omega) = \int\limits_0^1\mathrm{d}x \,x^{\omega-1} \Delta\overline{P}_{ij}(x) 
\end{align}
and subsequently constructs the eigenvalues of the anomalous dimension matrix $\frac{\lambda_1}{\ln(Q^2/\Lambda^2)}$ and $\frac{\lambda_2}{\ln(Q^2/\Lambda^2)}$ via Eqs.~\eqref{anomdim_eigenvalues}, one finds exactly the same perturbative expansion as that in \eq{eigenvalueexpansions}.\footnote{Note that we also have the freedom to interchange $\lambda_1 \leftrightarrow \lambda_2$ in Eqs.~\eqref{gammasareeigs}. On the basis of comparing Eqs.~\eqref{DGLAPgammaintegrals1} to Eqs.~\eqref{pdfsfromdglap}, without doing further calculations, we cannot uniquely identify the eigenvalues: the fact that the perturbative expansion \eqref{eigenvalueexpansions} matches the finite-order calculations confirms that our choice in Eqs.~\eqref{gammasareeigs} is correct.} Therefore, our solution constructed here passes this eigenvalue cross-check. 

Note that in \cite{Borden:2024bxa}, the polarized splitting functions were extracted from the most recent version of the small-$x$ helicity evolution equations \eqref{evoleqs_IR} order by order in $\as$, up to four loops. There, it was observed that the splitting functions predicted by the small-$x$ helicity evolution agree exactly with the full 3 loops of finite order calculations for $\Delta P_{qq}$ and $\Delta P_{GG}$, but disagree at the third loop for $\Delta P_{qG}$ and $\Delta P_{Gq}$ (though this disagreement was ultimately attributable to a scheme dependence \cite{Moch:2014sna}). However, at fixed coupling, the scheme transformations of the matrix of anomalous dimensions reduce only to rotations. Since rotations do not affect the eigenvalues of a matrix, the eigenvalues of the anomalous dimension matrix are scheme invariant, in the approximation where one can neglect the running of the coupling, such as our DLA. Indeed, an anomalous dimension matrix in any scheme can then be rotated to a scheme where the new matrix is diagonal (the `eigenscheme'), the diagonal entries being the eigenvalues. Thus, the anomalous dimension matrix eigenvalues calculated in any scheme ought to agree (again, at fixed coupling), and so indeed our \eq{eigenvalueexpansions} agrees with the equivalent expansion calculated in $\overline{\text{MS}}$, despite the fact that our splitting functions $\Delta P_{qG}(x)$ and $\Delta P_{Gq}(x)$ disagree at $\mathcal{O}(\as^3)$.

Returning to Eqs.~\eqref{DGLAPgammaintegrals1} and their comparison with Eqs.~\eqref{pdfsfromdglap}, we can make several additional identifications by matching the coefficients of the exponential structures,
\begin{subequations}\label{PDFids}
\begin{align}
    \label{PDFid1}
    &\frac{1}{\omega} = \left[ \lim_{\gamma\rightarrow\gw^{--}}\left(\gamma-\gw^{--}\right)G_{2\omega\gamma} + \lim_{\gamma\rightarrow\gw^{-+}}\left(\gamma-\gw^{-+}\right)G_{2\omega\gamma} \right] \,,\\
    \label{PDFid2}
    &\Delta\gamma_{qq}(\omega) - \Delta\gamma_{GG}(\omega) = -\omega\left(\gw^{--} - \gw^{-+} \right)\left[ \lim_{\gamma\rightarrow\gw^{--}}\left(\gamma-\gw^{--}\right)G_{2\omega\gamma} - \lim_{\gamma\rightarrow\gw^{-+}}\left(\gamma-\gw^{-+}\right)G_{2\omega\gamma} \right] \,,\\
    \label{PDFid3}
    & 0 = \frac{1}{2}\left[ \lim_{\gamma\rightarrow\gw^{--}}\left(\gamma-\gw^{--}\right)\widetilde{Q}_{\omega\gamma} + \lim_{\gamma\rightarrow\gw^{-+}}\left(\gamma-\gw^{-+}\right)\widetilde{Q}_{\omega\gamma} \right] \,,\\
    \label{PDFid4}
    &\Delta\gamma_{qG}(\omega) = \frac{N_f}{4N_c}\omega \left(\gw^{--} - \gw^{-+}\right)  \left[ \lim_{\gamma\rightarrow\gw^{--}}\left(\gamma-\gw^{--}\right)\widetilde{Q}_{\omega\gamma} - \lim_{\gamma\rightarrow\gw^{-+}}\left(\gamma-\gw^{-+}\right)\widetilde{Q}_{\omega\gamma} \right] \,,
\end{align}
\end{subequations}
where in obtaining Eqs.\eqref{PDFid2} and \eqref{PDFid4} we made use of the identities in Eqs.~\eqref{gammasareeigs}.

Eqs.~\eqref{PDFid1} and \eqref{PDFid3} can be verified to be true by explicit calculation using $G_{2\omega\gamma}$ and $\widetilde{Q}_{\omega\gamma}$ as written (for our particular choice of initial conditions) in Eqs.~\eqref{laplaceimagesfromics}. Meanwhile, we can supplement Eqs.~\eqref{PDFid2} and \eqref{PDFid4} with two additional equations. Using Eqs.~\eqref{anomdim_eigenvalues} and \eqref{gammasareeigs} we can write 
\begin{subequations}\label{sumanddiffofeigs}
\begin{align}
    \label{sumofeigs}
    &\Delta\gamma_{qq}(\omega) + \Delta\gamma_{GG}(\omega) = \gw^{--} + \gw^{-+} \quad\quad \text{and} \\
    \label{diffofeigs}
    & \Delta\gamma_{Gq}(\omega) = \frac{1}{4\Delta\gamma_{qG}(\omega)} \left[\left(\gw^{--} - \gw^{-+}\right)^2 - \left(\Delta\gamma_{qq}(\omega) -\Delta\gamma_{GG}(\omega)\right)^2   \right]\,.
\end{align}
\end{subequations}
Together, Eqs.~\eqref{PDFid2}, \eqref{PDFid4}, \eqref{sumofeigs}, \eqref{diffofeigs} form a system of four equations that we can solve for each of the four polarized anomalous dimensions. After considerable algebra we obtain
\begin{subequations}\label{alladims}
\begin{align}
    \label{alladimsqq}
    &\Delta\gamma_{qq}(\omega) = \frac{1}{2}\Bigg\{\gamma_{\omega}^{-+}+\gamma_{\omega}^{--} - \frac{\gamma_{\omega}^{-+}-\gamma_{\omega}^{--} }{\omega\left(2-\tfrac{N_f}{N_c}\right)\sqrt{s_2(\omega)} } \bigg[ 3\omega\left(6+\tfrac{N_f}{N_c}\right) - 2\left(8-\tfrac{N_f}{N_c}\right)\left(\delta_{\omega}^{++}+\delta_{\omega}^{+-}\right)  \\
    &  \hspace{9.2cm}+ 8\sqrt{4-\tfrac{2N_f}{N_c}}\left(\delta_{\omega}^{++}-\delta_{\omega}^{+-}\right) \bigg] \Bigg\} \,, \notag \\
    \label{alladimsGG}
    &\Delta\gamma_{GG}(\omega) = \frac{1}{2}\Bigg\{\gamma_{\omega}^{-+}+\gamma_{\omega}^{--}+ \frac{\gamma_{\omega}^{-+}-\gamma_{\omega}^{--} }{\omega\left(2-\tfrac{N_f}{N_c}\right)\sqrt{s_2(\omega)} } \bigg[3\omega\left(6+\tfrac{N_f}{N_c}\right) - 2\left(8-\tfrac{N_f}{N_c}\right)\left(\delta_{\omega}^{++}+\delta_{\omega}^{+-}\right) \\
    &\hspace{9.2cm}+ 8\sqrt{4-\tfrac{2N_f}{N_c}}\left(\delta_{\omega}^{++}-\delta_{\omega}^{+-}\right)   \bigg]     \Bigg\}\,, \notag \\
    \label{alladimsqG}
    &\Delta\gamma_{qG}(\omega) = -\frac{N_f}{N_c} \, \frac{\gamma_{\omega}^{-+}-\gamma_{\omega}^{--} }{4 \, \omega\left(2-\tfrac{N_f}{N_c}\right)\sqrt{s_2(\omega)} } \bigg[8\omega\left(2+\tfrac{N_f}{N_c}\right) - 16\left(\delta_{\omega}^{++}+\delta_{\omega}^{+-}\right) \\
    &\hspace{9.2cm}+ 8\sqrt{4-\tfrac{2N_f}{N_c}}\left(\delta_{\omega}^{++}-\delta_{\omega}^{+-}\right)   \bigg] \,, \notag\\
    \label{alladimsGq}
    &\Delta\gamma_{Gq}(\omega) = \frac{\gamma_{\omega}^{-+}-\gamma_{\omega}^{--} }{4 \, \omega\left(2-\tfrac{N_f}{N_c}\right)\sqrt{s_2(\omega)} } \bigg[8\omega\left(2+\tfrac{N_f}{N_c}\right) - 16\left(\delta_{\omega}^{++}+\delta_{\omega}^{+-}\right) \\
    &\hspace{9.2cm}+ 8\sqrt{4-\tfrac{2N_f}{N_c}}\left(\delta_{\omega}^{++}-\delta_{\omega}^{+-}\right)   \bigg] . \notag
\end{align}
\end{subequations} 
Eqs.~\eqref{alladims} are another important result of our work -- all-order in $\as$ expressions for the (small-$x$ and large-$N_c\&N_f$) polarized DGLAP anomalous dimensions. 

We can expand each of these anomalous dimensions in powers of $\as$ to obtain
\begin{subequations}\label{alladimexpansions}
    \begin{align}
    \label{ourdeltagammaqqexpanded}
    &\Delta\gamma_{qq}(\omega) = \left(\frac{\as N_c}{4\pi}\right)\frac{1}{\omega} + \left(\frac{\as N_c}{4\pi}\right)^2 \left(1-4\frac{N_f}{N_c}\right) \frac{1}{\omega^3} + \left(\frac{\as N_c}{4\pi}\right)^3\,2\left(1-20\frac{N_f}{N_c}\right)\frac{1}{\omega^5} \\
    &\hspace{3.7cm}+ \left(\frac{\as N_c}{4\pi}\right)^4 \left(5-748\frac{N_f}{N_c} + 80\frac{N_f^2}{N_c^2}\right)\frac{1}{\omega^5} + \mathcal{O}\left(\as^5\right) \notag\,,\\
    \label{ourdeltagammaggexpanded}
    &\Delta\gamma_{GG}(\omega) = \left(\frac{\as N_c}{4\pi}\right)\,8\,\frac{1}{\omega} + \left(\frac{\as N_c}{4\pi}\right)^2 \, 4\left(8-\frac{N_f}{N_c}\right)\frac{1}{\omega^3} + \left(\frac{\as N_c}{4\pi}\right)^3\,8\left(56-11\frac{N_f}{N_c}\right) \frac{1}{\omega^5} \\
    &\hspace{3.7cm} +  \left(\frac{\as N_c}{4\pi}\right)^4\, 4\left(1984-549\frac{N_f}{N_c}+20\frac{N_f^2}{N_c^2}\right)\frac{1}{\omega^7} + \mathcal{O}\left(\as^5\right) \notag \,, \\
    \label{ourdeltagammaqGexpanded}
    &\Delta\gamma_{qG}(\omega) = -\left(\frac{\as N_c}{4\pi}\right)\,2\frac{N_f}{N_c}\frac{1}{\omega} -\left(\frac{\as N_c}{4\pi}\right)^2\,10\frac{N_f}{N_c}\frac{1}{\omega^3}-\left(\frac{\as N_c}{4\pi}\right)^3\,4\frac{N_f}{N_c}\left(35-4\frac{N_f}{N_c}\right)\frac{1}{\omega^5}\\
    &\hspace{3.7cm}-\left(\frac{\as N_c}{4\pi}\right)^4\,2\frac{N_f}{N_c}\left(1213-224\frac{N_f}{N_c}\right)\frac{1}{\omega^7} + \mathcal{O}\left(\as^5\right) \notag \,, \\
    \label{ourdeltagammaGqexpanded}
    &\Delta\gamma_{Gq}(\omega) = \left(\frac{\as N_c}{2\pi}\right)\frac{2}{\omega} + \left(\frac{\as N_c}{2\pi}\right)^2 \frac{10}{\omega^3} + \left(\frac{\as N_c}{2\pi}\right)^3 \, 4\left(35-4\frac{N_f}{N_c}\right) \frac{1}{\omega^5} \\
    &\hspace{3.7cm}+ \left(\frac{\as N_c}{2\pi}\right)^4 \, 2\left(1213 - 224\frac{N_f}{N_c}\right) \frac{1}{\omega^7} + \mathcal{O}\left(\as^5\right)\,. \notag
\end{align}
\end{subequations}
All the expansions in Eqs.~\eqref{alladimexpansions} agree completely with those obtained by solving the small-$x$ evolution equations iteratively in \cite{Borden:2024bxa}. The comparisons with BER and finite-order calculations remain the same as discussed in that reference. Equations~\eqref{alladimexpansions} again exhibit full agreement with BER to three loops, with small disagreements beginning at four loops in all the polarized anomalous dimensions \cite{Bartels:1996wc,Blumlein:1996hb}, in analogy to the large-$N_c$ case \cite{Borden:2023ugd}. Equations~\eqref{alladimexpansions} are in full agreement with all 3 loops of finite-order calculation for $\Delta\gamma_{qq}$ and $\Delta\gamma_{GG}$, but disagree with finite-order results starting at three loops for $\Delta\gamma_{qG}$ and $\Delta\gamma_{Gq}$. This last disagreement (between the IREE results \cite{Bartels:1996wc,Blumlein:1996hb} and the finite-order calculations) was already known in \cite{Moch:2014sna} and was attributed to a scheme dependence there. The scheme transformation at low $x$ was explicitly constructed in Appendix B of \cite{Borden:2024bxa}.

Also of note is the fact that, as one can see from Eqs.~\eqref{alladimsqG} and \eqref{alladimsGq}, we predict that
\begin{align}\label{relationbtwqGandGqanomdims}
    \Delta\gamma_{Gq}(\omega) = -\frac{N_c}{N_f}\Delta\gamma_{qG}(\omega) \,,
\end{align}
which of course also implies
\begin{align}\label{relationbtwqGandGqsplittingfuncs}
    \Delta P_{Gq}(x) = -\frac{N_c}{N_f}\Delta P_{qG}.
\end{align}
This relationship between the $qG$ and $Gq$ polarized splitting functions was observed in \cite{Borden:2024bxa} where the splitting functions were constructed order by order in $\as$ up to four loops. However, here we have demonstrated that this prediction persists to all orders in the coupling (at small $x$ and large $N_c \& N_f$). The splitting functions obtained within the BER framework \cite{Blumlein:1996hb} obey the same property \eqref{relationbtwqGandGqsplittingfuncs} up to and including the four-loop level (the order the BER splitting functions in the existing literature are known up to at large $N_c \& N_f$), even though the splitting functions we obtain at four loops disagree with those from \cite{Blumlein:1996hb}.


\section{\texorpdfstring{Small-$x$ Asymptotics}{Small x Asymptotics}}\label{sec:asymptotics}

\subsection{The Intercept}
The leading small-$x$ asymptotic behavior of our polarized dipole amplitudes and thus also of the helicity distributions and $g_1$ structure function in Eqs.~\eqref{fullsolnpdfs} and \eqref{fullsolng1}, respectively, is governed by the singularity in the complex-$\omega$ plane with the largest real part. By studying our solution in Eqs.~\eqref{fullsoln} and \eqref{fullsoln1}, and assuming that the initial conditions contain no singularities in the complex-$\omega$ plane with a large real part of $\omega$, one can show that this leading singularity comes from a branch point of the large square root in the function $\gw^{--}$.\footnote{The right-most branch point in $\gw^{-+}$ does not have such a large real part as the right-most branch point of $\gw^{--}$; the branch points in $r_1^{\alpha\beta}$ and $r_2^{\alpha\beta}$ are not the branch points of ${\widetilde Q}_{\omega\gamma}$ and $G_{2\omega\gamma}$ because $r_1^{\alpha\beta}$ and $r_2^{\alpha\beta}$ enter the expressions for ${\widetilde Q}_{\omega\gamma}$ and $G_{2\omega\gamma}$ as $(\gamma - r_i^{\alpha\beta}) (\gamma - r_i^{-\alpha, \beta}) = \gamma^2 - \omega \gamma + r_i^{\alpha\beta} \, r_i^{-\alpha,\beta}$ in the numerator, for $i=1,2$. Since the large square roots in Eqs.~\eqref{fullsoln1r1alphabeta} and \eqref{fullsoln1r2alphabeta} disappear in the product $r_i^{\alpha\beta} \, r_i^{- \alpha, \beta}$, such terms do not generate new branch cuts. Being in the numerator, they do not generate new poles either.} Note that in this Section, we will work in the notation prior to the rescaling of \eq{resc}, so that $\gw^{--}$, $s_1 (\omega)$, and $s_2 (\omega)$ are given by Eqs.~\eqref{gammapmpmfull} and \eqref{gammapmpmfull12}. Equating the argument of the large (outer) square root of $\gw^{--}$ to zero (cf. \eq{gammapmpmfull}) in order to find the branch point, we readily see that the intercept $\omega_b$ of the helicity distributions satisfies the algebraic equation
\begin{align}\label{intercepteq}
    \omega_b^2 + s_1(\omega_b) - \sqrt{s_2(\omega_b)} = 0 ,
\end{align}
again with $s_1(\omega)$ and $s_2(\omega)$ defined in Eqs.~\eqref{gammapmpmfull12}. Due to the complicated functions involved, this is a challenging equation to solve analytically. However, we can solve it numerically for various values of $N_f$ and $N_c$. In Table~\ref{tab:intercepts} we show the numerical values of the intercept $\omega_b$ for several values of $N_f$ with $N_c = 3$ obtained by a numerical solution of \eq{intercepteq} and denoted $\omega_b^{(\text{us})}$. 

For comparison, we also show, for each $N_f$, the corresponding prediction for the intercept we obtained from the calculation by BER IREE \cite{Bartels:1996wc} in the large $N_c$ and $N_f$ approximation. The BER intercepts were calculated numerically by following the work of \cite{Bartels:1996wc} while taking the limit of large $N_c$ and $N_f$ in all the relevant formulas, after which we substituted $N_c = 3$ and the $N_f$ values indicated in Table~\ref{tab:intercepts}. For $N_f = 2, 3, 4$ the BER intercepts  were previously presented in \cite{Adamiak:2023okq}. The $N_f =6$ case has to be taken as a limit $N_f \to 6$. 
\begin{table}[tb]
\centering
\begin{tabular}{|c|c|c|c|}
    \hline
    $N_f$ & $\omega_b^{(\text{us})}$ & $\omega_b^{(\text{BER})}$ &  $\omega_b^{(\text{BER})} - \omega_b^{(\text{us})}$ \\
    \hline
    2 & 3.54523 &  3.54816 & 0.00293  \\
    3 & 3.47910 & 3.48182 & 0.00272 \\
    4 & 3.40514 & 3.40757 & 0.00243 \\
    5 & 3.32036 & 3.32237 & 0.00201 \\
    6 & $3.21930^{(*)}$ & 3.22062 & 0.00132 \\
    7 & 3.08946 & 3.08943 & -0.00003\\
    8 & 2.88228 & 2.87704 & -0.00524\\
    \hline
\end{tabular}
\caption{The intercepts $\omega_b$ for several values of $N_f$ with $N_c = 3$. $\omega_b^{(\text{us})}$ corresponds to our prediction based on the solution of the small-$x$ evolution in Eqs.~\eqref{evoleqs_IR} (that is, on numerically solving \eq{intercepteq}), while $\omega_b^{(\text{BER})}$ corresponds to the predictions of  the IREE formalism by Bartels, Ermolaev, and Ryskin \cite{Bartels:1996wc} obtained here and in \cite{Adamiak:2023okq} while employing the large-$N_c \& N_f$ approximation. Also shown in the last column are the differences between the predicted intercepts, which are quite small numerically in comparison to the intercepts' values. The asterisk for the $N_f = 6$ line denotes the case where an exact analytic expression for our intercept is available, given in \eq{interceptNfequals2Nc}. }
\label{tab:intercepts}
\end{table}

Comparing BER and our intercepts in Table~\ref{tab:intercepts} we conclude that the differences between the intercepts are numerically very minor. The intercepts of BER tend to be larger than ours for $N_f \leq 6$, while for $N_f \geq 7$ it appears that our intercepts become larger than BER intercepts, with the difference growing larger with increasing $N_f$ (indeed, the predictions for $N_f \geq 7$ should probably be taken as a purely theoretical exercise).

It is also worth noting that the intercepts we present here (that is, `our' intercepts), which are based on the most recent version of the small-$x$ evolution as published in \cite{Borden:2024bxa}, are slightly different than the intercepts obtained from the previous version of the large-$N_c \& N_f$ helicity evolution \cite{Cougoulic:2022gbk}. That is, the quark-to-gluon and gluon-to-quark transition operators which were incorporated into the small-$x$ evolution in \cite{Borden:2024bxa} modified the intercepts slightly, tending to make them a bit larger than their values obtained from the evolution without the transition operators. For example, our (unpublished) exact analytic solution of the previous version of the large-$N_c \& N_f$ helicity evolution equations from \cite{Cougoulic:2022gbk}, which were derived before \cite{Borden:2024bxa}, yielded an intercept of 3.31621 for $N_f=4$ and $N_c=3$ (also obtained in a numerical solution of the same equations in \cite{Adamiak:2023okq}), as compared to the updated value from Table~\ref{tab:intercepts} of 3.40514. A similar trend is observed when comparing `our' intercepts from Table~\ref{tab:intercepts} to those found in \cite{Adamiak:2023okq}, which were obtained by numerically solving the large-$N_c \& N_f$ helicity evolution equations from \cite{Cougoulic:2022gbk}.

The asterisk (*) in Table~\ref{tab:intercepts} denotes the fact that when $N_f = 2N_c$, \eq{intercepteq} becomes simple enough to solve analytically. The resulting right-most branch point is
\begin{align}\label{interceptNfequals2Nc}
    \omega_b^{(N_f = 2N_c)} = \sqrt{\frac{1}{57}\left(266 + 38\times2^{2/3}\,\text{Re}\left[\left(137 + 9\,i\,\sqrt{107} \right)^{1/3} \right]\right)} \approx 3.21930,
\end{align}
which has a qualitatively similar structure to the analytic intercept found from the large-$N_c$ evolution (Eq.~(61) in \cite{Borden:2023ugd}). Note also that taking $N_f = 0$ in \eq{intercepteq} and solving that equation, one obtains exactly the intercept from the large-$N_c$ helicity evolution (Eq. (61) in \cite{Borden:2023ugd}), as expected.

We conclude that all the helicity-dependent quantities involved grow with the same power of $1/x$ at small-$x$, driven by the leading branch point $\omega_b$ whose numerical values can be found by solving \eq{intercepteq} for any choice of $N_c$ and $N_f$. The power law for the asymptotic behavior is thus (see Eqs.~\eqref{fullsolnpdfs} and \eqref{fullsolng1})
\begin{align}\label{DSGg_asymptotics}
    \Delta\Sigma(x,Q^2) \sim \Delta G(x,Q^2) \sim g_1(x,Q^2) \sim \left(\frac{1}{x}\right)^{\alpha_h} \,,
\end{align}
where
\begin{align}\label{alphah}
    \alpha_h \equiv \sqrt{\bas} \, \omega_b \,.
\end{align}


\subsection{Asymptotic Behavior: Integral Around the Leading Branch Cut}

Let us now take a more detailed look at the asymptotic behavior of the helicity distributions. By considering the structure of our distributions in the complex-$\omega$ plane near the leading singularity (and not by just concentrating on the leading singularity itself, as was done in the previous Subsection), we will be able to better approximate the $\omega$-integrals in Eqs.~\eqref{fullsolnpdfs} and obtain a more detailed description of the behavior of the helicity distributions in their small-$x$ asymptotics. A similar analysis was done in Appendix~B of \cite{Kovchegov:2023yzd} for the large-$N_c$ version of the small-$x$ helicity evolution, based on the analytic solution to that evolution constructed in \cite{Borden:2023ugd}. Here, we follow the procedure in \cite{Kovchegov:2023yzd} very closely.

Based on the correspondence with polarized DGLAP established in Sec. \ref{sec:DGLAP} --- in particular using Eqs.~\eqref{pdfsfromdglap}, \eqref{anomdim_eigenvalues}, and \eqref{alladims} --- we can write the helicity PDFs for the same initial conditions as those chosen in \eq{DGLAPicsfordipoles} as
\begin{subequations}\label{preasy_pdfs}\allowdisplaybreaks
\begin{align}
    &\Delta\Sigma(y,t) = \frac{N_f}{\as 2\pi^2}\wint \, e^{\omega y} \, \frac{1}{\omega} \left(e^{t\gw^{--}} - e^{t\gw^{-+}} \right) \frac{8\omega\left(2 + \tfrac{N_f}{N_c}\right) - 16\left(\dw^{++}+\dw^{+-}\right) + 8\sqrt{4-\tfrac{2N_f}{N_c}}\left(\dw^{++}-\dw^{+-} \right)}{\omega\left(2-\tfrac{N_f}{N_c}\right)\sqrt{s_2(\omega)}} \,,\\
    &\Delta G(y,t) = \frac{N_c}{\as\pi^2}\wint \, e^{\omega y} \, \frac{1}{\omega} \, \Bigg[ e^{t\gw^{--}} + e^{t\gw^{-+}} \\
    &\hspace{3.75cm} - \left(e^{t\gw^{--}} - e^{t\gw^{-+}} \right) \frac{3\omega\left(6+\tfrac{N_f}{N_c}\right) - 2\left(8-\tfrac{N_f}{N_c}\right)\left(\dw^{++}+\dw^{+-}\right) + 8\sqrt{4-\tfrac{2N_f}{N_c}}\left(\dw^{++} - \dw^{+-}\right)}{\omega\left(2-\tfrac{N_f}{N_c}\right)\sqrt{s_2(\omega)}} \Bigg]\,. \notag
\end{align}
\end{subequations}
Note that here, as in the previous Subsection, we opt to work with the variables defined prior to the rescaling done in \eq{resc} in order to avoid factors of $\sqrt{\bas}$. We have defined $y \equiv \sqrt{\bas}\ln(1/x)$ and $t \equiv \sqrt{\bas}\ln(Q^2/\Lambda^2)$ (cf.~\cite{Kovchegov:2023yzd}).

\begin{figure}[ht]
\begin{subfigure}[b]{0.45\textwidth}
    \centering
    \includegraphics[width=0.9\linewidth]{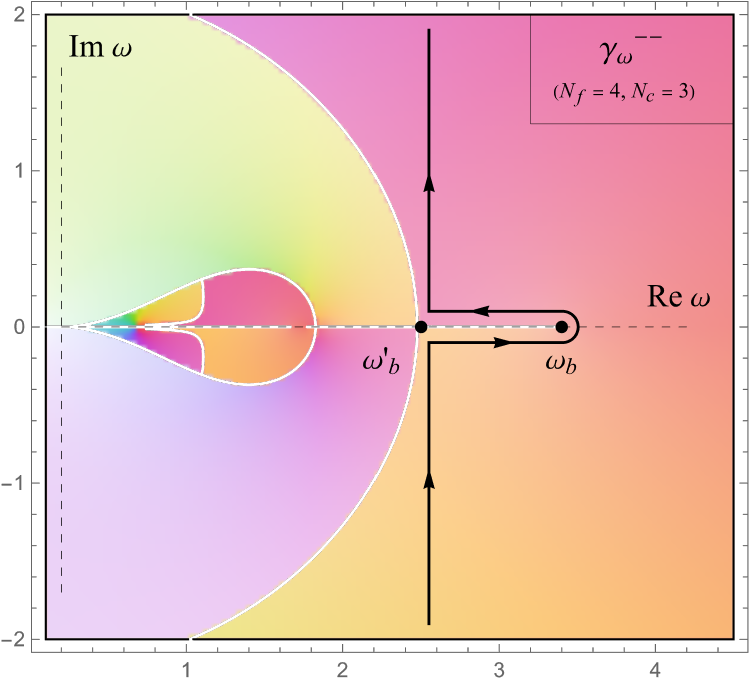}
    \label{fig:eigplotsmm}
    \caption{$\gw^{--}$}
\end{subfigure}
\begin{subfigure}[b]{.45\textwidth}
    \centering
    \includegraphics[width=0.9\linewidth]{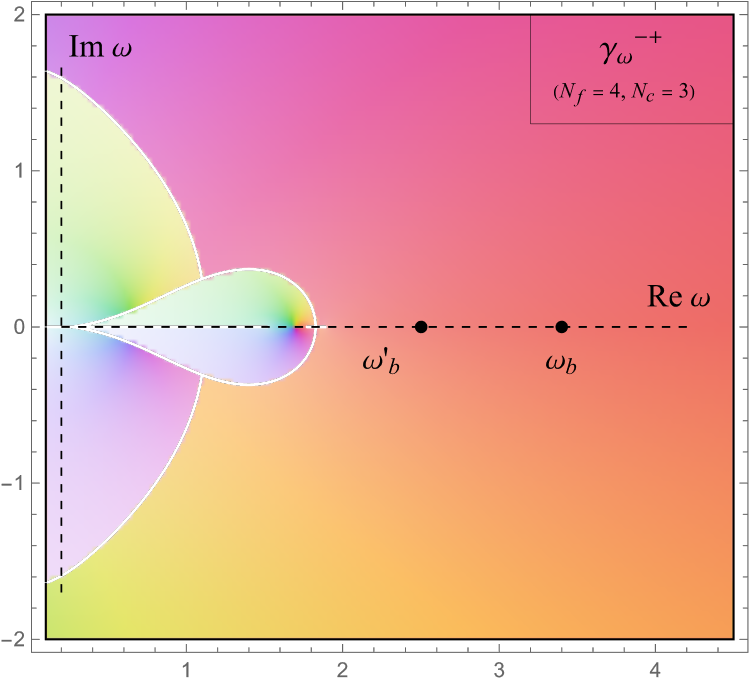}
    \label{fig:eigplotsmp}
    \caption{$\gw^{-+}$}
\end{subfigure}
    \caption{Complex $\omega$-plane structure of the eigenvalues of the anomalous dimension matrix for $N_f = 4$ and $N_c=3$. The left plot shows the structure of $\gw^{--}$ with the distorted inverse-Laplace integration contour overlaid (black solid line), while the right shows the structure of $\gw^{-+}$ without the integration contour. White lines indicate the branch cuts, while black dashed lines denote the axes. The two rightmost branch points of $\gw^{--}$ are denoted by $\omega_b$ and $\omega_b'$. For comparison, they are also shown on the right panel as well: we observe that $\gw^{-+}$ has no discontinuity in the $\omega_b' \le$~Re~$\omega \le \omega_b$ region or to the right of that region. Note that in these plots, colors correspond to the Arg of the plotted function, while the intensity of color corresponds to magnitude (paler color corresponds to larger magnitude).}
    \label{fig:eigplots}
\end{figure}

In order to obtain a more detailed description of the asymptotic behavior of these helicity distributions, we need to approximate the $\omega$ integrals in Eqs.~\eqref{preasy_pdfs} in the vicinity of the rightmost branch point $\omega_b$, which itself comes from the function $\gw^{--}$. In \fig{fig:eigplots} we show the graphs illustrating the structure of $\gw^{--}$ and $\gw^{-+}$ in the complex $\omega$-plane, concentrating on the region  around the right-most singularity $\omega_b$ of $\gw^{--}$. Branch cuts are denoted by white lines, while black dashed lines denote the axes and the solid black line in the left panel of \fig{fig:eigplots} denotes the integration contour, with $\omega'_b$ the sub-leading branch point of $\gw^{--}$. 

As shown in \fig{fig:eigplots}, we can wrap the integration contour around the leading branch point $\omega_b$. Then the vertical segments of the integration contour are sub-leading, and the $\omega$ integral can be approximated as the discontinuity across the leading branch cut on the real axis between $\omega_b'$ and $\omega_b$. Denoting the integrands, including pre-factors, of Eqs.~\eqref{preasy_pdfs} as $\Delta\Sigma_{\omega}$ and $\Delta G_{\omega}$, we can write (see Eqs.~(B6) in \cite{Kovchegov:2023yzd} with the overall sign corrected)
\begin{subequations}\label{preasy_disc}
\begin{align}
    &\Delta\Sigma(y,t) \approx - \lim_{\epsilon\rightarrow 0^+} \int\limits_0^{\infty}\frac{\mathrm{d}\xi}{2\pi i}\left(\Delta\Sigma_{\omega_b-\xi+i\epsilon} - \Delta\Sigma_{\omega_b-\xi-i\epsilon}\right) \,,\\
    &\Delta G(y,t) \approx - \lim_{\epsilon\rightarrow 0^+} \int\limits_0^{\infty}\frac{\mathrm{d}\xi}{2\pi i}\left(\Delta G_{\omega_b-\xi+i\epsilon} - \Delta G_{\omega_b-\xi-i\epsilon}\right) \,,
\end{align}
\end{subequations}
where we have defined $\omega = \omega_b - \xi$. Formally, we would integrate along the horizontal parts of the contour from $\omega = \omega_b$ to $\omega = \omega_b'$ or vice versa (equivalently, from $\xi = 0$ to $\xi = \omega_b - \omega_b'$). However, the factor $e^{\omega y}$ in Eqs.~\eqref{preasy_pdfs} ensures that the dominant contributions to the integrals come from larger values of $\omega$, so we can safely send $\omega_b'\rightarrow -\infty$ (equivalently, $\xi\rightarrow \infty$).

Since none of the other functions of $\omega$ involved in Eqs.~\eqref{preasy_pdfs} contain discontinuities in the immediate vicinity of the rightmost branch point $\omega_b$ and the connected branch cut, the relevant discontinuity here is only that in $\gw^{--}$. Using Eqs.~\eqref{preasy_pdfs} we can write the discontinuity in each helicity distribution across this branch cut as
\begin{subequations}\label{preasy_disc1}
\begin{align}
    &\Delta\Sigma_{\omega+i\epsilon} - \Delta\Sigma_{\omega-i\epsilon} = \frac{N_f}{\as 2\pi^2} e^{\omega y }\frac{1}{\omega} \left(e^{t\gamma^{--}_{\omega+i\epsilon}} - e^{t\gamma^{--}_{\omega-i\epsilon}} \right)\\
    &\hspace{5.5cm}\times\frac{8\omega\left(2 + \tfrac{N_f}{N_c}\right) - 16\left(\dw^{++}+\dw^{+-}\right) + 8\sqrt{4-\tfrac{2N_f}{N_c}}\left(\dw^{++}-\dw^{+-} \right)}{\omega\left(2-\tfrac{N_f}{N_c}\right)\sqrt{s_2(\omega)}} \,, \notag \\
    &\Delta G_{\omega+i\epsilon} - \Delta G_{\omega-i\epsilon} = \frac{N_c}{\as\pi^2} e^{\omega y }\frac{1}{\omega} \left(e^{t\gamma^{--}_{\omega+i\epsilon}} - e^{t\gamma^{--}_{\omega-i\epsilon}} \right) \\
    &\hspace{5.5cm}\times\left(1 - \frac{3\omega\left(6+\tfrac{N_f}{N_c}\right) - 2\left(8-\tfrac{N_f}{N_c}\right)\left(\dw^{++}+\dw^{+-}\right) + 8\sqrt{4-\tfrac{2N_f}{N_c}}\left(\dw^{++} - \dw^{+-}\right)}{\omega\left(2-\tfrac{N_f}{N_c}\right)\sqrt{s_2(\omega)}} \right) \,. \notag
\end{align}
\end{subequations}
Employing this in Eqs.~\eqref{preasy_disc}, we write
\begin{subequations}\label{preasy_disc2}
\begin{align}
& \Delta\Sigma(y,t) \approx - \frac{N_f}{\as 2\pi^2}\lim_{\epsilon\rightarrow 0^+} \int\limits_0^\infty\frac{\mathrm{d}\xi}{2\pi i} \frac{e^{(\omega_b-\xi)y}}{\omega_b-\xi}\left(e^{t\gamma^{--}_{\omega_b-\xi+i\epsilon}} - e^{t\gamma^{--}_{\omega_b-\xi-i\epsilon}} \right) \\
    &\hspace{3cm}\times\frac{8\left(\omega_b-\xi\right)\left(2 + \tfrac{N_f}{N_c}\right) - 16\left(\delta_{\omega_b-\xi}^{++}+\delta_{\omega_b-\xi}^{+-}\right) + 8\sqrt{4-\tfrac{2N_f}{N_c}}\left(\delta_{\omega_b-\xi}^{++}-\delta_{\omega_b-\xi}^{+-} \right)}{\left(\omega_b-\xi\right)\left(2-\tfrac{N_f}{N_c}\right)\sqrt{s_2(\omega_b-\xi)}} \notag \,,\\
& \Delta G(y,t) \approx - \frac{N_c}{\as \pi^2}\lim_{\epsilon\rightarrow 0^+} \int\limits_0^\infty\frac{\mathrm{d}\xi}{2\pi i} \frac{e^{(\omega_b-\xi)y}}{\omega_b-\xi}\left(e^{t\gamma^{--}_{\omega_b-\xi+i\epsilon}} - e^{t\gamma^{--}_{\omega_b-\xi-i\epsilon}} \right) \\
    &\hspace{3cm}\times\left(1 - \frac{3\left(\omega_b-\xi\right)\left(6+\tfrac{N_f}{N_c}\right) - 2\left(8-\tfrac{N_f}{N_c}\right)\left(\delta_{\omega_b-\xi}^{++}+\delta_{\omega_b-\xi}^{+-}\right) + 8\sqrt{4-\tfrac{2N_f}{N_c}}\left(\delta_{\omega_b-\xi}^{++} - \delta_{\omega_b-\xi}^{+-}\right)}{\left(\omega_b-\xi\right)\left(2-\tfrac{N-f}{N_c}\right)\sqrt{s_2(\omega_b-\xi)}} \right) \,. \notag
\end{align}
\end{subequations}
Because $y\sim \ln(1/x)$ is very large, the integrals in Eqs.~\eqref{preasy_disc2} are dominated by small values of $\xi$. Hence, we can first expand the integrands in powers of $\xi$, then integrate the resulting expression term by term over $\xi$. We employ the following expansion of $\gamma^{--}_{\omega}$ around its branch point $\omega_b$:
\begin{align}\label{preasy_gammaexpansion}
    \gamma^{--}_{\omega_b-\xi\pm i\epsilon} = \frac{\omega_b}{2} \mp i\frac{\omega_b}{2}\sqrt{C^{(1)}(\omega_b)} \, \xi^{1/2} - \frac{\xi}{2} \pm i \, \frac{\omega_b}{8}\frac{C^{(2)}(\omega_b)}{\sqrt{C^{(1)}(\omega_b)}} \, \xi^{3/2} + \mathcal{O}\left(\xi^{5/2}\right) \,,
\end{align}
where we define
\begin{subequations}\label{C1andC2}
\begin{align}
    \label{C1}
    &C^{(1)}(\omega_b) \equiv \frac{1}{\omega_b^2}\left[s_1'(\omega_b) - \frac{s_2'(\omega_b)}{2\left[s_1(\omega_b)+\omega_b^2\right]} \right] + \frac{2}{\omega_b} \,,\\
    \label{C2}
    &C^{(2)}(\omega_b) \equiv \frac{1}{\omega_b^2}\left[s_1''(\omega_b) - \frac{s_2''(\omega_b)}{2\left[s_1(\omega_b)+\omega_b^2\right]} + \frac{\left[s_2'(\omega_b)\right]^2}{4\left[s_1(\omega_b) + \omega_b^2\right]^3} \right] + \frac{2}{\omega_b^2} \,.
    \end{align}
\end{subequations}
In Eqs.~\eqref{C1andC2}, the primes denote differentiation, the functions $s_1(\omega)$ and $s_2(\omega)$ are those defined in Eqs.~\eqref{gammapmpmfull12}, and we have used \eq{intercepteq} in several places to replace $\sqrt{s_2(\omega_b)}$ with $s_1(\omega_b) + \omega_b^2$ since it is somewhat easier to numerically evaluate the latter.

Using the expansion in \eq{preasy_gammaexpansion} in Eqs.~\eqref{preasy_disc2}, and also expanding the rest of the integrands (the parts multiplying $e^{\left(\omega_b-\xi\right)y}$) in $\xi$, we obtain
\begin{subequations}\label{preasy_disc3}
\begin{align}
    &\Delta \Sigma(y,t) \approx - \frac{N_f}{\as 2\pi^2} t e^{\omega_b\tfrac{t}{2}}\int\limits_0^\infty \frac{\mathrm{d}\xi}{2\pi}e^{\left(\omega_b-\xi\right)y} \frac{\omega_b\sqrt{C^{(1)}(\omega_b)}}{s_1(\omega_b) + \omega_b^2} \Bigg\{-F^{(\Delta\Sigma)}_{\omega_b}\xi^{1/2} \\
    &\hspace{.7cm}+ \bigg[F_{\omega_b}^{(\Delta\Sigma)\prime} -\frac{s_2'(\omega_b)}{2\left[s_1(\omega_b)+\omega_b^2 \right]^2}F^{(\Delta\Sigma)}_{\omega_b} + \left(\frac{C^{(2)}(\omega_b)}{4 \, C^{(1)}(\omega_b)}  + \frac{t}{2} + \frac{\omega_b^2}{24} \, t^2 \, C^{(1)}(\omega_b)\right) F^{(\Delta\Sigma)}_{\omega_b}\bigg] \xi^{3/2} + \mathcal{O}\left(\xi^{5/2} \right) \Bigg\} \,,\notag \\
    &\Delta G(y,t) \approx - \frac{N_c}{\as \pi^2} t e^{\omega_b\tfrac{t}{2}}\int\limits_0^\infty \frac{\mathrm{d}\xi}{2\pi}e^{\left(\omega_b-\xi\right)y} \frac{\omega_b\sqrt{C^{(1)}(\omega_b)}}{s_1(\omega_b) + \omega_b^2} \Bigg\{-\frac{s_1(\omega_b)+\omega_b^2}{\omega_b} +F^{(\Delta G)}_{\omega_b}\xi^{1/2} \\
    &\hspace{.7cm}+ \bigg[-\frac{s_1(\omega_b)+\omega_b^2}{\omega_b^2} -F_{\omega_b}^{(\Delta G)\prime} +\frac{s_2'(\omega_b)}{2\left[s_1(\omega_b)+\omega_b^2 \right]^2}F^{(\Delta G)}_{\omega_b} \notag \\
    &\hspace{1cm}+ \left( \frac{C^{(2)}(\omega_b)}{4 \, C^{(1)}(\omega_b)}  + \frac{t}{2} + \frac{\omega_b^2}{24} \, t^2 \, C^{(1)}(\omega_b)\right)\left(\frac{s_1(\omega_b)+\omega_b^2}{\omega_b}  - F^{(\Delta G)}_{\omega_b} \right)\bigg] \xi^{3/2} + \mathcal{O}\left(\xi^{5/2} \right) \Bigg\} \,.\notag
\end{align}
\end{subequations}
In Eqs.~\eqref{preasy_disc3}, primes again denote derivatives and we have defined, for brevity,
\begin{subequations}\label{FDeltaSigmaandFDeltaG}
\begin{align}
    \label{FDeltaSigma}
    &F^{(\Delta\Sigma)}_{\omega} \equiv \frac{8\omega\left(2+\tfrac{N_f}{N_c}\right) - 16\left(\dw^{++} + \dw^{+-}\right) + 8\sqrt{4-\frac{2N_f}{N_c}}\left(\dw^{++}-\dw^{+-}\right)}{\omega^2\left(2-\tfrac{N_f}{N_c}\right)} \,,\\
    &F^{(\Delta G)}_{\omega} \equiv \frac{3\omega\left(6+\tfrac{N_f}{N_c}\right) - 2\left(8-\tfrac{N_f}{N_c}\right)\left(\dw^{++} + \dw^{+-}\right) + 8\sqrt{4-\frac{2N_f}{N_c}}\left(\dw^{++}-\dw^{+-}\right)}{\omega^2\left(2-\tfrac{N_f}{N_c}\right)}\,.
\end{align}
\end{subequations}
Next, we carry out the integrals over $\xi$ in Eqs.~\eqref{preasy_disc3} to obtain the full approximation for the asymptotic behavior of $\Delta\Sigma(y,t)$ and $\Delta G(y,t)$. The results can be written as (cf.~\cite{Kovchegov:2023yzd})
\begin{subequations}\label{preasymptoticexpansion}
\begin{align}
    \label{deltasigmapreasym}
    &\Delta \Sigma(y,t) \approx \left[\frac{d_{1,q}(t)}{y^{3/2}} + \frac{d_{2,q}(t)}{y^{5/2}} + \mathcal{O}\left(\frac{1}{y^{7/2}} \right) \right]e^{\omega_b y} \,,\\
    \label{deltagpreasym}
    &\Delta G(y,t) \approx \left[\frac{d_{1,G}(t)}{y^{3/2}} + \frac{d_{2,G}(t)}{y^{5/2}} + \mathcal{O}\left(\frac{1}{y^{7/2}} \right) \right]e^{\omega_b y} \,,
\end{align}
\end{subequations}
with the expansion coefficients given by
\begin{subequations}\label{expansioncoefsds}
\begin{align}
    \label{d1q}
    &d_{1,q}(t) = \frac{N_f}{\as 8 \pi^{5/2}}te^{\omega_b \tfrac{t}{2}} \,\frac{\omega_b\sqrt{C^{(1)}(\omega_b)}}{s_{1}(\omega_b) + \omega_b^2} F^{(\Delta\Sigma)}_{\omega_b} \,,\\
    \label{d2q}
    &d_{2,q}(t) = - \frac{3}{16}\frac{N_f}{\as \pi^{5/2}}te^{\omega_b\tfrac{t}{2}}\,\frac{\omega_b\sqrt{C^{(1)}(\omega_b)}}{s_{1}(\omega_b) + \omega_b^2} \bigg\{ F^{(\Delta\Sigma)\prime}_{\omega_b} - \frac{F^{(\Delta\Sigma)}_{\omega_b}}{2\left[s_1(\omega_b)+\omega_b^2\right]^2}s_2'(\omega_b) \\
    &\hspace{6cm} + F^{(\Delta\Sigma)}_{\omega_b}\left[ \frac{C^{(2)}(\omega_b)}{4 \, C^{(1)}(\omega_b)}  + \frac{t}{2} + \frac{\omega_b^2}{24}\, t^2 \, C^{(1)}(\omega_b)\right]  \bigg\} \notag \,,\\
    \label{d1G}
    &d_{1,G}(t) = \frac{N_c}{\as 4 \pi^{5/2}}te^{\omega_b \tfrac{t}{2}} \,\frac{\omega_b\sqrt{C^{(1)}(\omega_b)}}{s_{1}(\omega_b) + \omega_b^2}\left[\frac{s_1(\omega_b)+\omega_b^2}{\omega_b} - F^{(\Delta G)}_{\omega_b}\right] \,,\\
    \label{d2G}
    &d_{2,G}(t) = - \frac{3}{8}\frac{N_c}{\as \pi^{5/2}}te^{\omega_b\tfrac{t}{2}}\,\frac{\omega_b\sqrt{C^{(1)}(\omega_b)}}{s_{1}(\omega_b) + \omega_b^2} \bigg\{ -\frac{s_1(\omega_b)+\omega_b^2}{\omega_b^2} - F^{(\Delta G)\prime}_{\omega_b} + \frac{F^{(\Delta G)}_{\omega_b}}{2\left[s_1(\omega_b)+\omega_b^2\right]^2}s_2'(\omega_b) \\
    &\hspace{6cm} + \left[\frac{s_1(\omega_b)+\omega_b^2}{\omega_b} - F^{(\Delta G)}_{\omega_b} \right]
    \left[ \frac{C^{(2)}(\omega_b)}{4 \, C^{(1)}(\omega_b)}  + \frac{t}{2} + \frac{\omega_b^2}{24} \, t^2 \, C^{(1)}(\omega_b)\right]  \bigg\} \,, \notag
\end{align}
\end{subequations}
where, as a reminder, we have defined $y = \sqrt{\bas}\ln(1/x)$ and $t = \sqrt{\bas}\ln(Q^2/\Lambda^2)$. Thus in Eqs.~\eqref{preasymptoticexpansion} and \eqref{expansioncoefsds} we have obtained fully analytic expressions for the functional forms of $\Delta \Sigma$ and $\Delta G$ in the high energy asymptotic limit.

Note that Eqs.~\eqref{preasymptoticexpansion} and \eqref{expansioncoefsds} are valid for the rightmost branch point $\omega_b$, which can be found for any $N_f$ and $N_c$. We can use the values of $\omega_b$ we obtained in Table~\ref{tab:intercepts} to numerically compute the expansion coefficients in Eqs.~\eqref{expansioncoefsds}. We can also consider the asymptotic ratio of $\Delta G$ to $\Delta \Sigma$ by computing the ratio $d_{1,G}(t)/d_{1,q}(t)$ (this ratio was previously considered in \cite{Hatta:2018itc, Hatta:2016aoc, Boussarie:2019icw, Kovchegov:2023yzd}). Analytically, this ratio can be written using Eqs.~\eqref{d1q} and \eqref{d1G} in a relatively simple form,
\begin{align}\label{asymptratioanalytic}
    \left(\frac{\Delta G}{\Delta \Sigma}\right)^{(\text{asympt})} \equiv \frac{d_{1,G}(t)}{d_{1,q}(t)} = \frac{N_c}{4N_f}\,\frac{\tfrac{-4N_f}{N_c}\left(\delta_{\omega_b}^{+-}+\delta_{\omega_b}^{++}\right) + 2\omega_b\left(2+\tfrac{3N_f}{N_c}\right) -\omega_b^3\left(2-\tfrac{N_f}{N_c}\right)}{2\left(\delta_{\omega_b}^{+-} +\delta_{\omega_b}^{++}\right) + \left(\delta_{\omega_b}^{+-}-\delta_{\omega_b}^{++}\right)\sqrt{4-\tfrac{2N_f}{N_c}} - \omega_b\left(2+\tfrac{N_f}{N_c}\right)}\,.
\end{align}
In Table~\ref{tab:preasymptotics} we show, for various choices of $N_f$ with $N_c = 3$, the leading branch point $\omega_b$ (reproduced from Table~\ref{tab:intercepts}) along with the asymptotic ratio from \eq{asymptratioanalytic}, evaluated numerically for each choice of $N_f$.
\begin{table}[h!]
\centering
\begin{tabular}{|c|c|c|}
\hline
$N_f$ & $\omega_b$ & $(\Delta G/\Delta \Sigma)^{\text{(asympt)}}$  \\
\hline
2 & 3.5452 & -4.7871 \\
3 & 3.4791 & -3.0731 \\
4 & 3.4051 & -2.2075 \\
5 & 3.3204 & -1.6786 \\
6 & 3.2193 & -1.3143 \\
7 & 3.0895 & -1.0364 \\
8 & 2.8823 & -0.7872 \\
\hline
\end{tabular}
\caption{Table of small-$x$ intercepts and asymptotic ratios of $\Delta G$ to $\Delta \Sigma$ for values of $N_f$ with $N_c = 3$.}
\label{tab:preasymptotics}
\end{table}
For $N_f = 4$ and $N_c = 3$, it was predicted in \cite{Kovchegov:2023yzd} from the large-$N_c$ version of the small-$x$ helicity evolution that the asymptotic relation between the hPDFs is $\Delta G(y,t) \approx - 3 \, \Delta\Sigma(y,t)$. In \cite{Boussarie:2019icw} it was found, generalizing the formalism of BER (and working for any $N_c$ and $N_f$) that asymptotically $\Delta G(y,t) \approx - 2.29 \, \Delta\Sigma(y,t)$ when $N_f = 4$, $N_c=3$. As can be seen in Table~\ref{tab:preasymptotics}, our prediction here, based on the most recent version of the large-$N_c\&N_f$ small-$x$ helicity evolution, is the asymptotic relation $\Delta G(y,t) \approx - 2.21 \, \Delta\Sigma(y,t)$ for the same $N_f = 4$, $N_c=3$. We can see that considering the large-$N_c\&N_f$ version of the small-$x$ evolution has brought us closer to the predictions based on the BER formalism, although, as with the intercepts and the polarized DGLAP anomalous dimensions, small disagreements still persist and probably cannot be entirely attributed to us working in the large-$N_c\&N_f$ approximation with BER not employing this limit.


\subsection{Asymptotic Behavior: The Saddle Point Method}

As a complimentary cross-check for the results of the previous Section, we re-derive here the leading terms of the asymptotic expansion of the hPDFs in Eqs.~\eqref{preasymptoticexpansion}. In the previous Section we integrated the discontinuities of the integrands across the leading branch cut, but in this Section we will alternatively employ the saddle point method. The saddle point technique was recently used in \cite{Ermolaev:2025isl} to determine the small-$x$ asymptotics of hPDFs in the BER framework: that work inspired us to apply it here as well.\footnote{We would like to thank Boris Ermolaev for the correspondence which motivated us to explore the saddle point approach to hPDFs small-$x$ asymptotics.} Taking the same initial conditions from \eq{DGLAPicsfordipoles}, we begin with the hPDFs in Eqs.~\eqref{preasy_pdfs}. Making use of the notation in Eqs.~\eqref{FDeltaSigmaandFDeltaG}, we can write more compactly
\begin{subequations}\label{hpdfssaddlepoint}
\begin{align}
    \label{deltasigmacompact}
    &\Delta\Sigma(y,t) = \frac{N_f}{\as 2\pi^2}\wint \, e^{\omega y} \left(e^{\gamma^{--}_{\omega}\,t} - e^{\gamma^{-+}_{\omega}\,t} \right)\frac{F^{(\Delta\Sigma)}_\omega}{\sqrt{s_2(\omega)}}\,, \\
    \label{deltaGcompact}
    &\Delta G(y,t) = \frac{N_c}{\as\pi^2}\wint \, e^{\omega y} \, \frac{1}{\omega}\left[e^{\gamma^{--}_\omega\,t}\left(1 - \frac{\omega F^{(\Delta G)}_{\omega}}{\sqrt{s_2(\omega)}}   \right) + e^{\gamma^{-+}_\omega\,t}\left(1 + \frac{\omega F^{(\Delta G)}_{\omega}}{\sqrt{s_2(\omega)}}   \right)  \right]\,.
\end{align}
\end{subequations}
The saddle point of the terms containing the exponential $e^{\omega y + \gamma_{\omega}^{--}t}$ for large $y$ and $t$ is determined by
\begin{align}\label{saddlepointdiffeq}
    \frac{\mathrm{d}}{\mathrm{d}\omega}\left(\omega \, y + \gamma^{--}_\omega t\right) = 0 .
\end{align}
We will denote the saddle point (that is, the solution of \eq{saddlepointdiffeq}) by $\omega = \omega_{sp}$. One can show that a saddle-point evaluation of the terms containing the exponential $e^{\omega y + \gamma_{\omega}^{-+}t}$ leads to a sub-leading contribution at large $y$ in each of Eqs.~\eqref{hpdfssaddlepoint}, as compared to the terms containing $e^{\omega y + \gamma_{\omega}^{--}t}$: we can, therefore, discard the terms containing $e^{\omega y + \gamma_{\omega}^{-+}t}$ in our evaluation below.

One can further show that $\omega_{sp}$ defined by \eq{saddlepointdiffeq} lies in the vicinity (and to the right) of the leading branch point $\omega_b$ considered in the previous Section (as the ratio $y/t$ approaches infinity, $\omega_{sp}$ can be seen to approach $\omega_b$). Therefore, to employ the saddle method in order to approximate the integrals in Eqs.~\eqref{hpdfssaddlepoint} we can again employ \eq{preasy_gammaexpansion}, which contains our expansion of $\gamma^{--}_{\omega}$ near the leading branch point $\omega_b$. Employing \eq{preasy_gammaexpansion}, while now using $\omega = \omega_b + \xi$ (note the relative sign) and truncating the expansion earlier, we write
\begin{align}\label{gammammexpansionsaddle}
    \gamma^{--}_{\omega_b+\xi} = \frac{\omega_b}{2} - \frac{\omega_b}{2}\sqrt{C^{(1)}(\omega_b)} \, \xi^{1/2} + \frac{\xi}{2} + \mathcal{O}\left(\xi^{3/2}\right)\,.
\end{align}
Using \eq{gammammexpansionsaddle} in \eq{saddlepointdiffeq}, and defining $\omega_{sp} = \omega_b+\xi_{sp}$, we find
\begin{align}\label{xisp}
    \xi_{sp} = \frac{\omega_b^2 \, C^{(1)}(\omega_b)}{4 \, \left(1+\frac{2 \, y}{t}\right)^2} \approx \frac{\omega_b^2 \, C^{(1)}(\omega_b) \, t^2}{16 \, y^2}.
\end{align}
In the last step we have assumed that $y \gg t$, as is proper for high-energy asymptotics: in this limit, indeed, $\xi_{sp}$ is small, and the saddle point $\omega_{sp} = \omega_b + \xi_{sp}$ is close to the leading branch point $\omega_b$.

Neglecting the terms proportional to $e^{\gamma_\omega^{-+}t}$, we evaluate the hPDFs in Eqs.~\eqref{hpdfssaddlepoint} around the saddle point $\omega_{sp}$, obtaining
\begin{subequations}\label{hpdfssaddlepoint1}
\begin{align}
    &\Delta\Sigma(y,t) \approx \frac{N_f}{\as 2\pi^2} \wint \, e^{\omega_{sp}y + \gamma^{--}_{\omega_{sp}}\,t + \tfrac{1}{2}t\left(\omega-\omega_{sp}\right)^2 \left(\gamma^{--}_{\omega_{sp}}\right)''} \frac{F^{(\Delta\Sigma)}_{\omega_{sp}}}{\sqrt{s_2(\omega_{sp})}} \,,\\
    &\Delta G(y,t) \approx \frac{N_c}{\as \pi^2} \wint  \, e^{\omega_{sp}y + \gamma^{--}_{\omega_{sp}}\,t + \tfrac{1}{2}t\left(\omega-\omega_{sp}\right)^2 \left(\gamma^{--}_{\omega_{sp}}\right)''} \frac{1}{\omega_{sp}} \left(1 - \frac{\omega_{sp} F^{(\Delta G)}_{\omega_{sp}}}{\sqrt{s_2(\omega_{sp})}}\right)\,.
\end{align}
\end{subequations}
Now we evaluate the integrands around the saddle point. Employing the expansion in \eq{gammammexpansionsaddle} along with the saddle point $\omega_{sp} = \omega_b + \xi_{sp}$, with $\xi_{sp}$ in \eq{xisp}, we write
\begin{subequations}\label{hpdfssaddlepoint2}
\begin{align}
    &\Delta\Sigma(y,t) \approx \frac{N_f}{\as2\pi^2} \frac{F^{(\Delta \Sigma)}_{\omega_b}}{\sqrt{s_2(\omega_b)}} e^{\omega_b y +\tfrac{\omega_b}{2}t - \tfrac{\omega_b^2 C^{(1)}(\omega_b)}{16}\tfrac{t^2}{y} }
    \int\limits_{-\infty}^{\infty}\frac{\mathrm{d}\nu}{2\pi} \, \exp \left\{-\frac{4 \, y^3}{\omega_b^2 \, C^{(1)}(\omega_b) \, t^2} \nu^2\right\} \,,\\
    &\Delta G(y,t) \approx \frac{N_c}{\as\pi^2}\frac{1}{\omega_b}\left(1 - \frac{\omega_b F^{(\Delta G)}_{\omega_b}}{\sqrt{s_2(\omega_b)}}\right) e^{\omega_b y +\tfrac{\omega_b}{2}t - \tfrac{\omega_b^2 C^{(1)}(\omega_b)}{16}\tfrac{t^2}{y} }
    \int\limits_{-\infty}^{\infty}\frac{\mathrm{d}\nu}{2\pi} \, \exp \left\{- \frac{4 \, y^3 }{\omega_b^2 \, C^{(1)}(\omega_b) \, t^2} \nu^2\right\}\,,
\end{align}
\end{subequations}
where we integrate along the vertical contour $\omega = \omega_b + i\nu$. Since we are interested in reproducing the leading high-$y$ asymptotics here, we have neglected higher powers of $\xi_{sp}$ in the parts of the integrands multiplying the exponentials: these pre-factors are now outside the $\nu$-integrals.

Carrying out the Gaussian integrals we arrive at the small-$x$ asymptotics of hPDFs,
\begin{subequations}\label{hpdfssaddlepoint3}
\begin{align}
    &\Delta \Sigma(y,t) = \frac{e^{\omega_b y }}{y^{3/2}} \frac{N_f}{\as 8 \pi^{5/2}}\,t\,\,e^{\tfrac{\omega_b}{2}t - \tfrac{\omega_b^2 C^{(1)}(\omega_b)}{16}\tfrac{t^2}{y}}\,\, \frac{\omega_b\sqrt{C^{(1)}(\omega_b)}}{s_1(\omega_b) + \omega_b^2 } F^{(\Delta\Sigma)}_{\omega_b} \,,\\
    &\Delta G(y,t) = \frac{e^{\omega_b y}}{y^{3/2}} \frac{N_c}{\as 4\pi^{5/2}}\,t\,\,e^{\tfrac{\omega_b}{2}t - \tfrac{\omega_b^2 C^{(1)}(\omega_b)}{16}\tfrac{t^2}{y}}\,\,\frac{\omega_b\sqrt{C^{(1)}(\omega_b)}}{s_1(\omega_b) + \omega_b^2}\left(\frac{s_1(\omega_b) + \omega_b^2}{\omega_b} - F^{(\Delta G)}_{\omega_b}\right)\,,
\end{align}
\end{subequations}
where we employed the same substitution $\sqrt{s_2(\omega_b)} \rightarrow s_1(\omega_b) + \omega_b^2$ used in the previous Subsection, since the two quantities are equal at the branch point $\omega_b$, per \eq{intercepteq}. Note that in addition to the leading term in $y$ we have also obtained a diffusion term in the exponents of both of Eqs.~\eqref{hpdfssaddlepoint3},
\begin{align}\label{diffusionterm}
    \exp \left(-\frac{\omega_b^2C^{(1)}(\omega_b)}{16} \frac{t^2}{y}\right)\,.
\end{align}
This term is completely analogous to the similar diffusion term in the solution of the unpolarized Balitsky--Fadin--Kuraev--Lipatov (BFKL) \cite{Kuraev:1977fs,Balitsky:1978ic} evolution equation. If we neglect this diffusion term, putting the exponential in \eq{diffusionterm} equal to 1, Eqs.~\eqref{hpdfssaddlepoint3} would then exactly reproduce the first terms of the expansion in Eqs.~\eqref{preasymptoticexpansion} (with the relevant coefficients found in Eqs.~\eqref{d1q} and \eqref{d1G}). Furthermore, expanding the exponential in \eq{diffusionterm} to linear order in $t^2$ and using the result in Eqs.~\eqref{hpdfssaddlepoint3} yields exactly the last (order-$t^3$ in the pre-factor) term of each of $d_{2,q}(t)$ and $d_{2,G}(t)$ in Eqs.~\eqref{d2q} and \eqref{d2G}, showing that our method of integration across the leading branch cut in the previous Section also captured (parts of) this diffusion term. Conversely, the diffusion term appears to capture all the leading-power of $t$ terms in the coefficients of the $1/y$ expansion of the pre-factors in Eqs.~\eqref{preasymptoticexpansion}.  

The fact that, unlike the BFKL case, the helicity evolution allowed us to obtain the explicit expressions \eqref{alladims} for the corresponding anomalous dimensions (cf.~\cite{Borden:2023ugd}) enabled us here to perform the asymptotic analyses both in the branch-cut integral and saddle-point methods.


\section{Summary and Conclusions}\label{sec:conclusion}

In this paper we have analytically solved the most up-to-date version of the small-$x$ helicity evolution in the large-$N_c\&N_f$ limit, containing the quark-to-gluon and gluon-to-quark transition operators included in \cite{Borden:2024bxa}. Our solution is based on a double-inverse Laplace transform method and the complete solution of the evolution equations is presented in Eqs.~\eqref{fullsoln} and \eqref{fullsoln1}, yielding analytic expressions for all of the polarized dipole amplitudes. We have also explicitly constructed the corresponding analytic double-inverse Laplace transform expressions for the gluon and flavor singlet quark helicity PDFs, along with the $g_1$ structure function in Eqs.~\eqref{fullsolnpdfs} and \eqref{fullsolng1}. 

We have successfully cross-checked our solution against the known solution to the spin-dependent DGLAP equations, and in doing so we have extracted analytic predictions for the (fully resummed in powers of $\as/\omega^2$) eigenvalues of the matrix of DGLAP polarized anomalous dimensions, and subsequently obtained analytic predictions for all four individual polarized anomalous dimensions themselves ($\Delta\gamma_{GG}(\omega)$, $\Delta\gamma_{qq}(\omega)$, $\Delta\gamma_{qG}(\omega)$, and $\Delta\gamma_{Gq}(\omega)$): the results for the anomalous dimensions are given above in Eqs.~\eqref{alladims}.  All of these are fully consistent with the known finite-order results~\cite{Altarelli:1977zs,Dokshitzer:1977sg,Mertig:1995ny,Moch:2014sna}.

We have also obtained numerical values for the intercept of the helicity distributions in Table~\ref{tab:intercepts} (see also Eqs.~\eqref{DSGg_asymptotics} and \eqref{alphah}), along with a more detailed description of the asymptotic behavior of these distributions in Eqs.~\eqref{preasymptoticexpansion} and \eqref{hpdfssaddlepoint3}. In doing so, we found explicit analytic and numerical predictions (at large-$N_c\&N_f$) for the asymptotic ratio of the gluon helicity PDF to the flavor-singlet quark helicity PDF at small-$x$, presented in Table~\ref{tab:preasymptotics}.

Just as with the analytic solution constructed in \cite{Borden:2023ugd} for the large-$N_c$ version of the small-$x$ helicity evolution equations, we find here in the large-$N_c\&N_f$ limit the same general trend of full agreement with finite-order calculations and very minor disagreements with the predictions of the BER formalism \cite{Bartels:1996wc} (beyond the existing 3-loop precision of finite-order work). In particular, our polarized DGLAP anomalous dimensions are fully consistent with finite-order calculations to the existing three loops \cite{Altarelli:1977zs,Dokshitzer:1977sg,Mertig:1995ny,Moch:2014sna} (up to a scheme transformation \cite{Moch:2014sna, Borden:2024bxa}) and agree completely with those predicted in the BER formalism to three loops \cite{Bartels:1996wc}. However, our anomalous dimensions and those of BER \cite{Bartels:1996wc,Blumlein:1996hb} disagree beginning at the four-loop level. The difference was already present in the analytic solution of the large-$N_c$ helicity evolution equations \cite{Borden:2023ugd}, which explored the pure gluonic sector: therefore, it cannot be removed by a scheme transformation. We find a similarly small disagreement in the numerical values of the intercepts predicted by our solution and those predicted by BER, along with a small disagreement in the asymptotic ratio of $\Delta G$ to $\Delta \Sigma$ \cite{Bartels:1996wc, Boussarie:2019icw}.

Future calculations in the finite-order framework for the four-loop polarized DGLAP splitting functions can resolve the existing discrepancy between the predictions in this work and those of the BER IREE framework. Nevertheless, with the most general (large-$N_c\&N_f$) limit of the small-$x$ helicity evolution \cite{Kovchegov:2015pbl, Kovchegov:2016zex,  Kovchegov:2017lsr, Kovchegov:2018znm, Cougoulic:2019aja, Cougoulic:2022gbk, Borden:2024bxa} now completely solved, demonstrating agreement with finite-order calculations up to the existing three loops \cite{Altarelli:1977zs,Dokshitzer:1977sg,Mertig:1995ny,Moch:2014sna}, we hope the predictions herein convincingly demonstrate that the small-$x$ helicity formalism is a robust and accurate tool to further push the bounds of both theoretical and phenomenological work constraining the spin content of the proton at small $x$.


\vspace*{-6mm}
\section{Acknowledgments}

The authors are grateful to Boris Ermolaev, Ming Li, and Brandon Manley for informative discussions. 

This material is based upon work supported by
the U.S. Department of Energy, Office of Science, Office of Nuclear Physics under Award Number DE-SC0004286 and within the framework of the Saturated Glue (SURGE) Topical Theory Collaboration.


\vspace{5mm}


\begingroup\raggedright\endgroup

\end{document}